\documentclass[12pt]{article}
\pdfoutput=1
\usepackage[cmtip,arrow]{xy}\usepackage{pb-diagram,pb-xy}%

\input xy
\xyoption{all}
\input xy
\xyoption{all}
\usepackage{heck}
\usepackage{cite}
\usepackage{graphicx}
\usepackage{makeidx}
\usepackage{multicol}
\usepackage{amsfonts}
\usepackage{mathrsfs}
\usepackage{amssymb}
\usepackage{amsmath}%
\usepackage{array}

\usepackage{hyperref}
\hypersetup{colorlinks,%
citecolor=black,%
filecolor=black,%
linkcolor=black,%
urlcolor=black,%
pdftex}

\providecommand{\U}[1]{\protect\rule{.1in}{.1in}}
\numberwithin{equation}{section}

\def\a{\alpha}
\def\b{\beta}
\def\c{\gamma}
\def\d{\delta}

\def\l{\lambda}

\def\U{\Upsilon}

\def\ca{{\cal A}}
\def\cb{{\cal B}}
\def\cc{{\cal C}}
\def\cd{{\cal D}}

\def\cn{{\cal N}}
\def\co{{\cal O}}
\def\cp{{\cal P}}

\def\ct{{\cal T}}

\def\cw{{\cal W}}
\def\cx{{\cal X}}
\def\cy{{\cal Y}}

\def \Z {{\mathbb Z}}
\def \C {{\mathbb C}}

\newcommand{\bea}{\begin{equation}\begin{aligned}}
\newcommand{\eea}{\end{aligned}\end{equation}}
\newcommand{\beq}{\begin{eqnarray}}
\newcommand{\eeq}{\end{eqnarray}}
\newcommand{\be}{\begin{equation}}
\newcommand{\ee}{\end{equation}}
\newcommand{\bem}{\begin{pmatrix}}
\newcommand{\eem}{\end{pmatrix}}

\xyoption{arc}

\date{March, 2013}

\institution{SISSA}{\   Scuola Internazionale Superiore di Studi Avanzati, via Bonomea 265, Trieste, ITALY}
\institution{INFNTS}{\ INFN, Sezione di Trieste, via Valerio 2, Trieste, ITALY}
\institution{SNS}{\ Scuola Normale Superiore, piazza dei Cavalieri 7, Pisa, ITALY }
\institution{INFNP}{\ INFN, Sezione di Pisa, Largo Pontecorvo 3, Pisa, ITALY}

\title{More on the $\cn=2$ superconformal\\ systems of type $D_p(G)$}
\authors{Sergio Cecotti \worksat{\SISSA} \footnote{e-mail: {\tt cecotti@sissa.it}}, Michele Del Zotto \worksat{\SISSA,\INFNTS} \footnote{e-mail: {\tt eledelz@gmail.com}}, and Simone Giacomelli \worksat{\SNS,\INFNP} \footnote{e-mail: {\tt si.giacomelli@sns.it}}}

\abstract{A large family of 4d $\cn=2$ SCFT's was introduced in {\tt 1210.2886}. Its elements $D_p(G)$ are labelled by a positive integer $p\in\mathbb{N}$ and a simply--laced Lie group $G$; their flavor symmetry is at least $G$. 
In the present paper we study their physics in detail. 
We also analyze the properties of the theories obtained by gauging the diagonal symmetry of a collection of $D_{p_i}(G)$ models. 
In all cases the computation of the physical quantities reduces to simple Lie--theoretical questions. 

To make the analysis more functorial, we replace the notion of the BPS--quiver of the $\cn=2$ QFT by the more intrinsic concept of its \textsf{META}--quiver.

In particular: 1) We compute the SCFT central charges $a$, $c$, $k$, and flavor group $F$ for all $D_p(G)$ models. 
2) We identify the subclass of $D_p(G)$ theories which correspond to previously known SCFT's (linear $SU$ and $SO$-$USp$ quiver theories, Argyres--Douglas models, superconformal gaugings of Minahan--Nemeshanski $E_r$ models, \emph{etc.}), as well as to non--trivial IR fixed points of known theories. The $D_p(E_r)$ SCFT's with $p\geq 3$ \textit{cannot} be constructed by any traditional method. 3) We investigate the finite BPS chambers of some of the models. 4) As a by product, we prove three conjectures by Xie and Zhao, and provide new checks of the Argyres--Seiberg duality.
}

\begin{document}

\maketitle

\tableofcontents


\section{Introduction}

The four--dimensional $\cn=2$ supersymmetric theories constitute an important theoretical laboratory for the non--perturbative analysis of quantum field theories at strong coupling \cite{SW1,SW2,Gaiotto}. Of particular interest are the $\cn=2$ superconformal theories (SCFT) having a large global symmetry group $F$, especially when $F$ is an exceptional Lie group. Many interesting examples of such SCFT's are known, most of them without a weakly coupled Lagrangian formulation \cite{MN1,MN2,ganor,benini}.\smallskip

A big family of $\cn=2$ SCFT's with large flavor groups was constructed in ref.\!\cite{infinitelymany}. The models in the family, which we denote as $D_p(G)$, are labelled by an integer $p\in\mathbb{N}$, called the `period', and a simply--laced Lie group $G=ADE$. The SCFT $D_p(G)$ has flavor symmetry at least $G$, but the actual global symmetry may be larger, especially when the period $p$ and the Coxeter number $h$ of $G$ have many prime factors in common. For instance, $D_{30}(E_8)$ has flavor group $F=E_8\times H_8$, where $H_8\subseteq E_8$ is a rank $8$ subgroup. The models $D_p(SU(2))$ are just the well known Argyres--Douglas theories of type $D_p$ (whose flavor symmetry is at least $SU(2)$ \cite{CV11}), and the models $D_p(G)$ may be seen as their generalization to arbitrary (simply--laced) flavor group $G$.\smallskip

In this paper we continue the analysis of this family of theories and provide detailed proofs of many statements which were just sketched in \cite{infinitelymany}. We study both the $D_p(G)$ models \emph{per se}, and the asymptotically--free (resp.\! superconformal) theories which arise by gauging the diagonal $G$ symmetry of a collection of $D_{p_i}(G)$ models. In facts, as in \cite{infinitelymany}, our strategy is to start from the gauged asymptotically--free models, which have a simple quiver description, study their properties by elementary Lie--algebraic techniques, and then extract the physics of the $D_p(G)$ SCFT's by decoupling the $G$ SYM sector\footnote{\ In the categorical language of \cite{cattoy}, the $D_p(G)$ SCFT's are the \emph{simplest} examples of periodic $G$--tubes.}.\smallskip

Having isolated the $D_p(G)$ theories,
our first question is to determine which ones of them are simply related  to previously known SCFT's. Besides the Argyres--Douglas models for $G=SU(2)$, it turns out that $D_p(SU(N))$ with $p\mid N$ is a conformal linear quiver of $SU$ gauge groups, while
$D_p(SO(2n))$ with $p\mid (n-1)$ is a conformal linear quiver of $SO$-$USp$ gauge groups \cite{sousp}. More generally, $D_p(G)$ with $G$ a \textit{classical} group is identified with a non--trivial IR fixed point of a  quiver theory of classical gauge groups.\smallskip

More interestingly, $D_2(SO(8))$, $D_2(SO(10))$ and $D_2(E_6)$ are identified with superconformal gaugings  of a flavor subgroup of the Minahan--Nemeshanski theories with symmetry, respectively,  $E_6$, $E_7$, and $E_8$ \cite{MN1,MN2}. This identification leads to an independent check of some of the Argyres--Seiberg dualities \cite{argyresseiberg}. An amusing prediction is that the dimension $\Delta$ of the field parametrizing the Coulomb branch of the three Minahan--Nemeshanski theories is given by one--half the Coxeter number of, respectively, $SO(8)$, $SO(10)$ and $E_6$, which gives $\Delta=3$, $4$, and $6$. This is a nice example of the general fact that, in the $D_p(G)$ theories, the values of the physical quantities have simple Lie theoretical meanings. \smallskip

The \textit{exceptional} SCFT's $D_p(E_r)$ ($r=6,7,8$) with $p\geq 3$  appear not to be simply related to any previously known $\cn=2$ QFT and their construction with the methods of \cite{infinitelymany} has no known alternative. They are by far the most interesting models in the family: they should thought of as non--trivial IR fixed points of the 4d $\cn=2$ theories arising from the compactification of the  6d (2,0) theories of type $E_r$ on suitable (generalized) geometries with insertions of suitable defects. Here we fully characterize these exceptional SCFT's by specifying the dimensions of their Coulomb branch operators and many other physical quantities.  \smallskip

Next we compute the SCFT central charges $a$, $c$ and $k$ for all $D_p(G)$ models and describe properties which are common to all models in the family. We check that the values of $a$, $c$, $k$, and the monodromy order $r$ are consistent with our web of identifications and dualities.
\smallskip

Our construction produces an explicit quiver with superpotential for all theories of interest,  the $D_p(G)$'s as well as the asymptotically--free (resp.\! superconformal) theories obtained by gauging their $G$ symmetry. This gives enough data to determine, in principle, their BPS spectra in all chambers. In this paper we compute the BPS spectrum of some of these theories, with emphasis on the existence of finite BPS chambers. The BPS spectra of related $\cn=2$ models will be published elsewhere \cite{CDZtoappear}.

In the course of the analysis we realize that the standard notion of the BPS quiver of a $\cn=2$ QFT \cite{ACCERV2} is not the most convenient one, both conceptually and computationally. We replace it by the more intrinsic and elegant notion of \textsf{META}--quiver, which is an ordinary quiver (with relations) whose representations take values in a higher $\C$--linear Abelian category $\mathscr{C}$ and are twisted by an autoequivalence of $\mathscr{C}$. We expect that the \textsf{META}--quiver idea will have a lot of applications for the $\cn=2$ \textsc{susy} models with special reference to the classification program. 
\medskip

The rest of the paper is organized as follows. In section 2 we quickly review the small part of the 2d/4d correspondence \cite{CNV} relevant for this paper. In section 3 we discuss in detail the physics of the $\cn=2$ models $\widehat{A}(p,1)\boxtimes G$, which correspond to a single $D_p(G)$ SCFT coupled to $G$ SYM, adopting the quiver approach of \cite{ACCERV2,cattoy} as well as from the more modern \textsf{META}--quiver viewpoint which is introduced in \S.\,3.2. In section 4 we extend the construction to $G$ SYM coupled to a collection of $D_{p_i}(G)$ models such that the Yang--Mills coupling is either asymptotically--free or exactly marginal. In section 5 we give an informal geometric description of the relevant theories for $G$ a \emph{classical} group; we use it to identify the subset of $D_p(G)$ SCFT having a weakly coupled Lagrangian description. In section 6 we compute the central charges $a$, $c$, $k$, and other SCFT invariants, and perform a number of non--trivial checks. In section 7 we study in detail the subclass of SCFT's of period 2, $D_2(G)$. Indeed, the period 2 theories are expected to be much simpler than the ones with $p\geq 3$, since they are the generalization to arbitrary $G=ADE$ of the free theory $D_2$ for $G=SU(2)$. Technicalities, computations, and various generalizations are confined in the three appendices.

\section{A quick review of the 2d/4d correspondence}\label{2d4drev}

Let $\Gamma\simeq \Z^r$ be the lattice of all conserved charges (electric, magnetic, and flavor) of a 4d $\cn=2$ supersymmetric QFT. Following \cite{CV11,ACCERV2}, we say that the 4d $\cn=2$ theory has the \emph{BPS--quiver property} 
iff there is a set of generators $\{e_i\}$ of $\Gamma\simeq \oplus_i \Z\, e_i$ such that the charges $\gamma$ of all BPS--particles belong to the double cone
\begin{equation}
 \gamma\in \Gamma_+ \bigcup \left(-\Gamma_+\right),\qquad \text{where }\Gamma_+=\oplus_i \Z_{\geq 0}\, e_i.
\end{equation}
In this case \cite{CV11,CNV,ACCERV2} the BPS--states correspond to the representations of a quiver with superpotential $(Q,\cw)$ which are stable with respect to the central charge of the 4d $\cn=2$ superalgebra \cite{CV11,CNV,ACCERV2}. The 2d/4d correspondence \cite{CNV,CV11,cattoy} is the (conjectural) statement that, for each 4d $\cn=2$ supersymmetric quantum field theory with the BPS--quiver property, there is a two--dimensional $(2,2)$ system with $\hat{c} < 2$ such that the exchange matrix $B$ of the 4d quiver\footnote{\ Strictly speaking, this holds under the additional assumption that the 4d quiver $Q$ is $2$--acylic; the exchange matrix $B$ of a $2$--acylic quiver $Q$ has entry $B_{ij}$ equal to the signed number of arrows from node $i$ to node $j$ in $Q$. The subtleties which arise for 4d quivers which are \emph{not} $2$--acyclic \cite{ACCERV2} will not be relevant for the present paper. } $Q$ is
\begin{equation}
 B=S^t-S,
\end{equation}
where $S$ is the $tt^*$ Stokes matrix of the 2d system \cite{CV92}. The physical motivation of the conjecture arises from the world--sheet theory of the string engineering the 4d $\cn=2$ model. 

The inverse process of reconstructing the 2d theory from the 4d one (that is, of finding $S$ given $B$) involves some subtleties. In the lucky case that $Q$ is acyclic, one has simply
\begin{equation}\label{tree}
 S_{ij}=\delta_{ij}-\max\{B_{ij},0\},
\end{equation}
corresponding to the Euler form of the quiver $Q$. In this paper we need $S$ only for acyclic quivers, which correspond to (basic) \emph{hereditary} algebras, and for the slightly more general cases which are derived equivalent to a hereditary \textit{category} \cite{LE1}, the matrix $S$ still being given by the Euler form. Concretely, this means we shall use the $S$'s associated to the \emph{elliptic} Dynkin quivers besides the ordinary and affine ones \cite{CV11}.

An important observation is that the 2d/4d correspondence relates 2d superconformal theories to 4d superconformal ones. Indeed the scaling of the 2d theory may be seen as a scaling property of the Seiberg--Witten geometry, which in turn implies a scaling symmetry for the 4d theory.

The 2d quantum monodromy is $H=(S^t)^{-1} S$ \cite{CV92}, and the 2d theory is superconformal (in the UV) precisely when $H$ is semisimple of spectral radius $1$. The rank of the flavor group $F$ of the 4d system, \emph{i.e.}\! the dimension of the kernel of $B$ \cite{ACCERV2}, is equal to the dimension of the $+1$--eigenspace of $H$ (for $\hat c<2$ this happens to be the same as the multiplicity of  $+1$ as a root of the characteristic polynomial of $H$ \cite{cattoy}).
\medskip

Consider two 2d $(2,2)$ LG systems with superpotentials $W_1(X_i)$ and $W_2(Y_a)$ and (UV) Virasoro central charges $\hat{c}_{1}$ and $\hat{c}_{2}$. Their \emph{direct sum} is defined as the $(2,2)$ decoupled model with superpotential $$W(X_i,Y_a)=W_1(X_i) + W_2(Y_a).$$ It has $\hat c=\hat c_1+\hat c_2$. By the 2d/4d correspondence, it defines a 4d $\cn=2$ theory provided $\hat c<2$.
In ref.\!\cite{CNV} the four dimensional theories arising from the direct sum of two minimal $ADE$ models were considered; since a minimal model has $\hat c<1$, in that case the bound $\hat c<2$ is automatically satisfied. A more general application of the same strategy is to consider the direct sum of a 2d minimal model ($\hat c<1$) with a 2d model corresponding to a \emph{complete} 4d theory which has $\hat c\leq 1$ \cite{CV11}. Again the direct sum has automatically  $\hat c<2$. This construction is the starting point of the present paper.

The $tt^*$ Stokes matrix of the direct sum is the tensor product of the Stokes matrices of the summands \cite{CV92}
\begin{equation}
 S=S_1\otimes S_2.
\end{equation}
In our case $S_2$ is the Stokes matrix of an $ADE$ Dynkin quiver (which is a tree, so eqn.\eqref{tree} applies), while $S_1$ is the Stokes matrix of a $\hat c=1$ $(2,2)$ system. For most of the paper, we take $S_1$ to be the Stokes matrix of an \textit{acyclic} affine quiver. The construction may be extended to a general mutation--finite quiver, provided one knows the right Stokes matrix.

If the quivers $Q_1$, $Q_2$ associated to $S_1$, $S_2$ are acyclic, the quiver $Q$ with exchange matrix
\begin{equation}\label{rrrr34}
 B=S_1^t\otimes S_2^t-S_1\otimes S_2
\end{equation}
 is called the triangle tensor product of the two quivers, written $Q_1\boxtimes Q_2$ \cite{kellerP}\!\!\cite{CNV}. It is equipped with a unique superpotential $\cw_\boxtimes$ described in \cite{kellerP}\!\!\!\cite{arnold1, cattoy}. The 4d models studied in \cite{CNV} are then described by the tensor product of two Dynkin quivers; in this paper we are mainly interested in the case $Q_1\equiv \widehat{H}$, an acyclic affine quiver, and $Q_2\equiv G$ an ADE Dynkin quiver. With a slight abuse of notation, we shall use the symbol $\widehat{H}\boxtimes G$ to denote both the quiver (with superpotential) and the corresponding 4d $\cn=2$ theory. In addition we shall consider the $S_1$'s corresponding to the four \emph{elliptic} quivers $D^{(1,1)}_4$, $E^{(1,1)}_6$, $E^{(1,1)}_7$, and $E^{(1,1)}_8$ \cite{CV11}. However in this last case the quiver \eqref{rrrr34} is not simply a tensor product of the quivers of the 2d direct summands.

Since the 2d minimal models are conformal, the direct sum 4d theory will be a SCFT precisely when the $(2,2)$ theory corresponding to the factor $S_1$ is conformal. An acyclic affine quiver always corresponds to a 4d theory which is asymptotically--free with $\beta\neq 0$ \cite{CV92}\!\!\cite{CV11}, and the 4d models $\widehat{H}\boxtimes G$ are also asymptotically--free with a non--zero $\beta$--function. On the contrary, the direct sum of a minimal and an elliptic 2d theories leads automatically to a superconformal 4d model.

\section{$\widehat{A}(p,1)\boxtimes G$ models}\label{Ap1Gdetailed}

\begin{table}
\begin{center}
\begin{tabular}{|c|c|c|c|}
\hline
& $W_G(X,Y)$ & $(q_X,q_Y)$ & $h(G)$\\
\hline
$A_{n-1}$ & $X^n + Y^2$ & $(1/n,1/2)$ & $n$\\
\hline
$D_{n+1}$ & $X^n + XY^2$ &$ (1/n, (n-1)/2n)$ & $2n$ \\
\hline
$E_6$ & $X^3 + Y^4 $& $(1/3,1/4)$ & 12 \\
\hline
$E_7$ & $X^3 + X Y ^3 $& $(1/3,2/9)$ & 18\\
\hline
$E_8$ & $X^3 + Y^5 $& $(1/3, 1/5)$ & 30 \\
\hline
\end{tabular}
\end{center}
\caption{ List of the $ADE$ simple singularities, and some of the corresponding properties. }\label{ADEsings}
\end{table}

\subsection{The old viewpoint (quivers)}

The $\widehat{A}(p,1)\boxtimes G$ models are obtained by 2d/4d correspondence from the direct sum $(2,2)$ system with superpotential
\be\label{Ap1Geom}
W_{p,G}=e^{-Z} + e^{p \,Z} + W_G(X,Y) + U^2 + \text{lower terms},\qquad\ p\in\mathbb{N},
\ee
where $W_G(X,Y)$ is the superpotential of a type $G = ADE$ minimal model (see table \ref{ADEsings}), while the superpotential $e^{-Z} + e^{p\, Z}$, with the identification $Z \sim Z+ 2 \pi i$, describes an asymptotically--free $(2,2)$ system \cite{CV92}\!\!\cite{CV11} whose quiver--class contains an essentially unique acyclic representative given by the affine Dynkin graph $\widehat{A}_p$ oriented in such a way that $p$ arrows point in the positive direction and one in the negative one \cite{CV11}. This acyclic quiver will be denoted as $\widehat{A}(p,1)$, and we shall use the same symbol to refer to the corresponding QFT's. 
The ultraviolet 2d central charge of the direct sum system is
$\hat{c} = 1 + \hat{c}_G < 2$,
and we get a well--defined $\cn=2$ theory in 4d which  is again asymptotically--free. In the special case $p=1$ the direct sum 4d model reduces to pure SYM  with gauge group $G$ \cite{CNV,cattoy}; indeed, the hypersurface $W_{1,G}=0$ is the corresponding Seiberg--Witten geometry \cite{Tack}.


\begin{figure}
$$
\begin{gathered}
\xymatrix{\diamondsuit_1\ar[rr]^{\psi^{(p+1)}_1} && \diamondsuit_2 \ar[rr]^{\psi^{(p+1)}_2}&&\cdots\ar[rr]^{\psi^{(p+1)}_{n-2}}&& \diamondsuit_{n-1}\ar[rr]^{\psi^{(p+1)}_{n-1}}&& \diamondsuit_n\\
\\
&\bullet\ar[uul]_{A^{(1)}_p}\ar[rr]^{\psi^{(p)}_1\qquad}&&\bullet\ar[uul]_{A^{(2)}_p}\ar[rr]^{\psi^{(p)}_2}&&\cdots\ar[rr]^{\psi^{(p)}_{n-2}\qquad}&&\bullet\ar[uul]_{A^{(n-1)}_p}\ar[rr]^{\psi^{(p)}_{n-1}\qquad}&&\bullet\ar[uul]_{A^{(n)}_p}\\
\\
&\vdots\ar[uu]_{A^{(1)}_{p-1}}&&\vdots\ar[uu]_{A^{(2)}_{p-1}}&&\cdots&&\vdots\ar[uu]_{A^{(n-1)}_{p-1}}&&\vdots\ar[uu]_{A^{(n)}_{p-1}}\\
\\
&\bullet\ar[uu]_{A^{(1)}_2}\ar[rr]^{\psi^{(2)}_{1}\qquad}&&\bullet\ar[uu]_{A^{(2)}_2}\ar[rr]^{\psi^{(2)}_{2}}&&\cdots\ar[rr]^{\psi^{(2)}_{n-2}\qquad}&&\bullet\ar[uu]_{A^{(n-1)}_2}\ar[rr]^{\psi^{(2)}_{n-1}\qquad}&&\bullet\ar[uu]_{A^{(n)}_2}\\
\\
\spadesuit_1\ar[uuuuuuuu]_{B^{(1)}}\ar[uur]_{A^{(1)}_1}\ar[rr]^{\qquad\psi^{(1)}_{1}}&& \spadesuit_2\ar[uuuuuuuu]_{B^{(2)}}\ar[uur]_{A^{(2)}_1}\ar[rr]^{\qquad\psi^{(1)}_{2}}&&\cdots\ar[rr]^{\psi^{(1)}_{n-2}\qquad}&&\spadesuit_{n-1}\ar[uuuuuuuu]_{B^{(n-1)}}\ar[uur]_{A^{(n-1)}_1}\ar[rr]^{\qquad\psi^{(1)}_{n-1}}&&\spadesuit_n\ar[uuuuuuuu]_{B^{(n)}}\ar[uur]_{A^{(n)}_1}
}
\end{gathered}
$$
\caption{The tensor product quiver of $\widehat{A}(p,1)$ and $G=A_n$.\label{tenzorquivAp1An} }
\end{figure}
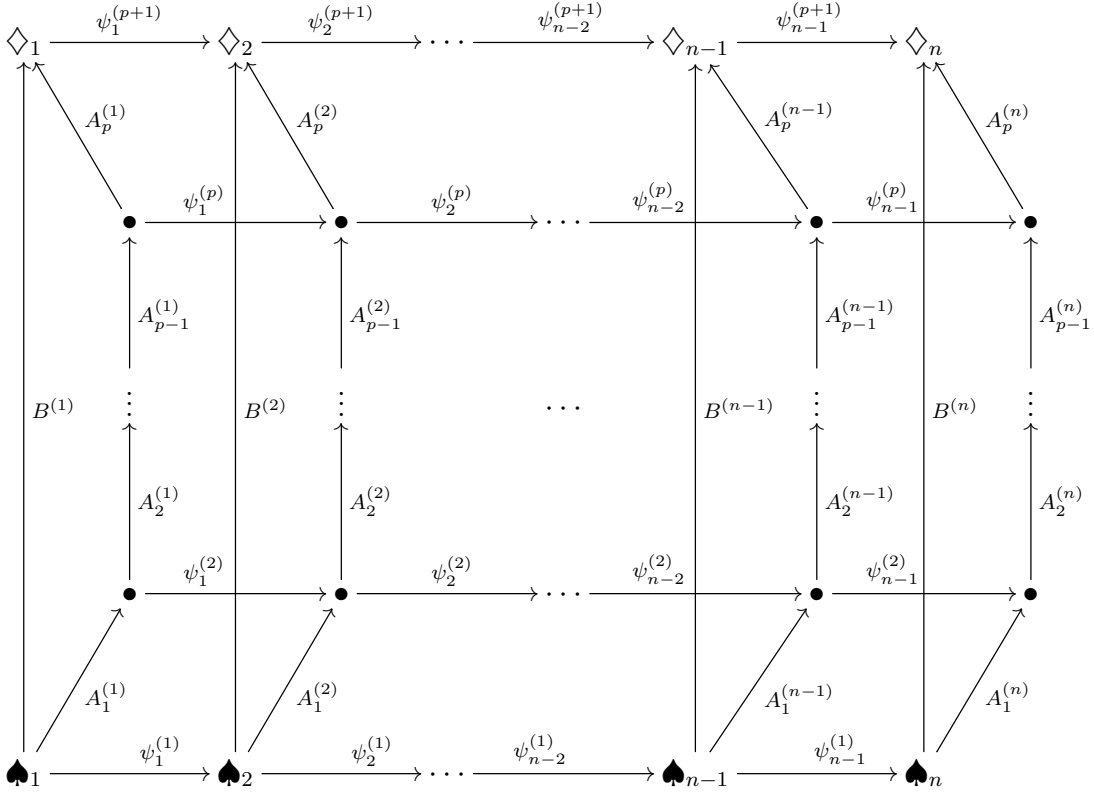

The quiver (with superpotential) of the 4d theory corresponding to \eqref{Ap1Geom} is $\widehat{A}(p,1)\boxtimes G$. It is obtained by $\Pi_3$ completion of the product algebra $\C\widehat{A}(p,1)\otimes \C G$ \cite{kellerP}. We illustrate the construction in the $G=A_n$ example.
The quiver of $\C\widehat{A}(p,1)\otimes \C A_n$ is depicted in figure \ref{tenzorquivAp1An}. To obtain the triangular tensor product quiver $\widehat{A}(p,1)\boxtimes A_n$, one adds to figure \ref{tenzorquivAp1An} one ``diagonal'' arrow, $\l^{(i,a)}$ resp.\! $\xi^{(a)}$, per each square subquiver according to the rules
\be
\begin{gathered}
\xymatrix{
\bullet\ar[rr]^{\psi^{(i+1)}_{a}}&&\bullet\ar[ddll]_{\l^{(i,a)}}\\
&&\\
\bullet\ar[rr]^{\psi^{(i)}_a}\ar[uu]^{A^{(a)}_i}&&\bullet\ar[uu]_{A^{(a+1)}_{i}}
}
\end{gathered}\qquad\qquad\begin{gathered}
\xymatrix{
\diamondsuit_a\ar[rr]^{\psi^{(p+1)}_{a}}&&\diamondsuit_{a+1}\ar[ddll]_{\xi^{(a)}}\\
&&\\
\spadesuit_a\ar[rr]^{\psi^{(1)}_a}\ar[uu]^{B^{(a)}}&&\spadesuit_{a+1}\ar[uu]_{B^{(a+1)}}
}
\end{gathered}
\ee
where $i=1,\dots, p$ and $a=1,\dots, n-1$. The commutativity relations of the tensor product algebra $\C\widehat{A}(p,1)\otimes \C G$ are then implemented by the superpotential
\be
\cw_{\boxtimes} \equiv \sum_{a=1}^{n-1} \sum_{i=1}^{p} \l^{(i,a)} \big( \psi^{(i+1)}_a \ A^{(a)}_i - A^{(a+1)}_i \psi^{(i)}_a\big) + \sum_{a=1}^{n-1} \xi^{(a)}\big( \psi^{(p+1)}_a B^{(a)} -  B^{(a+1)}\psi^{(1)}_a\big).
\ee

\subsubsection{Light subcategory of the $\widehat{A}(p,1)\boxtimes G$ model}\label{lightcatsAp1G}

The purpose of this section is to show, from the old perspective of \cite{cattoy},
that the $\cn=2$ model $\widehat{A}(p, 1)\boxtimes G$ has a corner in its parameter space where it behaves as SYM  with gauge group $G$ coupled to a $D_p(G)$ $\cn=2$ superconformal system\footnote{\ We adopt a different notation with respect to \cite{infinitelymany}: $D_p(G)\equiv D(G,p-1)$. In the new notation  $D_1(G)$ is the empty theory (zero degrees of freedom) for all $G$. }.

As a consequence we may extract the quivers with superpotential for the $D_p(G)$ systems from the known ones $\widehat{A}(p,1)\boxtimes G$. For notational simplicity, we give explicit expressions only for the case $G=A_n$,  the results being obviously valid for all simply--laced Lie algebras. 

 
In the relevant $S$--duality frame the simple $W$--bosons of the gauge group $G$ have charge vectors equal to the minimal imaginary root of each $\widehat{A}(p,1)$ ``vertical'' affine subquiver (cfr.\! figure \ref{tenzorquivAp1An}). Let us denote such subquivers as $\widehat{A}(p,1)_a$, and their generic representations with dimension the minimal imaginary roots $\delta_a$ as $W_a$ for $a=1,...,r(G)$. Seen as representation of the total quiver $\widehat{A}(p,1)\boxtimes G$, the representations $W_a$ are all mutually local \textit{i.e.}
\be
\langle\, \boldsymbol{\dim}\, W_a , \boldsymbol{\dim}\, W_b \,\rangle_\text{Dirac}\equiv (\boldsymbol{\dim}\,W_a)^t\,B\,\boldsymbol{\dim}\,W_b = 0 \qquad \forall \ a,b = 1, \dots, r(G).
\ee
The corresponding magnetic charges are simply \cite{cattoy}
\be\label{canmagnAp1G}
m_a(X) = \text{dim } X_{\spadesuit_a} - \text{dim } X_{\diamondsuit_a} \equiv \mathfrak{d}_{\widehat{A}(p,1)_a}(X) \in \mathbb{Z},
\ee
where $\mathfrak{d}_{\widehat{A}(p,1)}$ is the Dlab-Ringel defect of the algebra $\C\widehat{A}(p,1)$ \cite{CB,ringel}. The weak Yang--Mills coupling limit corresponds to taking the central charge function
\begin{equation}
Z(X)=-\frac{C_a}{g^2}\,m_a(X)+O(1), \qquad g\rightarrow 0,\ C_a>0,
\end{equation}
and the BPS particles which have bounded masses $|Z(X)|$ as $g\rightarrow 0$ correspond to stable objects in a special \textit{light} subcategory $\mathscr{L}$ \cite{cattoy}. As the Yang--Mills coupling $g\rightarrow 0$, $\mathscr{L}$ naturally splits in the SYM sector with gauge group $G$ (which contains vector multiplets making one copy of the adjoint of $G$) plus a `matter' sector.
For the above canonical choices, 
the light category $\mathscr{L}$ is the subcategory of the representations $X$ of $\widehat{A}(p,1)\boxtimes G$ which have zero magnetic charges $m_a(X)=0$ such that  all their submodules have non--positive magnetic charges \cite{cattoy}. $\mathscr{L}$ is characterized by the usual properties expressing the consistency of the weak coupling limit and the Higgs mechanism \cite{cattoy}. 

Before stating these properties, we recall some basic facts about the representations of the affine algebra $\C\widehat{A}(p,1)$. The module category of its \textit{regular} representations, $\ct_{\widehat{A}(p,1)}$, consists of direct sums of indecomposable representations of $\widehat{A}(p,1)$ such that either all their arrows are isomorphisms, or not all arrows are mono and also not all are epi \cite{ringel}. Equivalently, they are direct sums of indecomposables, $\oplus_iX_i$, all with vanishing defect $\mathfrak{d}_{\widehat{A}(p,1)}(X_i)=0$ (\textit{i.e.}\! zero magnetic charge) \cite{ringel,CB}. To each regular indecomposable representation $X$  of $\widehat{A}(p,1)_a$ one associates a point $\lambda\in\mathbb{P}^1$ as follows: if $B^{(a)}$ is not an isomorphism $\lambda=\infty$; otherwise $\lambda\in\C$ is the unique eigenvalue of the matrix $(B^{(a)})^{-1}A_p^{(a)}A_{p-1}^{(a)}\cdots A_1^{(a)}$. We write $\ct_{\widehat{A}(p,1)}(\lambda)$ for the category whose objects are direct sums of indecomposables of $\ct_{\widehat{A}(p,1)}$ with a fixed value of $\lambda$.

\medskip

{\bf Fact.}\footnote{\ We shall prove the \textbf{Fact} in \S.\,\ref{modern} from the \textsf{META}--quiver perspective.} \textit{Let $X$ be an \emph{indecomposable} representation of $\big(\widehat{A}(p,1)\boxtimes G,\cw_{\boxtimes}\big)$ which belongs to the light subcategory $ \mathscr{L}$ of the canonical $S$-duality frame. Then
\be\label{lemma222}
X \big|_{\widehat{A}(p,1)_a} \in \ct_{\widehat{A}(p,1)}(\lambda) \qquad \text{the \underline{same} } \l \text{ for all }a. 
\ee}

\vglue-6pt
 In particular, the arrows $B^{(a)}$ are isomorphisms\footnote{ \ This is true away from $\lambda=\infty$; the analysis at $\lambda=\infty$ is similar, and shows that there is no matter in the $\lambda=\infty$ subcategory. Hence, as far as we are interested in decoupling the matter sector from SYM, we may keep ourselves away from the point $\lambda=\infty$.} for all $a=1,...,r(G)$. In the quiver $\widehat{A}(p,1)\boxtimes G$ the  nodes $\spadesuit_a$ then get indentified with the nodes $\diamondsuit_a$, leading to an  effective quiver with $r(G)$ less nodes.  Around a pair of identified nodes the effective quiver $Q_\mathrm{eff}$ has the form 
\be
\begin{gathered}
\xymatrix{
&&&\vdots\\
&\vdots&&\bullet\ar@/_0.9pc/[ddll]_{\l^{(1,a)}}\ar[u]_{A_2^{(a+1)}}\\
&&&&\\
\dots&\spadesuit_a\ar[uu]^{A_1^{(a)}}\ar@/^1.5pc/[rr]^{\psi^{(p+1)}_{a}}\ar@/_1.5pc/[rr]_{\psi^{(1)}_{a}}&&\spadesuit_{a+1}\ar[uu]_{A_1^{(a+1)}}\ar[ll]_{\xi^{(a)}}\ar@/^0.9pc/[ddll]^{\l^{(p,a)}}&\dots\\
&&&&\\
&\bullet\ar[uu]^{A_p^{(a)}}&&\vdots\ar[uu]_{A_p^{(a+1)}}\\
&\vdots\ar[u]^{A_{p-1}^{(a)}}}
\end{gathered}
\ee
while its effective superpotential is obtained from that of $\widehat{A}(p,1)\boxtimes G$ by replacing the arrows $B^{(a)}$ with $1$'s: In particular, one obtains the relations
\beq\label{modpplus1}
\psi_a^{(p+1)} = \psi^{(1)}_a, \qquad
A_p^{(a)} \l^{(p,a)} = \xi^{(a)} = \l^{(1,a)} A_1^{(a+1)}, \qquad a = 1,...,n-1,
\eeq
which allow to eliminate the redundant arrows $\xi^{(a)}$ and $\psi_a^{(p+1)}$. This step corresponds to taking the \emph{reduced part} of the light effective quiver with superpotential $(Q_\mathrm{eff},\cw_\boxtimes\big|_{B^{(a)}=1})$ in the sense of \textbf{Theorem 4.6} of \cite{DWZ}.\smallskip

In conclusion, $\mathscr{L}$ is again a category of modules of a Jacobian algebra whose quiver
$Q_{p,G}$ is obtained by identifying the $\spadesuit_a$ and the $\diamondsuit_a$ nodes in $\widehat{A}(p,1)\boxtimes G$. Its (reduced) effective superpotential $\cw^{\prime}$ is obtained by replacing the $B^{(a)}$ arrows with 1's in the superpotential $\cw_{\boxtimes}$, and integrating out the `massive' arrows $\xi^{(a)}$ and $\psi_a^{(p+1)}$. For instance, the effective superpotential for $G=A_n$ is
\be\label{superAp1Glight}
\cw^\prime\equiv  \sum_{a=1}^{r(G)} \sum_{i=1}^{p-1} \l^{(i,a)} \big( \psi^{(i+1)}_a \ A^{(a)}_i - A^{(a+1)}_i \psi^{(i)}_a\big)+ \sum_{a=1}^{r(G)} \l^{(p,a)} \big( \psi^{(1)}_a A^{(a)}_p - A_p^{(a+1)} \psi^{(p)}_a\big).
\ee

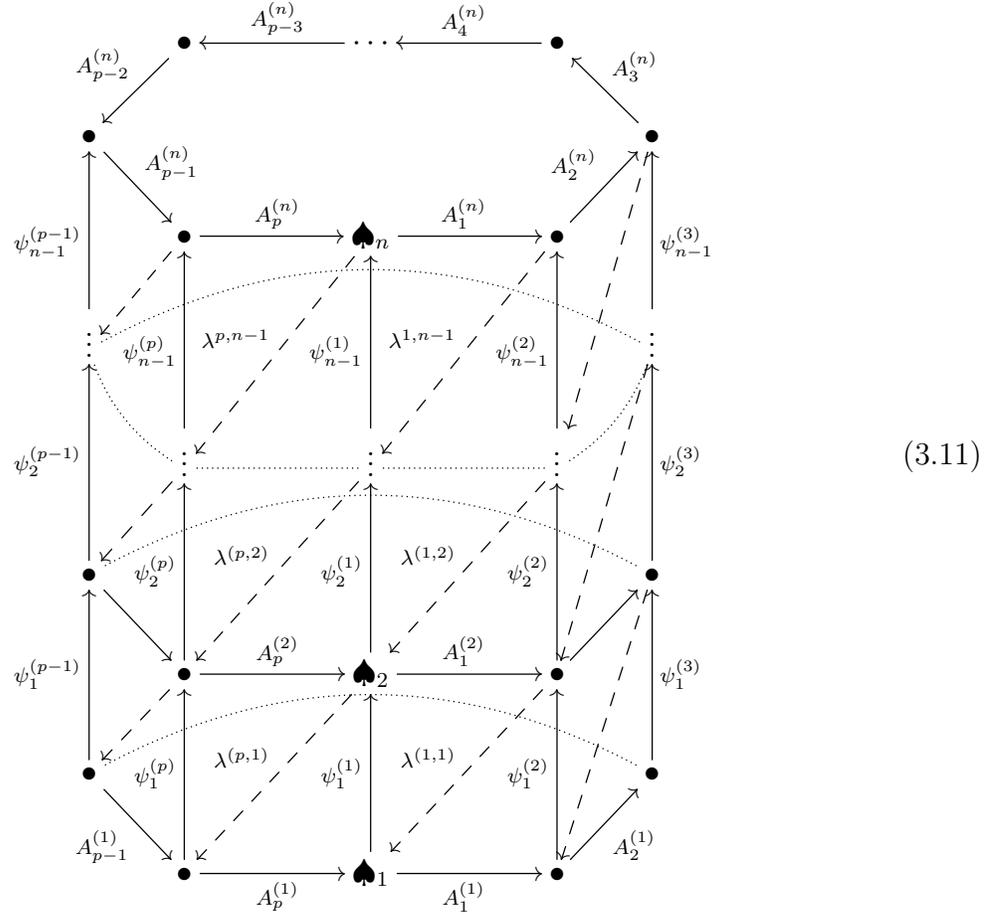
\begin{figure}
\begin{center}
\be
\begin{gathered}
\xymatrix{
&\bullet\ar[dl]_{A^{(n)}_{p-2}}&&\cdots\ar[ll]_{A^{(n)}_{p-3}}&&\bullet\ar[ll]_{A^{(n)}_{4}}&\\
\bullet\ar[dr]^{A^{(n)}_{p-1}}&&&&&&\bullet\ar[ul]_{A^{(n)}_3}\ar@{-->}[dddl]\\
&\bullet\ar[rr]^{A^{(n)}_p}\ar@{-->}[dl]&& \spadesuit_{n} \ar[rr]^{A^{(n)}_1}\ar@{-->}[ddll]_{\l^{p,n-1}}&&\bullet\ar[ur]^{A^{(n)}_2}\ar@{-->}[ddll]_{\l^{1,n-1}}\\
\vdots\ar[uu]^{\psi^{(p-1)}_{n-1}}\ar@/_0.5pc/@{..}[dr]\ar@/^2.5pc/@{..}[rrrrrr]&&&&&&\vdots\ar[uu]_{\psi^{(3)}_{n-1}}\ar@{-->}[dddl]\\
&\ar[uu]^{\psi^{(p)}_{n-1}}\vdots\ar@{-->}[dl]\ar@{..}[rr]&&\vdots\ar[uu]^{\psi^{(1)}_{n-1}}\ar@{-->}[ddll]_{\l^{(p,2)}}\ar@{..}[rr]&&\vdots\ar[uu]^{\psi^{(2)}_{n-1}}\ar@{-->}[ddll]_{\l^{(1,2)}}\ar@/_0.5pc/@{..}[ur]&\\
\bullet\ar[dr]\ar[uu]^{\psi^{(p-1)}_2}\ar@/^2.5pc/@{..}[rrrrrr]&&&&&&\bullet\ar[uu]_{\psi^{(3)}_2}\ar@{-->}[dddl]\\
&\bullet\ar[rr]^{A^{(2)}_p}\ar[uu]^{\psi^{(p)}_2}\ar@{-->}[dl]&& \spadesuit_2 \ar[rr]^{A^{(2)}_1}\ar[uu]^{\psi^{(1)}_2}\ar@{-->}[ddll]_{\l^{(p,1)}}&&\bullet\ar[ur]\ar[uu]^{\psi^{(2)}_2}\ar@{-->}[ddll]_{\l^{(1,1)}}\\
\bullet\ar[dr]_{A^{(1)}_{p-1}}\ar[uu]^{\psi^{(p-1)}_1}\ar@/^2.5pc/@{..}[rrrrrr]&&&&&&\bullet\ar[uu]_{\psi^{(3)}_1}\\
&\bullet\ar[uu]^{\psi^{(p)}_1}\ar[rr]_{A^{(1)}_p}&& \spadesuit_1\ar[uu]^{\psi^{(1)}_1} \ar[rr]_{A^{(1)}_1}&&\bullet\ar[ur]_{A^{(1)}_2}\ar[uu]^{\psi^{(2)}_1}
}
\end{gathered}
\ee
\end{center}
\caption{ The reduced quiver $Q_{p,A_n}$ of the light subcategory of the model $\widehat{A}(p,1)\boxtimes A_n$. The dashed arrows represents the relations inherited from the triangular tensor product.}\label{theloopquiverG}
\end{figure}


Eqn.\eqref{lemma222} implies that, for $X\in \mathsf{rep}(Q_{p,G},\cw^{\prime})$ and \textit{indecomposable}, the linear maps
\be\label{thisistheend}
\ell(j,a) \equiv  A^{(a)}_{j-1}\cdots A^{(a)}_2 A^{(a)}_1 A^{(a)}_p A^{(a)}_{p-1} \cdots A^{(a)}_{j+1} A^{(a)}_j \colon X_{s(A^{(a)}_j)} \longrightarrow X_{s(A^{(a)}_j)}
\ee
have the unique eigenvalue $\lambda$. We write $\mathscr{L}(\lambda)$ for the subcategory of modules in $\mathsf{rep}(Q_{p,G},\cw^{\prime})$ with a fixed value of $\lambda$.

%
%

\medskip

This shows that the models $\widehat{A}(p,1)\boxtimes G$ contain a $G$ SYM subsector. Indeed,
the light subcategory of the quiver $\widehat{A}(p,1)\boxtimes G$  is fibered over a $\mathbb{P}^1$ with the marked point $\lambda=0$. Away from $\lambda=0$, $\mathscr{L}(\lambda)$ is equivalent to the corresponding light category for pure SYM with gauge group $G$ 
\be
\mathscr{L}(\l) \simeq \mathscr{L}^{YM}\!(\l)_{G}
\ee
since, for $X\in\mathscr{L}(\l\neq0)$,  all maps $A^{(a)}_j$ are isomorphisms and allow to identify their source and sink nodes, reducing the quiver to the effective quiver for the YM light category  $\mathscr{L}^{YM}\!(\l)_{G}$ \cite{cattoy}.
The BPS particles, stable and light at weak coupling, which correspond to generic $\lambda$ representations are then vector multiplets forming precisely of one copy of the adjoint representation of $G$ (taking into account also the massless photons in the Cartan subalgebra).

The category $\mathscr{L}(\l=0)$, instead, contains other representations besides the ones obtained by taking the $\lambda\rightarrow 0$ limit of the  $\lambda\neq 0$ ones. In facts, one has
the inclusion
\be
\mathscr{L}^{YM}\!(\l=0)_G \subseteq \mathscr{L}(\l=0)
\ee
with equality if and only if $p=1$. For $p>1$ $\mathscr{L}(\l=0)$ contains in addition representations which (when stable) have the physical interpretation of \emph{matter} BPS particles charged under the $G$ gauge symmetry. By sending the Yang--Mills coupling to zero, we decouple from the SYM sector the matter system sitting in the category $\mathscr{L}(\lambda=0)$, which we call $D_p(G)$. Our next task is to characterize the non--perturbative physics of this system. Before doing that, we introduce a different perspective on $\widehat{A}(p,1)\boxtimes G$ and its subcategories of modules.

\subsection{The new viewpoint: \textsf{META}--quivers}\label{modern}

Usually, by a representation of a quiver $Q$ we mean the assignment of a vector space $X_i$ to each node $i$ of $Q$ and a linear map $X_\psi$ to each arrow $\psi$, that is, we assign to nodes resp.\! arrows objects resp.\! morphisms of the category $\mathsf{vec}$ of finite dimensional vector spaces.  Of course we may replace $\mathsf{vec}$ by any other category $\mathscr{C}$, getting a $\mathscr{C}$--valued representation of $Q$ where the concatenation of arrows along paths in $Q$ is realized as the composition of the corresponding morphisms in $\mathscr{C}$. If $\mathscr{C}$ is $\C$--additive, so that it makes sense to sum morphisms and to multiply them by complex numbers, we may even define $\mathscr{C}$--valued representations of quivers subjected to relations of the standard form. Thus it makes sense to speak of the \textit{category} of $\mathscr{C}$--valued representations of the quiver $Q$ bounded by an ideal $I$. If, in addition, $\mathscr{C}$ is Abelian, the category of $\mathscr{C}$--valued representations of a quiver with relations is again Abelian, and share most of the properties of the usual module categories. 

It is useful to extend the construction to \textit{twisted} $\mathscr{C}$--valued representations. Let $\{\sigma_\ell\}$  be a group of autoequivalences of the category $\mathscr{C}$, where $\sigma_0=\mathrm{Id}$. To each arrow in $Q$ we assign a valuation in $\{\sigma_\ell\}$.  Then a $\mathscr{C}$--valued representation of the \emph{valued} quiver assigns to an arrow $a$ with valuation $\sigma_{\ell(a)}$ a morphism  $\psi\in\mathrm{Hom}(\co_{s(a)},\sigma_{\ell(a)}\co_{t(a)})$. 
The twisted composition $\star$ of two arrows $a$ and $b$ with $s(b)=t(a)$ has valuation $\sigma_{\ell(a)}\sigma_{\ell(b)}$ and is given by\footnote{\ The fact that the $\{\sigma_a\}$ are taken to be just autoequivalences (and not automorphisms) introduces notorious subtleties; for the particular categories we are interested in, this will not be a problem, since we have an underlying concrete description in terms of elements of objects (the one given by the big BPS--quiver modules). The abstract viewpoint is, however, very useful since it allows to describe the physical phenomena in a unified way  for large classes of $\cn=2$ models, abstracting the essential physics from the intricate details of each particular example. } 
\begin{equation}
 \psi_b\star \psi_a\equiv \sigma_{\ell(a)}(\psi_b)\circ \psi_a\in\mathrm{Hom}\big(\co_{s(a)},\sigma_{\ell(a)}\sigma_{\ell(b)}\co_{t(b)}\big)
\end{equation}
where $\circ$ is the composition in $\mathscr{C}$.

Of course, by  this construction we are not introducing any \emph{real} generalization, the resulting Abelian category may always be seen (up to Morita equivalence) as a subcategory of the usual representations of some bigger (possibly infinite) quiver. However, in the case of $\cn=2$ QFT's working with twisted $\mathscr{C}$--valued representations turns out to be very convenient both conceptually and technically: we replace a messy BPS quiver with complicate Jacobian relations with a much smaller quiver having few nodes, few arrows and, typically, a higher symmetry which is almost never visible in the messy $\mathsf{vec}$ quiver $Q_\text{BPS}$. Besides, the messy $\mathsf{vec}$ quivers associated to QFT's have no simple universality property useful to characterize and classify them while, for a good choice of $\mathscr{C}$, the smaller $\mathscr{C}$--quivers have rather uniform behaviour. 

It is easy to introduce a notion of stability of the (twisted) $\mathscr{C}$--valued representations of $Q$ which is equivalent to the stability for the corresponding $\mathsf{vec}$--valued representation of the messy $Q_\text{BPS}$. One introduces a stability function (central charge) $\mathscr{Z}$ for the $\mathscr{C}$--valued representations $\cx$
\begin{gather}
\mathscr{Z}\equiv \Big\{Z_i\colon K_0(\mathscr{C})\rightarrow \C,\ i\in \text{(nodes of $Q$)}\Big\}\\
\mathscr{Z}(\cx)=\sum_i Z_i(\cx_i)\in \C,
\end{gather}
where the homomorphism of Abelian groups $Z_i$ coincides with the usual central charge for the subcategory of the representations of the messy quiver $Q_\text{BPS}$ which map into $\mathscr{C}$--valued representations having support on the $i$--th node of $Q$. A $\mathscr{C}$--valued representation $\cx$ is $\mathscr{Z}$--stable iff, for all non--zero proper sub--objects $\cy$, 
$\arg \mathscr{Z}(\cy)<\arg\mathscr{Z}(\cx)$.\smallskip

Given a $\cn=2$ QFT  $T$ we say that the quadruple $(Q, I, \mathscr{C},\nu)$ --- $Q$ being a finite connected quiver, $I$ a bilateral ideal of relations in $\C Q$, $\mathscr{C}$ a $\C$--additive Abelian category, and $\nu$ a valuation of $Q$ in the autoequivalences of $\mathscr{C}$ --- is \emph{a \textsf{META}--quiver for $T$} iff the stable, $\nu$--twisted, $\mathbb{C}$--valued representations of $Q$, subjected to the relations in $I$, give the BPS spectrum of $T$ (in some chamber).

\subsubsection{The \textsf{META}--quiver for the light category of the $\widehat{A}(p,1)\boxtimes G$ model}

Let us construct  \textsf{META}--quivers for the light category of $\cn=2$ models at hand.
We consider the following very canonical quivers with relations (whose physical interpretation is the Higgs branch of an auxiliary $\cn=2$ system). Given a (finite, connected) graph $L$, we define its \emph{double quiver} $\overline{L}$ by replacing each edge $\xymatrix{\ar@{-}[r]^a&}$ in $L$ with a pair of opposite arrows $\xymatrix{\ar@<0.3ex>[r]^{a}&\ar@<0.3ex>[l]^{a^*}}$. The quotient of the path algebra of $\overline{L}$ by the ideal generated by the relations
\begin{equation}\label{pgcp}
\sum_{a\in L}(aa^*-a^*a)=0,
\end{equation}
is called the \emph{preprojective algebra} of the graph $L$ \cite{pre1,pre2}, which we write as $\cp(L)$. A basic result is that $\cp(L)$ is finite dimensional if and only if $L$ is an $ADE$ Dynkin graph $G$.

Let us chose $\mathscr{C}=\ct\equiv\ct_{\widehat{A}(p,1)}$, \textit{i.e.}\! the category of regular representations of the affine quiver $\widehat{A}(p,1)$. Let $\tau$ be its autoequivalence given by the Auslander--Reiten translation \cite{ringel,CB}. We consider the double quiver $\overline{G}$ with the following valuation: direct arrows $a$ are valued by Id, while inverse arrows $a^*$ by $\tau$. The \textsc{lhs} of eqn.\eqref{pgcp} has then valuation $\tau$.
The category of the \emph{twisted} $\mathscr{C}$--valued modules of $\cp(G)$ is equivalent to the light category of $\widehat{A}(p,1)\boxtimes G$. Indeed, the constraints $\partial_\psi\cw_\boxtimes=\partial_\lambda\cw_\boxtimes=0$ just state that the arrows $a,a^*$ of $\overline{G}$ are morphisms of $\mathscr{C}$, twisted by the appropriate autoequivalence, while $\partial_A\cw_\boxtimes=\partial_B\cw_\boxtimes=0$ give the $\cp(G)$ constraints \eqref{pgcp}.

 The tube subcategories $\ct_{\widehat{A}(p,1)}(\lambda)$ and $\ct_{\widehat{A}(p,1)}(\mu)$ are preserved by $\tau$, and Hom--orthogonal for $\lambda\neq \mu$ \cite{ringel,CB} \textit{i.e.}
\begin{equation}
 \mathrm{Hom}(X,Y)=0\quad \text{if }X\in \ct_{\widehat{A}(p,1)}(\lambda),\ Y\in \ct_{\widehat{A}(p,1)}(\mu),\quad \lambda\neq\mu.
\end{equation}
Since the arrows $a, a^*$ take value in the Hom groups, they automatically vanish between objects in different tubes. Hence if
$\cx$ is a $\ct_{\widehat{A}(p,1)}$--valued \emph{indecomposable} representation of $\cp(G)$
\begin{equation}
 \cx_i\in \ct_{\widehat{A}(p,1)}(\lambda)\quad \text{with the \underline{same} $\lambda$ for all nodes $i$ of $\overline{G}$},
\end{equation}
which is a much simpler way to get the physical consistency condition in eqn.\eqref{lemma222}.

For $\lambda\neq 0$ we have the equivalence $\ct_{\widehat{A}(p,1)}(\lambda)\simeq \ct_{\widehat{A}(1,1)}(\lambda)$ (the homogeneous tube), so the category of $\ct_{\widehat{A}(p,1)}(\lambda)$--representations of $\cp(G)$ coincides in this case with the corresponding light subcategory of SYM. In facts, as we are going to show, the category of twisted $\ct$--valued representations of $\cp(G)$ contains a canonical subcategory isomorphic to the  light Yang--Mills one (for gauge group $G$).

\subsubsection{The SYM sector}
There is a more canonical way of looking to the SYM sector. For $\lambda\neq 0$, $\ct(\lambda)$ is a homogeneous tube, so $\tau \co=\co$ for all its indecomposable objects. Fix an indecomposable $\co_0\in\ct(\lambda\neq0)$; we have a functor from the category of modules (in the standard sense) of the preprojective algebra $\cp(G)$ to the one of $\ct(\lambda\neq0)$--valued twisted module --- that is, to $\mathscr{L}(\lambda\neq0)$ --- given by
\begin{gather}\label{oootimes}
\cx_i= \co_0\otimes X_i, \quad \cx_a=\mathrm{Id}\otimes X_a, \qquad \cx_{a^*}=\mathrm{Id}\otimes X_{a^*}.
\end{gather}
In particular, the bricks of $\mathscr{L}(\lambda\neq0)$ are obtained by taking as $\co_0$ the unique regular brick in $\ct(\lambda)$, \textit{i.e.}\! the $SU(2)$ $W$--boson representation $W(\lambda)$; one gets
\begin{equation}\label{eee2231}\text{\bf bricks of }\mathscr{L}(\lambda\neq0) \ \ \equiv\ \ \ W(\lambda)\otimes X,\qquad X\ \text{a brick of }\mathsf{mod}\,\cp(G).\end{equation}
The bricks of $\mathsf{mod}\,\cp(G)$
have dimension vectors equal to the positive roots of $G$ and are \emph{rigid} (this is an elementary consequence of \cite{CBlemma}, see \cite{cattoy}), that is, the family of representations \eqref{eee2231} correspond to BPS vector multiplets with the quantum numbers of the $W$--bosons of $G$ (the SYM sector).

$\ct(\lambda=0)$ is a tube of period $p$. Then the \begin{equation}\label{socle1}\cx_i=\bigoplus_{s=1}^{m_i} R_{i,s}\end{equation} with $R_{i,s}$ indecomposable regular modules characterized uniquely by their (regular) socle $\tau^{k_{i,s}} S$ and lenght $r_{i,s}$ \cite{CB} ($S$ being a reference regular simple).
The $\lambda=0$ $SU(2)$ $W$--boson representations are the indecomposables of regular lenght $p$; we write $W(k)$ for the length $p$ regular indecomposable with socle $\tau^kS$ ($0\leq k \leq p-1$). Since
\begin{equation}\label{socle2}
 \dim\mathrm{Hom}(W(k),W(\ell))=\delta_{k,\ell},\qquad \quad \dim\mathrm{Hom}(W(k),\tau W(\ell))=\delta_{k,\ell+1},
\end{equation}
we may promote each ordinary representation $X$ of $\cp(G)$ to a representation of the SYM sector of $\mathscr{L}(\lambda=0)$ by replacing the basis vectors $v_{i,s}$ of $X_i$ by $W(k_{i,k})$ in such a way that two basis vectors $v,v^\prime$ related by $v^\prime=a(v)$ (resp.\! $v^\prime=a^*(v)$) have the $k^\prime=k$ (resp.\! $k^\prime=k-1$). The assignement of $k$'s
may be done consistently since, $\cp(G)$ is finite--dimensional (here it is crucial that $G$ is Dynkin) and hence all closed cycles are nilpotent. Again, the bricks of $\mathscr{L}(\lambda=0)$ with $r_{i,k}=p$ for all $i,k$ have the quantum numbers of the $W$--bosons of $G$.
\subsubsection{Non--perturbative completion}

The twisted $\ct$--valued representations of $\cp(G)$ give just the light category $\mathscr{L}$ of the $\widehat{A}(p,1)\boxtimes G$ model. Physically, one is interested to a \textsf{META}--quiver interpretation of the total non--perturbative category, which includes, besides the light objects, also heavy ones carrying non--zero magnetic charge. In the language of \cite{cattoy}, this corresponds to the non--perturbative completion of $\mathscr{L}$.  The naive choice $\mathscr{C}=\mathsf{mod}\,\C \widehat{A}(p,1)$ will \textit{not} work, since the Auslander--Reiten translation $\tau$ is not an autoequivalence for this module category\footnote{\ Because of the presence of projective and injective modules.}. This problem may be fixed by recalling the derived equivalence \cite{LE1}
\begin{equation}
D^b\big(\mathsf{mod}\,\C \widehat{A}(p,1)\big)=D^b\big(\mathsf{Coh}(X_p)\big),
\end{equation}
where  $\mathsf{Coh}(X_p)$ is the Abelian category of coherent sheaves on $X_p$, the projective line with a marked point (say the origin $\lambda=0$) of weight $p$ (this means that the skyskraper sheaf at $\lambda=0$ has length $p$ in $\mathsf{Coh}(X_p)$). 
In the category $\mathsf{Coh}(X_p)$ $\tau$ is an autoequivalence given by the tensor product with the dualizing sheaf $\omega_{X_p}$. Then the category of the $\tau$--twisted $\mathsf{Coh}(X_p)$--valued $\cp(G)$ representations makes sense, and gives the non--perturbative closure of the light category $\mathscr{L}$ in the sense of \cite{cattoy} (the case $G=A_1$ is discussed in  that paper). 

This construction may be generalized by taking the ($\tau$--twisted) representations of $\cp(G)$ valued in the category
$\mathsf{Coh}(X_{p_1,p_2,\dots,p_s})$ of the coherent sheaves over a weighted projective line with $s$ marked points of weights $p_1,p_2,\dots, p_s$ (a marked point of weight 1 being equivalent to an unmarked one) \cite{LE1}.

As we shall discuss in section \ref{rrrrmore}, the  ($\tau$--twisted) representations of $\cp(G)$ valued in the category
$\mathsf{Coh}(X_{p_1,p_2,\dots,p_s})$ will correspond to the (non--perturbative) category of SYM gauging the diagonal symmetry group $G$ of a collection of $s$ decoupled $D_{p_i}$ systems the ranks $p_i$ being equal to the weights of the marked points.  

\subsubsection{The product $\circledast$}

We saw above that, in order to capture the non--perturbative physics of the 4d $\cn=2$ model corresponding to the direct sum of an $ADE$ 2d minimal model and the $W_1=e^{pZ}+e^{-Z}$ one (or any other affine (2,2) theory \cite{CV92}), we may consider the $\tau$--twisted representations of $\cp(G)$ valued in the coherent sheaves of the geometry associated to $W_1$. This procedure is a kind of product, which `morally' is the same as the triangle tensor product $\boxtimes$. We shall denote it by the symbol $\circledast$.
So, if $A$ stands for the 4d model whose quiver with potential has the property
\begin{equation}
D^b\big(\mathsf{rep}(Q,\cw)\big)=D^b\big(\mathsf{Coh}(X_{p_1,p_2,\dots,p_s})\big),
\end{equation}
we write 
$$A\circledast G$$
to denote both the \text{META}--quiver
\begin{equation}
\big(\cp(G), \mathsf{Coh}(X_{p_1,\dots,p_s}),\tau\big),
\end{equation}
as well as the corresponding 4d $\cn=2$ QFT associated to the direct sum 2d theory. We shall use this construction in section 4.

\subsection{The $D_p(G)$ category}\label{xxx324}

The matter $\cn=2$ theory $D_p(G)$ has its own quiver with superpotential $(Q_\text{mat.},\cw_\text{mat.})$. They are characterized by the property that there is a matter functor $\mathscr{M}$
\begin{equation}
 \mathscr{M}\colon \mathsf{rep}(Q_\text{mat.},\cw_\text{mat.})\rightarrow \mathsf{rep}(Q_{p,G},\cw^\prime),
\end{equation}
which preserves indecomposable modules and iso--classes (in the RT jargon, one says that the functor $\mathscr{M}$ \emph{insets} indecomposable modules) as well as the quantum numbers
\begin{equation}
\boldsymbol{\dim}\,\mathscr{M}(X)=\boldsymbol{\dim}\,X,
\end{equation}
and such that, if
  $X$ is an indecomposable module of $\mathsf{rep}(Q_\text{mat.},\cw_\text{mat.})$, then $\mathscr{M}(X)$ is a $\lambda=0$ indecomposable module of $\mathsf{rep}(Q_{p,G},\cw^\prime)$ which is \emph{rigid} in the $\lambda$ direction, that is, $\mathscr{M}(X)$ cannot be continuously deformed to a $\lambda\neq 0$ module.

Since the simple representations with support at a single node of $Q_{p,G}$ are rigid, and there are no other simple repr.'s, the Gabriel quiver \cite{ASS1} of the matter module category has the same nodes as $Q_{p,G}$. It has also the same arrows, since $Q_{p,G}$ is simply--laced and the modules with support on the (full) $A_2$ subquivers are obviously rigid. Thus the matter quiver is simply
\begin{equation}Q_\text{mat.}\equiv Q_{p,G}.\end{equation}

The superpotential, however, should be modified to rigidify the parameter $\lambda$. In the case $G=A_1$, where the quiver $Q_{p,A_1}$ is just the $\widehat{A}_{p-1}$ affine Dynkin graph with the \emph{cyclic} orientation, $\lambda$--rigidity is achieved by taking
$\cw_\mathrm{mat.}=A_pA_{p-1}\cdots A_1$ \cite{cattoy}; this is consistent with the physical identifications of \cite{CV11} since $A_pA_{p-1}\cdots A_1$ is  indeed the right superpotential for the cyclic form of the $D_p$ Argyres--Douglas quiver \cite{ACCERV1}. Considering the representations of  $\mathsf{rep}(Q_{p,G},\cw^\prime)$ with support in a single affine cyclic subquiver $\widehat{A}(p,1)_a$,  and comparing with the $A_1$ case, we deduce that we have to add to the superpotential $\cw^\prime$ at least the extra term
\begin{equation}
\delta\cw=\sum_a A^{(a)}_pA^{(a)}_{p-1}\cdots A^{(a)}_1.
\end{equation}
One may wonder whether this modification is enough, or we need to add additional higher order corrections corresponding to cycles not supported in single affine cyclic subquivers. We claim that this is not the case (up to terms which do not modify the universality class of $\cw$, and hence may be ignored as far as the BPS spectrum is concerned).

To substantiate the claim, the first thing to check is that the modified superpotential $\cw_\text{mat.}=\cw^{\,\prime}+\delta\cw$ \textit{does} rigidify $\lambda$ to zero. For notational convenience we write down the case $G=A_n$, but the argument goes trough if the $A_n$ Dynkin quiver is replaced by any tree. First observe that the maps in eqn.\eqref{thisistheend} still define an element of $\mathrm{End}(X)$. Hence, for an \textit{indecomposable} module $X\in\mathsf{rep}(Q_{p,G},\cw^{\,\prime}+\delta\cw)$, we have
\begin{multline}
\lambda\cdot \mathrm{Id}_{X_{(j,a)}}+N_{(j,a)}= A^{(a)}_{j-1}A^{(a)}_{j-2}\cdots A_1^{(a)}A_p^{(a)}\cdots A^{(a)}_{j}=\\
=\partial_{A^{(a)}_j}(\cw^\prime+\delta\cw)\,A^{(a)}_j-\lambda^{(j,a)}\psi_a^{(j+1)}A_j^{(a)}+\psi^{(j)}_{a-1}\lambda^{(j,a-1)}A^{(a)}_j\equiv\\
\equiv -\lambda^{(j,a)}\partial_{\lambda^{(j,a)}}(\cw^\prime+\delta \cw)-\lambda^{(j,a)}A_j^{(a+1)}\psi^{(j)}_a+\psi^{(j)}_{a-1}\lambda^{(j,a-1)}A^{(a)}_j,
\end{multline}
where $N_{i,a}$ is nilpotent.
Taking traces we get
\begin{equation}
 \lambda\cdot\dim X_{(j,a)}= \mathrm{tr}\big(\psi^{(j)}_{a-1}\lambda^{(j,a-1)}A^{(a)}_j\big)-\mathrm{tr}\big(\psi^{(j)}_a\lambda^{(j,a)}A_j^{(a+1)}\big).
\end{equation}
Summing this relation over the nodes of $Q_{p,G}$, we get for all indecomposable module $X$
\begin{equation}\label{lambdarigidity}
 \lambda\cdot\sum_{j,a}\dim X_{(j,a)}=0
\end{equation}
which implies $\lambda=0$ rigidly. Since all its non--zero indecomposables are rigid in the $\lambda$ direction, the Abelian category $\mathsf{rep}(Q_{p,G},\cw^\prime+\delta\cw)$ is rigid in that direction. By this we mean that any object  of the category $\mathsf{rep}(Q_{p,G},\cw^{\,\prime}+\delta\cw)$ restricts in each cyclic affine subquiver to a module of the uniserial self--injective Nakayama algebra \cite{ASS1} given by the quotient of the path algebra of the \textit{cyclic} $\widehat{A}_{p-1}$ quiver by the bilateral ideal generated by all cyclic words $A^{(a)}_{i}A^{(a)}_{i+1}\cdots A^{(a)}_{i-1}$ for $i=1,\dots, p$.
Note that the module category of this Nakayama algebra is \emph{strictly larger} than the module category of the Nakayama algebra  which is the Jacobian algebra of the $D_p$ Argyres--Douglas system, the difference being that the maximal length of the composition series is now $p$ instead of $p-1$. Thus, $D_p(G)$ is not, in any sense, the tensor product of $D_p$ and $G$.

This rigidity result implies, in particular, that the limit as $\lambda\rightarrow 0$ of a $
\lambda\neq 0$ brick of $\mathsf{rep}(Q_{p,G},\cw^{\,\prime})$ does not satisfy the modified relations, and hence the superpotential $\cw^\prime+\delta\cw$ has the effect of `projecting out' the SYM sector.

To get the claim, it remains to construct the functor $\mathscr{M}$. It should be such that, for all  representations $X$ of the quiver $Q_{p,G}$ which satisfy the relations $\partial(\cw^\prime+\delta\cw)=0$, $\mathscr{M}(X)$ is a representation of the same quiver satisfying $\partial\cw^\prime=0$. We define $\mathscr{M}$ as follows: for $X$ an indecomposable $D_p(G)$ module, let $X\big|_a$ be its restriction to the $a$--th affine cyclic subquiver. Write
\begin{equation}\label{www3}
 X\big|_a=Y_a\oplus Z_a
\end{equation}
where $Y_a$ is a direct sum of indecomposables of lenght $p$, while the direct summands of $Z_a$ have lengths $\leq p-1$. Then $\mathscr{M}(X)$ is obtained from $X$ by the arrow replacement
\begin{equation}
 \lambda^{(\ast,a-1)}\rightarrow \lambda^{(\ast,a-1)}P_{Z_a},\qquad\quad \psi_a^{(\ast)}\rightarrow \psi_a^{(\ast)}P_{Z_a}, 
\end{equation}
 where $P_{Z_a}$ is the projection on the second summand in eqn.\eqref{www3}. One checks that $\mathscr{M}$ has the desired properties, and that $\dim \mathscr{M}(X)\neq \sum_a n_a \dim W_a$ for all bricks $X$. A more intrinsic way of stating these properties is described in the next subsection.

\subsubsection{Deformed preprojective \textsf{META}--algebras
\textit{vs.}\! $D_p(G)$ SCFT's}

In the context of the standard $\mathsf{vec}$--valued representations, the preprojective algebra $\cp(L)$ has a generalization, called the \textit{deformed preprojective algebra of weight $\lambda$}, written $\cp(L)^\lambda$, which is defined by the same double quiver $\overline{L}$ as $\cp(L)$ and the deformed relations  \cite{defr1,defr2,defr3,rump}
\begin{equation}\label{fayettt}
\sum_a (a\,a^*-a^*\,a)=\sum_i\lambda_i\,e_i\equiv \lambda
\end{equation}
where $e_i$ is the lazy path at the $i$--th node of $\overline{L}$, and the fixed complex numbers $\lambda_i$'s are the weights (also called Fayet--Illiopoulos terms). In more abstract terms we may say that the \textsc{rhs} of \eqref{fayettt} is a sum over the nodes $i$ of $\overline{L}$ of \underline{fixed} \emph{central} elements of $\mathrm{End}(X_i)$
(the endomorphism ring of the object at node $i$, not to be confused with the End for the representation of the total quiver). 

The \textsf{META} counterpart of this construction is to consider the category of the representations of $\overline{L}$ valued in some $\C$--linear Abelian category $\mathscr{C}$ satisfying the relation \eqref{fayettt}, where the \textsc{rhs} is replaced by a sum of prescribed central endomorphisms of the objects at each node.
In addition, the representation may be twisted by autoequivalences of $\mathscr{C}$ as in sect.\ref{modern}.

It is convenient to restrict ourselves to categories $\mathscr{C}$ having a `trace' map which generalizes the usual trace of $\mathsf{vec}$. That is, for each object $\co\in\mathscr{C}$ we require the existence of  a map
\begin{equation}
\mathrm{Tr}\colon \mathrm{End}(\co)\rightarrow \C,
\end{equation} 
which is invariant under the adjoint action of $\mathrm{Aut}(\co)$ and has the trace property
\begin{equation}\label{tracepro}
\mathrm{Tr}(\ca\cb)=\mathrm{Tr}(\cb\ca), \qquad \forall\;\ca\in\mathrm{Hom}(\co_1,\co_2),\ \ \cb\in \mathrm{Hom}(\co_2,\co_1),
\end{equation}
(the trace in the \textsc{lhs}, resp.\! \textsc{rhs}, being taken in $\mathrm{End}(\co_2)$, resp.\! $\mathrm{End}(\co_1)$).
\smallskip

We are particularly interested in the following family of categories. $\mathscr{V}(p)$ is the category whose objects are the pairs $\co\equiv (V,A)$, where $V$ is a $\Z_p$--graded vector space and $A\colon V\rightarrow V$ is a \emph{degree} 1 linear map. Its morphisms $\Psi\colon \co_1\rightarrow \co_2$ are given by $\Z_p$--graded linear maps $\psi\colon V_1\rightarrow V_2$ which satisfy the compatibility condition
\begin{equation}\label{compatibility}
\psi(A_1 z)=A_2\,\psi(z)\qquad \forall\, z\in V_1,
\end{equation}
with the obvious compositions and identities. Equivalently, $\mathscr{V}(p)$ is the category of the finite--dimensional representations of the \emph{cyclic} affine quiver $\widehat{A}(p,0)$ (no relations).
It follows from \eqref{compatibility} that the degree $k\;\mathrm{mod}\,p$ endomorphisms $E^k\colon (V,A)\rightarrow(V,A)$ given by the maps $A^k\colon V\rightarrow V$ ($k\in \Z_+$) belong to the center of $\mathrm{End}(V,A)$. It makes sense, therefore, to consider the $\mathscr{V}(p)$--valued representations of the deformed $\cp(L)$ with graded weights of the form $\sum_i \lambda_{i}\,E^s_i$, where $E_i$ stands for $E$ acting on the object $\co_i\in\mathscr{C}$ sitting at the node $i$ of $\overline{L}$. 

$\mathscr{V}(p)$ has a natural autoequivalence $\sigma$ which acts on objects as $([1],\mathrm{Id})$ where [1] is the operation of shifting the degree by 1. Clearly $\sigma^p=1$.
$\mathscr{V}(p)$ has also a natural trace map which on the degree zero endomorphism $\Psi=(\psi)$ is simply
\begin{equation}
\mathrm{Tr}(\Psi)\equiv \mathrm{tr}(\psi),
\end{equation}
which clearly satisfies the trace property \eqref{tracepro}. We generalize the trace to the  endomorphisms of degree $-k\;\mathrm{mod}\,p$
by replacing $\mathrm{Tr}(\Psi)$ with 
\begin{equation}
 \mathbb{T}\mathrm{r}(\Psi)= \mathrm{Tr}(E^k\Psi),\qquad \Psi\in \text{End}(\co)\ \text{of degree }-k\!\!\mod p,\ \ 0\leq k<p.
\end{equation}
$\mathbb{T}\mathrm{r}(\cdot)$ still satisfies the trace property since $E$ is central in the endomorphism ring.

The Abelian category of (ordinary) representations of the quivers $Q_{p,G}$ with superpotential $\cw_\text{matter}$, that is, the Abelian category of the $D_p(G)$ theories, is then manifestly equivalent to the category of \emph{twisted} $\mathscr{V}(p)$--valued representations of the deformed preprojective algebra $\cp(G)^\lambda$ with the degree $-1$ weight
\begin{equation}\label{555ggg}
\lambda=\sum_iE_i^{p-1}
\end{equation}
where the direct arrows of $\overline{G}$ have valuation Id and the inverse ones $\sigma^{-1}$.

If $t$ is a non--zero complex number, the weights $\lambda$ and $t\,\lambda$ produce equivalent representation categories. In the case of $\mathsf{vec}$--valued representations the basic result  on the deformed preprojective algebras of a Dynkin graph $G$ \cite{defr1,defr2,rump} says that an indecomposable module $X_0$ of the undeformed preprojective algebra (at $t=0$) may be continuously deformed to a module $X_t$ of the $t\neq 0$ one if and only if the trace of the weight $\lambda$ vanishes on $X_0$ (the trace of the weight being defined, of course, as the sum of the traces of the endomorphisms at each node). That this is necessary follows from taking the trace of the two sides of eqn.\eqref{fayettt}. If the trace obstruction vanishes, one constructs $X_t$ order by order in $t$, the procedure stopping since the arrows of $X_t$ have a polynomial dependence on $t$ \cite{rump}.

After replacing ordinary $\mathsf{vec}$--valued representations by twisted $\mathscr{V}(p)$--valued ones with the weight \eqref{555ggg}, the corresponding statement is that the (twisted) trace of the weight $\lambda$, seen as a (sum of) degree $-1$ endomorphism(s), is an obvious obstruction to the deformation of the representations for $t=0$ to $t\neq 0$. This is exactly the $\lambda$--rigidity result of eqn.\eqref{lambdarigidity}.
It is not true, however, that all indecomposable $t=0$ $\mathscr{V}(p)$--valued representations with
$\mathbb{T}\mathrm{r}(\lambda)=0$ may be deformed to $t\neq 0$ ones. In facts, as we argued in the previous subsection, only the $\lambda$--rigid one may be deformed. Let us check that there is an obstruction to the deformation at $t\not= 0$ of the representations\footnote{\ We identity the category $\ct(\lambda=0)$ with the category of nilpotent representations of $\widehat{A}(p,0)$.} corresponding to the $W$--bosons of $G$ (cfr.\! eqns.\eqref{socle1}\eqref{socle2}). The weight at node $i$ is (cfr.\! eqn.\eqref{555ggg})
\begin{equation}
0\neq \lambda_i\equiv E_i^{p-1}= \bigoplus_s A^{p-1}\Big|_{W(k_{i,s})}\in \bigoplus_s \mathrm{End}\big(W(k_{i,s})\big),
\end{equation}
that is, $\lambda_i$ is block--diagonal and \emph{non--zero},
 while
\begin{equation}
 (aa^*-a^*a)\Big|_{i\ \text{node}}\in \bigoplus_s \mathrm{Hom}\big(W(k_{i,s}),W(k_{i,s}-1)\big),
\end{equation}
\textit{i.e.}\! it is block \emph{off}--diagonal. Hence the constraint
\begin{equation}
 \sum_a(aa^*-a^*a)=t\sum_i\lambda_i,
\end{equation}
cannot be satisfied for $t\neq0$.
Of course, this just says that the deformation $t\lambda$ projects out the $W$--bosons from the light category $\mathscr{L}(\lambda=0)$; what remains is the correct $D_p(G)$ BPS category we were looking for.

\subsection{Quantum monodromies, beta function, and flavor charges}\label{DpGflavor}

\begin{table}
\begin{center}
\begin{tabular}{|c| c |}
\hline
 & $\chi_G(X)$\\
\hline
$A_n$ & $\phantom{\Bigg|}\prod_{d \mid (n+1) \atop d \neq 1}{\bf \Phi}_d(X)$\\
\hline
$D_n$ & $\phantom{\Bigg|}{\bf \Phi}_2(X) \prod_{d  \mid  2(n-1) \atop d  \nmid  (n-1) } {\bf \Phi}_d(X)$\\
\hline
$E_6$ & $\phantom{\Big|}{\bf \Phi}_3(X){\bf \Phi}_{12}(X)$\\
\hline
$E_7$ & $\phantom{\Big|}{\bf \Phi}_2(X){\bf \Phi}_{18}(X)$\\
\hline
$E_8$ & $\phantom{\Big|}{\bf \Phi}_{30}(X)$\\
\hline
\end{tabular}
\end{center}
\caption{Factorization of the ADE characteristic polynomials $\chi_G(X)$.}\label{ADEcharacteristic}
\end{table}

The 2d quantum monodromy of the $\widehat{A}(p,1)\boxtimes G$ system is the tensor product of the monodromies of the factors. Then \cite{CV92,cattoy}
\be\label{Coxtensor}
H_{p,G}\equiv \Phi_{\widehat{A}(p,1)} \otimes \Phi_G.
\ee
where $\Phi_Q$ stands for the Coxeter element of the acyclic quiver $Q$.
Let us denote by ${\bf \Phi}_d(X)$ the $d$-th cyclotomic polynomial. The characteristic polynomial of the Coxeter element of $\widehat{A}(p,1)$ is \cite{LE1}
\be\label{characteristicAp1}
\text{det}\big[z - \Phi_{\widehat{A}(p,1)}\big] = {\bf \Phi}_1(z) \prod_{d | p} {\bf \Phi}_d(z).
\ee
For each $G=ADE$ we define a function $\d(d;G): \mathbb{N} \to \mathbb{Z}_{\geq 0}$ by the formula
\be\label{deltaG}
\chi_G(z) \equiv \text{det}\big[z - \Phi_G\big] = \prod_{d\in\mathbb{N}} {\bf \Phi}_d(z)^{\d(d;G)},
\ee
see table \ref{ADEcharacteristic}. 
Putting together the last two equations we get
\begin{equation}\label{pppp1}
\det[z-H_{p,G}]=\chi_G(z)\,\prod_{d\in\mathbb{N}}\prod_{\ell\mid d}\prod_{j\mid p}\big(z^{\mathrm{lcm}(\ell,j)}-1\big)^{\gcd(\ell,j)\,\mu(d/\ell)\,\delta(d;G)},
\end{equation}
where $\mu(d)$ is the M\"obius function.
In section \ref{2d4drev} we saw that the rank $f(p;G)$ of the flavor group $F$ of the $\widehat{A}(p,1)\boxtimes G$ model is the multiplicity of $z=1$ as a root of the \textsc{rhs} of \eqref{pppp1}, that is,
\begin{equation}\label{pppp2}
f(p,G)=\sum_{d\in\mathbb{N}}\sum_{\ell\mid d}\sum_{j\mid p}\gcd(\ell,j)\,\mu(d/\ell)\,\delta(d;G)\equiv \sum_{d\mid p}\phi(d)\,\delta(d;G),
\end{equation}
where $\phi(d)$ is the totient function and we used again the M\"obius sum formula.

Then, after decoupling the $G$ Yang--Mills sector, we remain with a matter system, $D_p(G)$, which has flavor symmetry $G\times F$ whose rank is
\begin{equation}
f(D_p(G))=r(G)+f(p,G).
\end{equation}

From the 2d monodromy $H_{p,G}$ it is easy to compute the $\beta$ function of the Yang--Mills coupling $g$ for the model $\widehat{A}(p,1)\boxtimes G$. Indeed, since the matter $D_p(G)$ is superconformal, in the weak YM coupling limit, $g\rightarrow 0$, the trace of the energy--momentum tensor is proportional to the YM $\beta$--function. Then, by $\cn=2$ supersymmetry, the coefficient $b$ of the $\beta$--function
$$\mu\frac{\partial\, \tau}{\partial\mu}=\frac{i}{2\pi}\,b,$$
is the same as the coefficient of the chiral $U(1)_R$ anomaly which counts the net chiral number of Fermi zero--modes in the instanton background. A $U(1)_R$ rotation by $2\pi n$ ($n\in\Z$) is equivalent to a shift of the vacuum angle $\theta$ by $2\pi b\,n$, which has the effect of changing the electric/magnetic charges of a BPS dyon as 
\cite{WI}
\begin{equation}\label{tttrr}
(e,m)\rightarrow (e+b\,n\,m,m).
\end{equation}
Thus, if we know the action of a chiral $2\pi\,n$ rotation on the charge lattice $\Gamma$ (that is, on the dimensions of the corresponding modules), we may extract the coefficient $b$.

The 2d monodromy $H_{p,G}$ acts on the CY $3$--form $\Omega$ of the geometry $W_{p,G}=0$ (cfr.\! eqn.\eqref{Ap1Geom}) as \cite{CNV}
\begin{equation}\label{oomega}\begin{split}
\Omega=\frac{dX\wedge dY\wedge dU}{\partial_Z W} &\mapsto \exp\!\big(2\pi i(q_X+q_Y-1/2)\big)\: \frac{dX\wedge dY\wedge dU}{\partial_Z W}\equiv\\
&\equiv \exp(2\pi i/h(G))\:\frac{dX\wedge dY\wedge dU}{\partial_Z W},
\end{split}\end{equation}
where $h(G)$ is the Coxeter number of $G$. Hence the action of the chiral $2\pi\,n$ rotation, $\Omega\mapsto e^{2\pi i\,n}\,\Omega$, on the dimension/charge lattice $\Gamma$ is given by the matrix
\begin{equation}
H_{p,G}^{\,n\,h(G)}=\Phi_{\widehat{A}(p,1)}^{\,n\,h(G)}\otimes \Phi_{G}^{\,n\,h(G)}\equiv \Phi_{\widehat{A}(p,1)}^{\,n\,h(G)}\otimes 1.
\end{equation}
It is convenient to take $n=p$; we have (see \textit{e.g.}\! \cite{cattoy})
\begin{equation}
\Phi_{\widehat{A}(p,1)}^{\,p\,h(G)}=\mathrm{Id}+(p+1)\,h(G)\,\delta\otimes \langle\delta,\cdot\rangle_E
\end{equation}
where $\delta$ is the minimal imaginary root of $\widehat{A}_p$, which corresponds to a purely electric charge, while the form
$\langle\delta,\cdot\rangle_E$ measures the magnetic charge \cite{cattoy}. Comparing with eqn.\eqref{tttrr}, taking care of the appropriate normalizations, we get
\begin{equation}
p\,b=(p+1)\,h(G),
\end{equation}
that is, the $\beta$ coefficient $b$ of the $\widehat{A}(p,1)\boxtimes G$ model is
\begin{equation}\label{eeeee6}
b=\frac{p+1}{p}\,h(G),
\end{equation}
whose sign implies asymptotic freedom. $b$ receives a contribution $2h(G)$ from the SYM sector and a negative contribution $-b_{p,G}=-k_{p,G}/2$ from the matter $D_p(G)$ SCFT. We use eqn.\eqref{eeeee6} to extract the central charge of the $G$--current algebra of the SCFT $D_p(G)$
\begin{equation}
k_{p,G}=2\,\frac{p-1}{p}\,h(G).
\end{equation}

On the other hand, a $U(1)_R$ rotation by $2\pi$ defines the 4d quantum monodromy $\mathbb{M}$ \cite{CNV}. For the model $\widehat{A}(p,1)\boxtimes G$, which is just asymptotically--free, $\mathbb{M}$ has not finite order; however, once we decouple the SYM sector, we remain with  the SCFT $D_p(G)$ whose 4d quantum monodromy has a finite order $r(p,G)$. As we saw above, the action of a $U(1)_R$ rotation by $2\pi$ on the charges of $D_p(G)$ is given by the semi--simple part of the $h(G)$ power of $H_{p,G}$, that is, by
\begin{equation}
\Phi_{\widehat{A}(p,1)}^{h(G)}\Big|_\text{semi--simple}\otimes 1.
\end{equation}
The order of the 4d quantum monodromy $\mathbb{M}$ is just the order of this operator. Comparing with eqn.\eqref{characteristicAp1} we get \cite{infinitelymany}
\begin{equation}\label{eeeee7}
r(p,G)=\frac{p}{\gcd\{p,h(G)\}}.
\end{equation}

\section{Several $D_p(G)$ matter subsectors}\label{rrrrmore}
One could ask what happens if we couple more than one $D_p(G)$ system to a $G$ SYM subsector.  Suppose we gauge the diagonal group $G$ of $\ell$ subsystems of type $D_{p_i}(G)$. The requirement of no Landau poles gives
\be\label{UVcomplete}
b = \Big(2 - \sum_{i=1}^{\ell} \frac{p_i-1}{p_i}\Big)h(G) \geq 0
\ee
The solutions to this condition are listed in table \ref{moreThanOne}. The allowed $\ell$-uples of $p_i$ are well--known in representation theory: the numbers we obtain for asymptotically--free (resp.\! for conformal) theories are precisely the tubular types of the $\mathbb{P}^1$ families of regular representations of Euclidean (resp.\! tubular) algebras \cite{LE1,ringel}.

\begin{table}
\begin{center}
\begin{tabular}{|c|c|c|c|}
\hline
$\ell$ & allowed $p_i$ & UV behavior & $\cc$\\
\hline
2 & $(p,q)$ : $p\geq q\geq 1$ & AF & $A(p,q)$\\
\hline 
3 & $(2,2,p)$ : $p \geq 2$ & AF & $\widehat{D}_{p+2}$\\
&(2,3,3)&AF&$\widehat{E}_6$\\
&(2,3,4)&AF&$\widehat{E}_7$\\
&(2,3,5)&AF&$\widehat{E}_8$\\
&(3,3,3)&CFT& $E^{(1,1)}_6$\\
&(2,4,4)&CFT&$E^{(1,1)}_7$\\
&(2,3,6)&CFT&$E^{(1,1)}_8$\\
\hline
4&(2,2,2,2)&CFT&$D^{(1,1)}_4$\\
\hline
\end{tabular}
\end{center}
\caption{Solutions to eqn.\eqref{UVcomplete} and corresponding algebras.}\label{moreThanOne}
\end{table}

Therefore the asymptotically--free model one gets are precisely the $\widehat{H}\boxtimes G$ models of \cite{infinitelymany}. Such models consist of a $G$ SYM subsector weakly gauging the $G$-flavor symmetry of several $D_{p_i}(G)$ matter systems according to the following table:
\be\label{pqrnumbs}
\begin{tabular}{|c c|c|}
\hline
&& superconformal system\\
\hline
$A(p,q)\boxtimes G$ & $p\geq q \geq 1$ & $D_p(G) \oplus D_q(G) \oplus D_1 (G)$\\
\hline
$\widehat{D}_r\boxtimes G$ & $r \geq 4$ & $D_2(G)\oplus D_2(G) \oplus D_{r-2}(G)$\\
\hline
$\widehat{E}_r \boxtimes G$ & $r=6,7,8$ & $D_2(G)\oplus D_3(G) \oplus D_{r-3}(G)$\\
\hline
\end{tabular}
\ee

The superconformal models correspond to the 2d theories which are the direct sums of the minimal $G$--models with the ones associated to the four elliptic complete SCFT's \cite{CV11}
\begin{equation}
D_4^{(1,1)},\quad E_6^{(1,1)},\quad E_7^{(1,1)},\quad E_8^{(1,1)}.
\end{equation}
Equivalently, they may be defined as the models
\begin{equation}
D_4^{(1,1)}\circledast G,\quad E_6^{(1,1)}\circledast G,\quad E_7^{(1,1)}\circledast G,\quad E_8^{(1,1)}\circledast G.
\end{equation}
Their `messy' $\mathsf{vec}$--quivers $Q_\text{BPS}$
may also be easily written down since $D_4^{(1,1)}$ is a Lagrangian theory ($SU(2)$ SQCD with $N_f=4$) while \cite{CV11}
\begin{equation}
E_6^{(1,1)}=D_4\boxtimes A_2,\quad E_7^{(1,1)}=A_3\boxtimes A_3,\quad E_8^{(1,1)}=A_2\boxtimes A_5,
\end{equation}
 which allows to write their elliptic Stokes matrices $S_\mathrm{ell}$ as tensor products of Dynkin ones. The BPS quiver of the SCFT model $H^{(1,1)}\circledast G$ is then given by the exchange matrix
\begin{equation}
B=S_\mathrm{ell}^t\otimes S_G^t-S_\mathrm{ell}\otimes S_G.
\end{equation} 
The periods $p_i$'s of the $D_{p_i}$ matter subsectors may be read from the characteristic polynomial of the Coxeter of the corresponding affine/toroidal Lie algebra \cite{LE1}
\begin{equation}
\det[z-\Phi]=(z-1)^2\prod_i \frac{z^{p_i}-1}{z-1}
\end{equation}
see table \ref{ellip}. This equation also implies that the rank of the flavor group of a $\widehat{H}\boxtimes G$ QFT (resp.\! $H^{(1,1)}\circledast G$ SCFT) is additive with respect to the matter subsectors
\begin{equation}
\mathrm{rank}\,F=\sum_i f(p_i,G),
\end{equation}
where $f(p,G)$ is the function defined in \eqref{pppp2}.

From the point of view of section \ref{modern}, the module category for $G$ SYM coupled to $\oplus_{i=1}^\ell D_{p_i}(G)$ may be more conveniently realized as the ($\tau$--twisted) representations of the preprojective algebra $\cp(G)$ valued in the  Abelian category $\mathsf{Coh}(X_{p_1,\dots,p_\ell})$.

\begin{table}\centering
\begin{tabular}{|c|c|}
\hline
& superconformal system\\
\hline
$D^{(1,1)}_4\circledast G$ & $D_2(G)\oplus D_2(G)\oplus D_2(G)\oplus D_2(G)$\\ 
\hline
$E^{(1,1)}_6\circledast G$ & $D_3(G)\oplus D_3(G)\oplus D_3(G)$\\
\hline
$E^{(1,1)}_7\circledast G$ & $D_2(G)\oplus D_4(G)\oplus D_4(G)$\\
\hline
$E^{(1,1)}_8\circledast G$ & $D_2(G)\oplus D_3(G)\oplus D_6(G)$\\
\hline
\end{tabular}
\caption{\label{ellip} Matter content of the $G$--tubular SCFT's.}
\end{table}

%
%
%
%
%
%

\subsection{The BPS spectrum of $\widehat{H}\boxtimes G$ models at strong coupling}
All the models of type $\widehat{H}\boxtimes G$ admit a finite BPS chamber containing only hypermultiplets with charge vectors
\be
e_a \otimes \a \in \Gamma_{\widehat{H}} \otimes \Gamma_G, \qquad \a \in \Delta_+(G),
\ee
that is, a copy of the positive roots of $G$ per each simple root (node) of $\widehat{H}$. We get a finite chamber with
\be
\#\{ \text{ hypermultiplets } \} = \frac{1}{2} \, r(G)\,h(G)\,r(\widehat{H}).
\ee
This result follows from the mutation algorithm of \cite{ACCERV1,ACCERV2}. The details of the computation are rather technical: The mutation sequences corresponding to these finite BPS-chambers are constructed in appendix \ref{BPSpec}.

\section{Geometry of $D_p(G)$ SCFT's for $G$ a \emph{classical} group}
In the previous section we have defined the four-dimensional $\cn=2$ superconformal systems $D_p(G)$ from the study of the light subcategory of the $\widehat{A}(p,1)\boxtimes G$ models. The underlying $(2,2)$ system of the $\widehat{A}(p,1)\boxtimes G$ was given in \eqref{Ap1Geom} by
$$
e^{-Z} + e^{p Z} + W_G(X,Y) + U^2 = \text{ lower terms in } X,Y. 
$$
In particular we have shown that the $\widehat{A}(p,1)\boxtimes G$ system has a $G$ SYM subsector. Geometrically, by the scaling arguments of \cite{Tack}, we would expect that the size of the cylinder in the $Z$ coordinates of \eqref{Ap1Geom} is related to the size of the $G$ SYM coupling. Thus, formally, the $D_p(G)$ system could be described by the engineering of the Type II B superstring on the limit $Z \to \infty$ of the geometry \eqref{Ap1Geom}:
\be\label{conformalgeom}
W_{p,G} \equiv e^{pZ} + W_G(X,Y) + U^2 + \text{ lower terms }=0,
\ee
with holomorphic top form given as in eqn.\eqref{oomega}.
%
%
In this section, we explore the consequences of this geometric picture when $G$ is a (simple, simply--laced) classical Lie group.

\subsection{The identification of the $D_p(G)$ Lagrangian subclass}\label{DpGlagra}
The chiral operators of a 4d $\cn=2$ SCFT with a weakly coupled Lagrangian formulation have integral dimension. Hence the order of its 4d quantum monodromy is necessarily $1$ \cite{CNV} which --- in view of eqns.\eqref{eeeee6}\eqref{eeeee7} --- is equivalent to $b\in\mathbb{Z}$. For the $D_p(G)$ models this statement has a partial converse.
Indeed, we claim that:
\begin{itemize}\item{A model $\widehat{A}(p,1)\boxtimes A_{N-1}$ is Lagrangian iff $b_{p,SU(N)}$ is an integer.} \item{A model $\widehat{A}(p,1)\boxtimes D_N$ is Lagrangian iff $b_{p,SO(2N)}$ is an \textit{even} integer.} \item {No model $\widehat{A}(p,1)\boxtimes E_r$ is Lagrangian.}
\end{itemize}

\medskip

The statement that $\widehat{A}(p,1)\boxtimes E_r$ has no Lagrangian formulation is elementary.  The theories of type $D_p(G)$ have $b_{p,G} < h(G)$: If a $D_p(G)$ theory is Lagrangian, its contribution to the YM $\beta$--function should be equal to the $U(1)_R$ anomaly coefficient $b(R)$ of a free hypermultiplet in some (generally reducible) representation $R$ of $G$. Since $E_8$ has no non--trivial representation with $b(R)<30$ this is impossible. For $E_7$ the only representation with $b(R)<18$ are the $\tfrac{k}{2}\,\mathbf{56}$, $k=1,2$ with $b = 6k$. Then, in order to have a Lagragian model,
\be
b_{p,E_7} = \frac{p-1}{p} 18 = 6 k \Longleftrightarrow (3 - k )p = 3 \Longrightarrow k = 2, p = 3.
\ee
A Lagrangian theory with two half--hypers would have flavor symmetry at least $SO(2)$; but the theory $A(3,1)\boxtimes E_7$ has $f(3,E_7)=0$ by \eqref{pppp2}, and therefore the model $D_3(E_7)$ cannot be Lagrangian. Finally, the only $E_6$ representation with $b(R)<12$ is the {\bf 27} with $b=6$; to have a Lagrangian model
\be
b_{p,E_6} \equiv \frac{p-1}{p} 12 = 6 \Longleftrightarrow p = 2,
\ee
which would imply $F$ at least $U(1)$, while \eqref{pppp2} gives $f(2,E_6)=0$.

Let us now proceed to show the claim for $G=SU(N), SO(2n)$.

\subsubsection{The case $G=SU(N)$}\label{SUlagR}
We have to show that
\be\label{lagraSUN}
\begin{gathered}D_p(SU(N)) \text{ admits } \\ \text{ a Lagrangian formulation}\end{gathered} \ \Longleftrightarrow \ h(SU(N)) = N = m p.
\ee
Consider the $(2,2)$ superpotentials of type $W_{p,SU(N)}$: Only if $N=m p$ the corresponding $(2,2)$ system admits, at the conformal point, several \emph{exactly marginal} deformations. By 2d/4d correspondence, we known that under such deformations the quiver mutation class is invariant. Since the models of type $D_p(G)$ are defined only by the mutation class of their quivers with superpotential, properly speaking, the 2d/4d correspondence associates to a 4d model the \emph{universal} 2d superpotential $W_{p,G}(t_\alpha)$ over the space of exactly marginal/relevant deformations. The dimensions of the generators of the 2d chiral ring $\mathcal{R}$ are
\be
q(X) = \frac{1}{m p} \qquad q(e^Z) = \frac{1}{p} \qquad q(Y) = q(U) = \frac{1}{2}.
\ee
The marginal deformations of $W_{p,SU(mp)}$ correspond to the operators $X^\alpha e^{\beta Z}\in\mathcal{R}$ such that
\be\label{marginaldefs}
q(X^{\a} e^{\b Z}) = 1 \Longleftrightarrow \quad \left[\begin{aligned} &\a = m (p - k)\\ &\b = k \end{aligned}\right. \quad k = 0, \dots, p.
\ee
The universal superconformal family of superpotentials is then
\be
W_{p,SU(mp)} = e^{pZ} + X^{mp} + \sum_{k=1}^{p-1} t_k\, X^{m(p-k)}e^{kZ} + Y^2 + U^2.
\ee
The Seiberg--Witten geometry that corresponds to the $\widehat{A}(p,1) \boxtimes SU(mp)$ model is
\be
e^{-Z} + W_{p,SU(mp)} = 0.
\ee
The corresponding Seiberg--Witten curve can be written as
\be\label{SWcAp1SUN}
e^{-Z} + e^{pZ} + X^{mp} + \sum_{k=1}^{p-1}t_k\,  X^{m(p-k)}e^{kZ} = \text{ relevant deformations of } W_{p,SU(mp)}
\ee
with canonical Seiberg--Witten differential $\l_{SW} = X\, dZ$. Let us change variables as follows 
\be
s = e^Z \quad v = X \Longrightarrow \l_{SW} = v\, \frac{ds}{s}.
\ee
Multiplying by $s$ the equation \eqref{SWcAp1SUN} in the new variables, we obtain the Seiberg--Witten curve
\be
1 + p_{mp}(v) s + p_{m(p-1)}(v) s^2 + \dots + p_{2m}(v)s^{p-1} + p_m(v)s^p + s^{p+1}=0, 
\ee
where the $p_i(v)$ are polynomials of degree $i$ in $v$. This is a well-known Seiberg--Witten curve (see for example eqn.(2.41) of \cite{WittenMth}), and therefore we conclude that all theories $\widehat{A}(p,1)\boxtimes A_{mp-1}$ have a Lagrangian $S$-duality frame in which they are described as the quiver gauge theory \footnote{Here, as usual, an edge $-$ denotes a bifundamental hypermultiplet.}
\be\label{lagraexplSU}
SU(mp) - SU(m(p-1)) - SU(m(p-2))- \dots -SU(2m) - SU(m).
\ee
Decoupling the first $SU(mp)$ SYM sector, we get that the only Lagrangian theories of type $D_p(SU(N))$ are
\be
D_p(SU(mp)) =\left\{\begin{gathered} \widehat{A}(p,1)\boxtimes A_{m(p-1)-1}\\ \text{ coupled to } mp \text{ fundamental } \\ SU(m(p-1)) \text{ hypers}\end{gathered}\right\}
\ee
which is indeed a SCFT as expected.
Let us perform a couple of consistency checks:
\begin{enumerate}
\item The rank of the flavor group $F$ of the theory $\widehat{A}(p,1)\boxtimes A_{mp-1}$ is
$$
\d(d,SU(mp)) = \begin{cases} 1 \text{ if } d \mid m p \text{ and } d \neq 1 \\ 0 \text{ else } \end{cases} \Longrightarrow  f(p,SU(mp)) = \sum_{d \mid p \atop d \neq 1} \varphi(d) = p - 1,
$$
which is precisely the number of bifundamentals  in the linear quiver \eqref{lagraexplSU}.
\item The rank of the gauge group of the theory \eqref{lagraexplSU} is
\be
r(G) = \sum_{k=1}^{p} (m k - 1) = \frac{m p (p+1)}{2} - p
\ee
In addition we have the $f = p-1$ hypermultiplets. The rank of the charge lattice then matches the number of nodes of the quiver $\widehat{A}(p,1)\boxtimes A_{mp-1}$:
\be
2 r(G) + f = m p (p+1) - 2 p + p - 1 =  (p+1)(mp -1). 
\ee
\end{enumerate}

\subsubsection{Lagrangian subclass for $G=SO(2N)$}\label{SOSplagR}
The necessary condition that $b_{p,SO(2N)}$ must be an even integer follows from the fact that all representations with $b<h$ of $SO(2N)$ have even Dynkin-index $b$. So,
\be
b_{p,SO(2N)} = \frac{p-1}{p}2(N-1) \in 2 \mathbb{N} \Longleftrightarrow N = m p + 1.
\ee
Again, this is precisely the case in which the corrisponding $(2,2)$ superpotential at the superconformal point admits  marginal deformations. Indeed, 
\be\label{WpD}
W_{p,D_{mp+1}} = e^{pZ} + X^{m p} + X Y^2 + U^2,
\ee
while the dimensions of the generators of $\mathcal{R}$ are
\be
q(X) = \frac{1}{mp} \qquad q(Y) = \frac{mp-1}{2 mp}  \qquad q(e^Z) = \frac{1}{p} \qquad q(U)=\frac{1}{2}.
\ee
The marginal deformations of $W_{p,D_{mp+1}}$ are those in eqn.\eqref{marginaldefs}, and the universal family of superpotentials for $\widehat{A}(p,1)\boxtimes D_{mp+1}$ is
\be
e^{-Z} + e^{pZ} + X^{mp} + \sum_{k=1}^{p-1}t_k\, X^{m(k-p)}e^{kZ} + X Y^2 + U^2 = \text{ `lower terms' }.
\ee
The generic `lower terms' have the form $2 \l Y+\text{`independent of }Y'$ for some non--zero $\lambda$. Integrating out $Y$ we obtain the equivalent geometry
\be
e^{-Z} + e^{pZ} + X^{mp} + \sum_{k=1}^{p-1}t_k\, X^{m(k-p)}e^{kZ} - \frac{\l^2}{X} + U^2 = \text{lower terms }, 
\ee
which is the Seiberg--Witten (SW) curve of the 4d theory with differential $\l_{SW} = X d Z$. Now we change variables $X = v^2$, $s = e^{Z}$, and we multiply the resulting curve by $s v^2$. The final form of the SW curve for the model $\widehat{A}(p,1)\boxtimes D_{mp+1}$ is
\be
v^2 + v^2\, s^{p+1} + \sum_{k=1}^{p} p_{m(p-k+1)+1}(v^2)\, s^k = 0,
\ee
where the $p_i(v^2)$ are polynomials of degree $2i$ in $v$. These Seiberg-Witten curves are well-known (see \textit{e.g.}\! section 3.6 of \cite{SOP}): they are part of the family
\be
v^2 + v^2\, s^{p+1} + \sum_{k=1}^{p} p_{2\ell_k + 1 + (-1)^k} (v)\, s^{k} = 0,
\ee
that corresponds to linear quiver theories of type
\be
SO(2 \ell_1) - USp(2 \ell_2) -\cdots-SO(2 \ell_{k-1})-USp(2\ell_k)-SO(2 \ell_{k+1}) - \cdots 
\ee
where the edges represents \emph{half}-hypermultiplets in the bifundamental repr. In our case
\be
\begin{gathered}
2 \ell_k + 1 + (-1)^k = 2 (m(p-k+1) + 1) \Longrightarrow \ \left[\begin{aligned}&k \text{ even} \ \colon \ \ell_k = m (p-k+1)\\ & k \text{ odd} \ \colon \ \ell_k = m(p-k+1) + 1\,\end{aligned}\right.
\end{gathered}
\ee
for $k=1,\dots,p$. In conclusion: all theories of type $\widehat{A}(p,1)\boxtimes D_{mp+1}$ have a Lagrangian description (in a suitable region of their parameter space) as the linear quiver theory
\be
\begin{footnotesize}
\begin{aligned}\label{ttt789}
SO(2mp+2)-USp(2m(p-1))&-SO(2m(p-2)+2)-USp(2m(p-3))-\cdots\\
&\cdots-SO(2m(p-2\ell)+2)-USp(2m(p-2\ell-1))-\cdots
\end{aligned}
\end{footnotesize}
\ee
The linear quiver have two possible ends, depending on the parity of $p$:
\be\label{SOSpEnds}
\begin{aligned}
& p \text{ odd } \colon \qquad \cdots - USp(4m) - SO(2m + 2)\\
& p \text{ even } \colon \qquad \cdots - SO(4m+2) - USp(2m) - \framebox{$SO(2)$}
\end{aligned}
\ee
where the box represents an ungauged flavor group. Consequenctly the only Lagrangian theories of type $D_p(SO(N))$ are the theories:
\be
D_p(SO(2mp+2)) = \left\{
{\footnotesize
\begin{aligned}
&\boxed{ \ SO(2(mp+1)) \ } - USp(2m(p-1)) - \cdots\\
&\cdots - SO(2m(p-2\ell)+2)-USp(2m(p-2\ell-1)) - \cdots\\
&\quad\text{{with the same ends as in eqn.\eqref{SOSpEnds}}}
\end{aligned}}\right\}
\ee
where again the box represents an ungauged flavor group.
A few checks are in order:
\begin{enumerate}
\item The rank of the flavor group of $\widehat{A}(p,1)\boxtimes D_{mp+1}$ is
\be
f(p,D_{mp+1}) = \begin{cases} 0 \qquad p\text{ odd }\\ 1 \qquad p\text{ even}\end{cases}
\ee
which is consistent with \eqref{ttt789}: half-hypermultiplets carry no flavor charge.
\item The rank of the gauge group of the $\widehat{A}(p,1)\boxtimes D_{mp+1}$ theory is
\be
r(G) =\frac{1}{2} \sum_{k=1}^{p} \Big( 2 (m(p-k)) + 1 - (-1)^k \Big)= \frac{1}{2}\Big(p(mp+m+1) +\frac{1}{2}\big(1-(-1)^p\big)\Big).
\ee
Therefore
\be
2 r(G) + f = p(mp+m+1) + 1 = (p+1)(mp+1)
\ee
which is the number of nodes for the quiver $\widehat{A}(p,1)\boxtimes D_{mp+1}$.
\item The beta function contribution to the $SO(2mp+2)$ gauge group is
\be
4 m p - 2 m ( p-1)  = b_{\widehat{A}(p,1)\boxtimes D_{mp+1}}.
\ee
\end{enumerate}

\section{Computing the 4d $a$, $c$, SCFT central charges}\label{accharges}
%
%
%


\subsection{4d quantum monodromy and the SCFT central charge $c$}
It is well known that for a $\cn=2$ 4d SCFT the corresponding topological theory in curved spacetime develops a superconformal anomaly which is sensitive to the topology of the background manifold \cite{ShapereTachikawa,Witten95}. If the Euler characteristic $\chi$ of the background manifold is zero, such an anomaly is proportional to the central charge $c$ of the $\cn=2$ 4d SCFT. 

The trace of the 4d quantum monodromy operator $\mathbb{M}(q)$ 
is a particular instance of topological partition function on the $R$--twisted Melvin cigar $MC_q \times_g S^1$ \cite{CNV} which has $\chi=0$. In principle, $\mathrm{Tr}\,\mathbb{M}(q)$ is uniquely fixed once we give the quiver and superpotential of the 4d theory, $(Q,\cw)$, \emph{via} the BPS spectrum (computed in any chamber) \cite{CNV}. 

It turns out that $\mathrm{Tr}\,\mathbb{M}(q)$ is equal to a Virasoro character of a 2d CFT \cite{CNV}; the effective 2d CFT central, $c_\mathrm{eff}\equiv c_\mathrm{eff}(Q,\cw)$, then measures an anomaly of the topological partition function which should correspond to the 4d SCFT one. It follows that the 2d effective central charge $c_\mathrm{eff}$ should be identified with the 4d central charge $c$, up to normalization.   
However one has to take into account, in addition, the contribution of the 
massless sector, which is omitted in the usual definition of $\mathbb{M}(q)$ \cite{CNV}. For the models of interest in this paper the massless sector consists just of the free photon multiplets, since there are no hypermultiplets which are everywhere light on the Coulomb branch. The number of the free photon multiplets is equal to the dimension of the Coulomb branch, which is $\mathrm{rank}\,B/2$, where $B$ is the exchange matrix of the quiver of the theory. Since one free vector multiplet contributes $+1/6$ to the 4d SCFT central charge $c$, we get
\begin{equation}\label{cfirst}
c=\alpha\cdot  c_\text{eff}(Q,\cw)+\frac{\mathrm{rank}\,B}{12},
\end{equation}  
where $\alpha$ is a universal normalization constant still to be determined. To fix $\alpha$ we apply this formula to a free hypermultiplet.

In ref.\!\cite{CNV} $c_\mathrm{eff}$ was computed for the models $G\boxtimes G^\prime$ ($G,G^\prime$ being Dynkin quivers)
\begin{equation}\label{csecond}
c_\mathrm{eff}(G\boxtimes G^\prime)=\frac{r(G)\,r(G^\prime)\,h(G)\,h(G^\prime)}{h(G)+h(G^\prime)}.
\end{equation}
A free hypermultiplet corresponds to $G=G^\prime=A_1$; since it has $c=1/12$, eqns.\eqref{cfirst}\eqref{csecond} give
\begin{equation}
\frac{1}{12}= \alpha\cdot \frac{1\cdot 1\cdot 2\cdot 2}{2+2}\equiv \alpha,
\end{equation}
and the final formula for $c$ is

%
%
\begin{equation}\label{cfromqumon}
c=\frac{1}{12}\Big(c_\mathrm{eff}(Q,\cw)+\mathrm{rank}\,B\Big).
\end{equation}

Comparing our analysis with \cite{ShapereTachikawa} we see that the 2d CFT central charge $c_{\text{eff}}(Q,\cw)$ captures the scale dimension of the discriminant of the SW curve at the conformal point. Our formula then may be seen as an expression for the discriminant scale in terms of Lie--theoretical invariants of $Q$.
%
%

A few examples are in order.

\medskip

{\bf Example 1}: $A_{N-1}$ Argyres--Douglas corresponds to $G=A_{N-1}$, $G^\prime=A_1$ in eqn.\eqref{csecond}. Then
\begin{equation}
c_\mathrm{eff}=\frac{2N(N-1)}{N+2}
\end{equation} 
and $\mathrm{rank}\,B=2[(N-1)/2]$, so
\begin{align}
&c=\frac{\ell(6\ell+5)}{6(2\ell+3)} \quad\text{for}\quad N=2\ell+1\\
&c=\frac{(2\ell-1)(3\ell+1)}{12(\ell+1)} \quad\text{for}\quad N=2\ell
\end{align}
which are the values reported in \cite{Xie:generalAD}.
\medskip

{\bf Example 2}: $D_{N}$ Argyres--Douglas
corresponds to $G=D_N$, $G^\prime = A_1$.  Thus
\be
c_\text{eff}=\frac{2 N\cdot  2(N-1)}{2 + 2(N-1)} = 2 (N-1)
\ee
and $\mathrm{rank}\,B=2[(N-2)/2]$, so
\begin{align}\label{wwwrt}
& c = \frac{1}{6}(3\ell - 2)\quad\text{for}\quad  N = 2 \ell\\
& c = \frac{1}{2}\,\ell \quad\text{for}\quad N = 2 \ell + 1,
\end{align}
in agreement with \cite{Xie:generalAD}.

\medskip

{\bf Example 3}: Nore generally, for all models $A_{k-1}\boxtimes A_{N-1}$ our formula reproduces the value of $c$ conjectured by Xie \cite{Xie:generalAD}.

\subsection{Generalization to $\widehat{H}\boxtimes G$}

Unfortunately no one has computed $c_\mathrm{eff}$ for $\cn=2$ models more general than the $G^\prime\boxtimes G$ ones. However in this paper we are interested only in the slightly more general case where the finite--dimensional Lie algebra $G^\prime$ is replaced by  
the infinite dimensional Kac--Moody Lie algebra $\widehat{H}$, our prime application being to $\widehat{H}=\widehat{A}(p,1)$. The similarity with the case analyzed in \cite{CNV} suggests that $c_\mathrm{eff}(\widehat{H}\boxtimes G)$ is still expressed in terms of Lie--theoretic invariants of the two algebras $G$ and $\widehat{H}$, in facts by eqn.\eqref{csecond} where the invariants $r(G^\prime)$ and $h(G^\prime)$ of $G^\prime$ are replaced by the appropriate invariants of $\widehat{H}$.

In order to make the correct replacements, we have to return to the computation leading to \eqref{csecond} was computed, and to track the origin of each Lie--theoretical quantity in the \textsc{rhs} of \eqref{csecond}. The Coxeter numbers in the numerator arise as (twice) the number of quiver mutations we need to perform to get the complete BPS spectrum (in each one of the two canonical chambers) by the mutation algorithm of \cite{ACCERV2,ACCERV1}. In other words, it is (twice) the number of mutations after which the mutation algorithm (in those chambers) \emph{stops.} One may see this number also as (twice) the number of mutations we need to perform to collect the contributions to the quantum monodromy $\mathbb{M}(q)$ from \emph{all} the BPS particles. Mathematically, the factor $r(G^\prime)\,h(G^\prime)$ may be seen as (twice) the size of preprojective component of the AR quiver of $\C G$, generated by repeated application of the inverse AR translation $\tau^-$. On the other hand, the denominator of \eqref{csecond} may be understood in terms of the identification of physical observables computed at a chiral phase $\phi$ with the physical observables at $\phi+2\pi$ twisted by the action of $\mathbb{M}$.

Now, all the these viewpoints about the origin of the $h(G^\prime)$'s appearing in the \textsc{rhs} of  \eqref{csecond} lead to the conclusion that the proper value of $h$ for an affine Lie algebra $\widehat{H}$ is $\infty$: in one of the two canonical chambers the mutation algorithm will be go on forever visiting particle after particle in the infinite towers of (preprojective) dyons.

Therefore, our educated guess is
\begin{equation}\label{edguess}
c_\mathrm{eff}(\widehat{H}\boxtimes G)=\lim_{h\rightarrow\infty}
\frac{r(\widehat{H})\,r(G)\,h(G)\,h }{h+h(G)}\equiv r(\widehat{H})\,r(G)\,h(G),
\end{equation}
For $\widehat{H}\boxtimes G$ one has \begin{equation}\mathrm{rank}\,B= r(G)\left(\sum_{i=1}^3p_i-1\right)-\sum_{i=1}^3f(p_i,G)
\end{equation}
where  $\{p_1,p_2,p_3\}$ are the three periods of $\widehat{H}$ (listed for each acyclic affine quiver $\widehat{H}$ in the second column of the table in eqn.\eqref{pqrnumbs}) and $f(p,G)$ is the function defined in eqn.\eqref{pppp2}
Then eqn.\eqref{cfromqumon} gives the following expression for
 $c$ of the $\widehat{H}\boxtimes G$ QFT
\begin{equation}\label{whatchg}
c\big(\widehat{H} \boxtimes G\big)=\frac{1}{12}\left\{\left(\sum_{i=1}^3 p_i-1\right)r(G)\Big(h(G)+1\Big)-\sum_{i=1}^3 f(p_i,G)\right\}.\end{equation}

Let us check that this expression has the right physical properties. First of all, it should be additive in the following sense. Let $\widehat{H}$ be an acyclic affine quiver and $\{p_1,p_2,p_3\}$ its periods (cfr.\! table in eqn.\eqref{pqrnumbs}). Since our formula for $c$ refers to the value at the UV fixed point, and the YM coupling is asymptotically free, $c(\widehat{H}\boxtimes G)$ should be the sum of the $c$'s of the four UV decoupled sectors: $G$ SYM, $D_{p_1}(G)$,  $D_{p_2}(G)$, and $D_{p_3}(G)$, while $c(D_1(G))\equiv 0$ for all $G$ since it corresponds to the empty matter sector
\begin{equation}\label{wwwxz}
c(\widehat{H}\boxtimes G)= \frac{1}{6}\,\dim G+\sum_i c(D_{p_i}(G)),
\end{equation} This gives two conditions which need to be satisfied by the expression \eqref{whatchg}
\begin{gather}
c(\widehat{A}(1,1)\boxtimes G)=( \,c\ \text{of pure SYM with group $G$\,)}\equiv \frac{1}{6}\,\dim G\\
c\big(\widehat{H}\boxtimes G\big)= \sum_{i=1}^3 c\big(\widehat{A}(p_i,1)\boxtimes G\big)-2\,c\big(\widehat{A}(1,1)\boxtimes G\big).
\end{gather}
They are both true and corroborate our educated guess 
\eqref{edguess}.

From eqn.\eqref{wwwxz} we extract the value of the central charge $c$ for the SCFT $D_p(G)$. It is given by the function
\be\label{cpG}
c(p,G)=\frac{1}{12}\left\{(p-1)\,r(G)\Big(h(G)+1\Big)- f(p,G)\right\}.
\ee

%

\medskip

{\bf Example 4}: the model $\widehat{A}(2,1)\boxtimes A_2$ is $SU(3)$ SYM gauging the $SU(3)_\mathrm{flavor}$ of the Argyres--Douglas of type $D_4$ \cite{infinitelymany}. From eqn.\eqref{wwwrt} $c$ of $D_4$ is $2/3$, and then the $c$ of the gauged model should be $c=8/6+2/3\equiv 2$
in agreement with eqn.\eqref{whatchg}
\begin{equation}
c(\widehat{A}(2,1)\boxtimes A_2)=\frac{3\cdot 2\cdot 4}{12}=2.
\end{equation}

In \S.\,\ref{furtherex} we shall present a number of additional examples of the formula \eqref{cpG}, which always produces the correct physical results. See also \S.\,\ref{exceptttt} for further applications of the formula.

\subsection{Computing $a$}

The formula for $c$ also determines the central charge $a$ in view of the 2d/4d correspondence. Indeed, consider a $\cn=2$ $4d$ SCFT, and let $\mathcal{R}$ be the chiral ring of primary operators \cite{VafaR} of the $(2,2)$ superconformal system associated to it. For an element $\psi \in \mathcal{R}$, let us denote with $q(\psi)$ its 2d $R$-charge. Deforming the $2d$ superpotential
\be
W \longrightarrow W + \sum_{\psi \in \mathcal{R}} u_{\psi} \psi \qquad u_{\psi} \in \C
\ee
we induce a massive deformation of the 4d theory with primary operators $\co_{\psi}$ that have dual parameters $u_{\psi}$. Their scaling dimensions are $D[\co_{\psi}] = 2 - D[u_{\psi}]$. Let $\Omega$ be the Seiberg--Witten form of the theory. 
At the conformal point the Seiberg--Witten geometry has a holomorphic scaling symmetry ($i.e.$ the 2d $R$-=symmetry) under which the Seiberg-Witten form transforms as
\be
\Omega \longrightarrow \l^{q(\Omega)} \Omega.
\ee
The 4d scaling dimensions of the mass parameters $u_{\psi}$ are fixed by requiring that the Seiberg-Witten form have dimension equal to 1
\be
D[\Omega] \equiv 1 \Longrightarrow \begin{cases} D[\psi] = q(\psi) / q(\Omega)  \\
 D[u_{\psi}] = (1-q(\psi))/ q(\Omega) \end{cases}
\ee
It is a known fact \cite{ShapereTachikawa} that
\be\label{knownfact}
4(2a-c) = \sum_{i=1}^{d} (2 D[u_i] - 1)
\ee
where $d$ is the dimension of the Coulomb branch of the model and the $u_i$ are the physical deformations that parametrizes it, $i.e.$ the deformations with scaling dimensions $D[u_i]>1$. Since $d \equiv \text{rank} \, B/2$, we have
\be
a = \frac{1}{2} \Big( c + \frac{1}{4} \cdot \sum_{i=1}^{d} (2 D[u_i] - 1) \Big)= \frac{1}{2} \cdot c  + \frac{1}{16} \cdot \text{rank} \, B + \frac{1}{4}\cdot\sum_{i=1}^{d} ( D[u_i] - 1).
\ee
We define
\be
u(\mathcal{R}) \equiv \frac{1}{4} \cdot \sum_{\psi \in \mathcal{R}} \Bigg[ \frac{1-q(\psi)}{q(\Omega)}-1\Bigg]_+
\ee
where $[x]_+$ is the function
\begin{equation}
[x]_+=\begin{cases} x & \text{for }x\geq 0\\
0 &\text{for }x<0.
\end{cases}
\end{equation}
This gives the final formula for $a$
\be
a = \frac{1}{2} \cdot c  + \frac{1}{16} \cdot \text{rank} \, B + u(\mathcal{R}).
\ee

For the $D_p(G)$ SCFT this formula may be rewritten in a simpler way. We write $E(G)$ for the set of exponents of $G$ (equal to the degrees of the fundamental Casimirs of $G$ \emph{minus 1}). Then $u(\mathcal{R})$ is given by the function
\be\label{upG}
u(p , G) = \frac{1}{4} \sum_{s = 1}^{p-1} \sum_{j \in E(G)} \Bigg[ j - \frac{h(G)}{p} s \Bigg]_+.
\ee
This formula is very obvious if you recall that $q(\Omega)\equiv 1-\hat c/2\equiv 1/h(G)$ (eqn.\eqref{oomega}). Hence $4\, u(\mathcal{R})$ is simply $\sum_{RR} [h(G)\,q_{RR}]_+$, where the $q_{RR}$'s are the $U(1)_R$ charges of the RR vacua whose set is precisely \begin{equation}\{q_{RR}\}\equiv\{j/h(G)-s/p\ \colon j\in E(G),\ 1\leq s\leq p-1\}.\end{equation}


Putting everything together we obtain that the central charge $a$ for the $D_p(G)$ models is
\be\label{apG}
a(p;G)=u(p;G)+\frac{1}{48}\Big((p-1)\,r(G)\big(2\, h(G)+5)-5 f(p;G)\Big).
\ee
Correspondingly, {at the UV fixed point}, the $a$--central charge for the $\widehat{H}\boxtimes G$ models is
\begin{equation}\label{yyyyiiii}
a(\widehat{H}\boxtimes G)\big|_\mathrm{UV}=\frac{1}{4}\sum_{j\in E(G)}j+\frac{1}{24}r(G)\big(2\,h(G)+5)+\sum_{i=1}^3 a(p_i;G).
\end{equation}

\medskip

{\bf Example 5}: $SU(2)$ SQCD with $N_f\leq 3$ flavors. $N_f=1,2,3$ correspond, respectively, to the quiver $A(2,1)\boxtimes A_1$, $A(2,2)\boxtimes A_1$, and $\widehat{D}_4\boxtimes A_1$. One has
$c_\mathrm{eff}(\widehat{H}\boxtimes A_1)= 2\,r(\widehat{H})\equiv 2(2+N_f)$,
while $\mathrm{rank}\,B=2$ for all $N_f$. Then
\begin{gather}
c=\frac{1}{12}\big(4+2N_f+2)= \frac{3}{6}+\frac{2N_f}{12},\\
a=\frac{1}{4}+\frac{9}{24}+N_f\, a(2,A_1)= \frac {5\dim SU(2)}{24}+\frac{2 N_f}{24}
\end{gather}
consistent with $\dim SU(2)=3$ vector--multiplets and $N_f$ hyper {doublets}.

\subsection{Further examples and checks}\label{furtherex}
\subsubsection{Example 6: $SU(N)$ linear quivers}
For all $m,p\in \mathbb{N}$ we consider the linear quiver theory
\begin{gather}
SU(m p)-SU(m(p-1))-SU(m(p-2))-\cdots-SU(2m) - SU(m).
\end{gather}
As we discussed in \S.\,\ref{SUlagR}, such theory has quiver $\widehat{A}(p,1)\boxtimes A_{mp-1}$. 
The weak coupling computation of $c$ is
\begin{equation}
c=\frac{1}{6}n_v+\frac{1}{12}n_h=\frac{1}{6}\sum_{k=1}^p(m^2k^2-1)+\frac{1}{12}\sum_{k=1}^{p-1} (m k) \, m (k+1)
=\frac{1}{12}p(m^2 p(p+1)-2).
\end{equation}
The computation from the quiver $\widehat{A}(p,1)\boxtimes A_{mp-1}$ is
\be
\begin{aligned}
\frac{1}{12}&\Big(\big((p+1)(mp-1)mp\big)+\big((p+1)(mp-1) - p+1\big)\Big)=\frac{1}{12}p(m^2 p(p+1)-2),
\end{aligned}
\ee
in perfect agreement with the weak coupling computation. Now consider the central charge $a$. Let us start computing
\begin{equation}
4\,u(p;A_{mp-1})=\sum_{s=1}^{p-1}\sum_{j=1}^{mp-1}[j-ms]_+=\sum_{s=1}^{p-1}\Delta(ms),
\end{equation}
where we have set
\begin{equation}
\Delta(N)=\sum_{j=1}^{N-1}j=\frac{N(N-1)}{2}.
\end{equation}
We already know that
\begin{equation}
f(p;A_{mp-1})=\gcd\{p,mp\}-1=p-1.
\end{equation}
so that \eqref{yyyyiiii} gives
\begin{equation}\begin{split}
a(p;A_{mp-1})&= \frac{1}{4}\sum_{s=1}^{p-1}\Delta(ms)+\frac{1}{48}\Big((p-1)(mp-1)(2mp+5)-5(p-1)\Big)=\\
&=\frac{1}{48}(p-1)(4m^2p^2-m^2p-10)
\end{split}
\end{equation}
On the other hand, let us compute $a$ using weakly coupled QFT in the UV; $a$ for $SU(N)$ SYM is
\begin{equation}
a(SU(N))=\frac{5}{24}\dim SU(N)=\frac{5}{24}(2\Delta(N)+N-1)
\end{equation}
while the hypers in the bifundamental $(N_1,\overline{N}_2)$ contribute with $N_1N_2/24$.
Then
\begin{equation}
\begin{split}
a(\text{linear quiver})\Big|_\mathrm{QFT}&=\frac{5}{24}\sum_{s=1}^{p-1}\Big(2\Delta(ms)+ms-1\Big)+\frac{m^2}{24}\sum_{s=1}^{p-1}s(s+1)=\\
&=\frac{1}{48}(p-1)(4m^2p^2-m^2p-10),
\end{split}
\end{equation}
in perfect agreement.

\subsubsection{Example 7: $SO/USp$ linear quivers}

For all $m,p\in \mathbb{N}$ we consider the linear quiver theory
\be
\begin{footnotesize}
\begin{aligned}
SO(2mp+2)-USp(2m(p-1))&-SO(2m(p-2)+2)-USp(2m(p-3))-\cdots\\
&\cdots-SO(2m(p-2\ell)+2)-USp(2m(p-2\ell-1))-\cdots
\end{aligned}
\end{footnotesize}
\ee
(the two lines are meant to be concatenated). The edges now stand for \emph{bifundamental} HALF hypermultiplets. The linear quiver have two possible ends, depending on the parity of $p$. 
\be
\begin{aligned}
& p \text{ odd } \colon \qquad \cdots - USp(4m) - SO(2m + 2)\\
& p \text{ even } \colon \qquad \cdots - SO(4m+2) - USp(2m) - \framebox{$SO(2)$}
\end{aligned}
\ee
where the box means an ungauged flavor group. As we discussed in \S.\,\ref{SOSplagR}, the BPS quiver of this theory is $\widehat{A}(p,1)\boxtimes D_{mp+1}$.

\medskip

Let us compute $c$ from weak coupling; we specialize to $p=2\ell$ even: 
\begin{multline}
c=\frac{1}{6}\,n_v+\frac{1}{24}\,n_{\tfrac{1}{2}h}=\frac{1}{6}\left[\sum_{k=1}^\ell \dim SO(4m k+2)+\sum_{k=1}^\ell \dim USp(2m(2k-1)\right]+\\
+\frac{1}{24}\sum_{k=1}^\ell \big[2m(2k-1)\big]\big[2m(2k)+2+2m(2k-2)+2\big]=\\
=\frac{1}{6}\left[\sum_{k=1}^\ell (2mk+1)(4mk+1)+\sum_{k=1}^\ell m(2k-1)[2m(2k-1)+1]\right]+\\
+\frac{1}{24}\sum_{k=1}^\ell \big[2m(2k-1)\big]\big[2m(2k)+2+2m(2k-2)+2\big]=\\
=\frac{\ell}{6}\big(8 m^2\ell^2+4m^2\ell+6 m\ell+3m+1\big),
\hskip2cm\end{multline}
while from the quiver
\begin{equation}
c=\frac{1}{12}\big((2\ell+1)(2\ell m+1)4\ell m+(2\ell+1)(2m \ell+1)-1\big)\equiv \frac{\ell}{6}\big(8 m^2\ell^2+4m^2\ell+6 m\ell+3m+1\big),
\end{equation}
with perfect agreement. Let us consider now $p$ odd $=2\ell+1$.
\begin{multline}
c=\frac{1}{6}\,n_v+\frac{1}{24}\,n_{\tfrac{1}{2}h}=\frac{1}{6}\left[\sum_{k=0}^\ell \dim SO(2m (2k+1)+2)+\sum_{k=1}^\ell \dim USp(4m k)\right]+\\
+\frac{1}{24}\sum_{k=1}^\ell 4 m k\big[2m(2k+1)+2+2m(2k-1)+2\big]=\\
=\frac{1}{6}\left[\sum_{k=0}^\ell (m(2k+1)+1)(2m(2k+1)+1)+\sum_{k=1}^\ell 2m k [4m k+1]\right]+\\
+\frac{1}{24}\sum_{k=1}^\ell 4 m k\big[2m(2k+1)+2+2m(2k-1)+2\big]=\\
=\frac{1}{6}(4m\ell+2m+1)(2m\ell+m+1)(\ell+1)
\end{multline}
while from the quiver
\begin{multline}
c=\frac{1}{12}\Big((2\ell+2)\big((2\ell+1)m+1\big)\big(2m(2\ell+1)\big)+(2\ell+2)\big((2\ell+1)m+1\big)\Big)=\\
=\frac{1}{6}(4m\ell+2m+1)(2m\ell+m+1)(\ell+1),
\end{multline}
in complete agreement.

If one considers the $D_p(SO(2mp+2))$ models, one has
\begin{gather}
f(p;D_{mp+1})=\begin{cases}1 & p\ \text{even}\\
0 &\text{otherwise}\end{cases}\quad\equiv p+1-2\left[\frac{p+1}{2}\right]\end{gather}
\begin{equation}\begin{split}
u(p;D_{mp+1})&=\frac{1}{4}\sum_{s=1}^{p-1}\left(\sum_{k=1}^{mp}[2k-1-2ms]_++[mp-2ms]_+\right)=\\
&=\frac{1}{4}\sum_{s=1}^{p-1}\sum_{k=1}^{m(p-s)}(2k-1)+\frac{m}{4}\sum_{s=1}^{[p/2]}(p-2s)=\\
&=\frac{m^2}{24}(2p-1)p(p-1)+\frac{m}{4}\,[(p-1)/2]\,[p/2]
\end{split}
\end{equation}
and then 
\begin{multline}
a(p;D_{mp+1})=\frac{m^2}{24}(2p-1)p(p-1)+\frac{m}{4}\,[(p-1)/2]\,[p/2]\,+\\
+\frac{1}{48}\Big\{(p-1)(mp+1)(4mp+5)-5(p+1)+10[(p+1)/2]\Big\}\end{multline}
for $p=2q$ even this is 
\begin{equation}\label{for1}\frac{1}{24}\Big(32 m^2 q^3-20 m^2q^2+24m q^2+2m^2 q-15 m q+5q-5\Big)
\end{equation}
while for $p=2q+1$ odd 
\be\label{oddtestSUSP}
\frac{1}{24}\,q\big(32 m^2q+28m^2q+24m q+6m^2+9m+5\big) 
\ee
We have already computed the number of vector multiplets and half--hypers in the linear quiver $\widehat{A}(p,1)\boxtimes D_{mp+1}$: the computation for the theory $D_p(SO(mp+1))$ changes just $n_v$ by the $SO(mp+1)$ contribution; for $p=2q$
\begin{equation}
n_v=\frac{1}{3}(16m^2q^3-12 m^2q^2+12mq^2+2m^2q-9m q+3q-3)
\end{equation}
\begin{equation}
n_{\tfrac{1}{2}h}=\frac{8}{3}m q(4m q^2+3q-m)
\end{equation}
and therefore
\begin{equation}
a=\frac{5}{24}n_v+\frac{1}{48}n_{\tfrac{1}{2}h}=\frac{1}{24}\Big(32 m^2 q^3-20 m^2q^2+24m q^2+2m^2 q-15 m q+5q-5\Big)
\end{equation}
in agreement with the formula \eqref{for1}.

For $p=2q+1$, we have
\be
n_v =\frac{1}{3} q ( 16 m^2 q^2 + 12 m^2 q + 12 m q + 2 m^2 + 3 m + 3)
\ee
\be
n_{\tfrac{1}{2}h}=\frac{8}{3} m q ( 4 m q ^2 + 6 m q + 3 q+ 2m+3)
\ee
and therefore
\be
a=\frac{5}{24}n_v+\frac{1}{48}n_{\tfrac{1}{2}h}= \frac{1}{24} q \left(32 m^2 q^2+ 28 m^2 q + 24 m + 6 m^2+9 m+5\right)
\ee
in agreement with formula \eqref{oddtestSUSP}.

\subsubsection{A comment about the theories of type $(I_{k,N},F)$ of \cite{XIE2}}\label{XiZ}

With the change of coordinates $t = e^{Z}$, the SW curve associated to the $D_p(SU(N))$ theory (at the conformal point) and SW differential are 
\be
t^{p}+x^{N}=0;\quad\lambda= 
\frac{x}{t}dt.
\ee
With the trivial substitution $y=x/t$ the SW differential becomes $ydt$ and we recognize (at least for $p>N$) 
the SW curve for the $(I_{k,N},F)$ theories recently studied by Xie and Zhao in \cite{XIE2}. The parameters $(k,N)$ in that paper correspond to $(N,p-N)$ in the present notation. We thus propose to identify $D_{N+k}(SU(k))$ with $(I_{k,N},F)$. This observation gives in particular a realization of our models in terms of the 6d $\mathcal{N}=(2,0)$ theory of type $A_{k-1}$ compactified on a sphere with two punctures; one irregular of type I (in the language of \cite{XIE2}) and one maximal. It is then natural to propose that $D_p(SO(2N))$ and $D_p(E_N)$ theories can be constructed ``compactifying'' on a two punctured sphere the 6d theories of type $D_N$ and $E_N$ respectively.


The authors of \cite{XIE2} have been able to compute the $a$ and $c$ central charges for $(I_{k,N},F)$ theories when $N$ is a multiple 
of $k$ exploiting the fact that in this case the mirror dual of the $\mathcal{N}=4$ 3d theory obtained compactifying $(I_{k,N},F)$ 
on $S^1$ is Lagrangian. By identifying explicitly this theory they can determine the dimension of its Coulomb branch, which in turn 
coincides with the dimension of the Higgs branch of the parent 4d theory. This allows to extract the value of $c-a$ and combining 
this with eqn.\eqref{knownfact} one can determine both $a$ and $c$. They also conjecture a formula for $N$ generic using the results of 
\cite{ShapereTachikawa}: in that paper the authors derive a formula (valid for any $\mathcal{N}=2$ SCFT) for $a$ and $c$ which depends on the R-charge 
of the discriminant $\Delta$ of the SW curve (more precisely they consider the R-charge of $B=\Delta^{1/8}$). Using the result 
for $a$ and $c$ derived from the 3d mirror they extract a formula for $R(B)$ and then propose that it is valid for general $k$ and $N$. 
We will now see that our formula for $c$ agrees perfectly with the result obtained using mirror symmetry. At the same time this will 
support our result and confirm the conjecture of \cite{XIE2}. 

Notice first of all that since the SW curve and differential for $D_{N+k}(SU(k))$ and $(I_{k,N},F)$ are the same, the scaling dimension 
of the various operators cannot differ. We thus learn that the rank of the theory (\textit{i.e.}\! the dimension of the 
Coulomb branch) and the value of $2a-c$ (see \eqref{knownfact}) necessarily coincide as well. It is now convenient to use the 
formula given in \cite{ShapereTachikawa} for $c$: \begin{equation}\label{centr} c=\frac{1}{3}R(B)+\frac{r}{6},\end{equation} where $r$ is the rank of the theory. This clearly implies that 
$a$ and $c$ coincide for $D_{N+k}(SU(k))$ and $(I_{k,N},F)$, since in principle $R(B)$ can be computed from the curve. However, it is 
hard, in general, to determine it explicitly (as we will do in the case $p=2$). The knowledge of the BPS quiver allows to bypass this
difficulty: as we have seen before the $c$ central charge is given by the formula $$c=\frac{1}{12}(c_{eff}(Q,\mathcal{W})+\text{rank B}),$$ 
where $c_{eff}(Q,\mathcal{W})$ is equal to the scaling dimension of the discriminant of the SW 
curve at the conformal point, as we have noticed above. This is precisely what we need, since $R(B)$ is just the scaling dimension of the discriminant divided 
by four. Using now our formula for $c_{eff}$ we find immediately the answer for generic $p,N$ and any $G=ADE$:
\begin{equation}
R(B)=\frac{1}{4}(p-1)r(G)h(G).
\end{equation}
Specializing to the case $G=SU(k)$ (and setting $p=N+k$ as before) we find $$R(B)=\frac{1}{4}k(k-1)(N+k-1),$$ in perfect agreement 
with equation (2.44) of \cite{XIE2}.  
\smallskip

{\bf Remark.} In the above mentioned paper the authors also analyze theories on the sphere with only one irregular 
puncture. In particular they study the so called $(A_N,A_k)$ theories introduced in \cite{CNV} and propose a formula for $R(B)$ also in this case. Their argument relies on 3d mirrors as before. Using our formula for the $c$ central charge we are able to confirm their conjecture in this case as well: as we have just seen $R(B)$ should be identified with $\frac{1}{4}c_{eff}(Q,\mathcal{W})$, which is in turn equal to
$$\frac{1}{4}\frac{r(G)r(G')h(G)h(G')}{h(G)+h(G')}$$
for $(G,G')$ theories. If we now set $G=A_{N-1}$ and $G'=A_{k-1}$ we find
\begin{equation}
R(B)=\frac{N(N-1)k(k-1)}{4(N+k)}.
\end{equation}
This is precisely eqn.(2.28) of \cite{XIE2}.

\subsection{General properties of the $D_p(G)$ SCFT's}

We have obtained quite a precise physical picture of the SCFT $D_p(G)$. We know:
\begin{itemize}
\item the rank of the flavor group
\be
f=r(G)+f(p,G)
\ee
\item the dimension of the Coulomb branch
\be\label{rrvb}
d=\frac{1}{2}\Big((p-1)r(G)-f(p,G)\Big)
\ee
\item the order of the 4d quantum monodromy
\be
r=\frac{p}{\gcd\{p,h(G)\}}
\ee
\item the dimension of the discriminant of the SW curve at the UV CFT point 
\be
\propto (p-1)r(G)h(G)
\ee
\item the SCF central charge $a$ is $a(p;G)$ defined in eqn.\eqref{apG}
\item the SCF central charge $c$ is $c(p;G)$ defined in eqn.\eqref{cpG}
\item the $G$--current algebra central charge $k_G$  is
\be
k_G \equiv 2\,b_{p,G} = \frac{2(p-1)}{p}\,h(G)
\ee
\item the set of the dimensions of the operators parametrizing the Coulomb branch
\begin{equation}\label{dddddmx}\Big\{\Delta_1,\Delta_2,\dots,\Delta_d\Big\}=
\Big\{\;j-\frac{h(G)}{p}s+1\ \Big|\ j>\frac{h(G)}{p}s,\ j\in E(G),\ s=1,\dots,p-1\;\Big\}.
\end{equation}
\end{itemize}


From these expressions we may extract some  general properties of the $D_p(G)$ SCFT which are \emph{typical} of this class of theories. For instance, as we are going to show, these theories have Coulomb branches of large dimension, $d =O(p)$ (eqn.\eqref{rrvb}), while, for fixed $G$, the dimension of their Higgs branches is bounded above by $2\,r(G)\, h(G)^2$ (a sharper inequality holds for $p\geq h(G)$, see below).

We list a few such properties:
\begin{description}
\item[The Coulomb branch operator of maximal dimension:] Since $h(G)-1\in E(G)$ for all $G$'s, we see from \eqref{dddddmx} that the maximal dimension of the Coulomb branch operators is equal to the $\beta$--function coefficient $b_{p,G}$,
\begin{equation}
\Delta_d\equiv \frac{p-1}{p}\,h(G)=b_{p,G}\equiv \frac{1}{2}\,k_G,
\end{equation}
and is \emph{always less} that the maximal dimension $\Delta_\text{max}(G)\equiv h(G)$ for SYM with gauge groups $G$. 
\item[The rank of the flavor group $f$] is always $\leq 2\, r(G)$ with equality iff $h(G)\mid p$.
\item[The dimensions $\Delta_i\in\mathbb{N}$]$\Longleftrightarrow r\in\mathbb{N}\Longleftrightarrow b_{p,G}\in\mathbb{N}$.
\item[Asymptotic behavior of $a$ and $c$:] For fixed $G$ and large $p$ the asymptotics of the SCFT central charges $a$ and $c$ are
\begin{gather}\label{th1}
a(p;G)\approx \frac{\dim G}{12}\,p+O(1)\\
c(p;G)\approx \frac{\dim G}{12}\,p+O(1).\label{th2}
\end{gather}
In particular, $c(p;G)-a(p;G)$ is constant for large $p$ up to a few percent  Number--Theoretic modulation (see next item). 
\item[Dimension of the Higgs branch:] Assume $p\geq h\equiv h(G)$. Then ($r\equiv r(G)$)
\begin{multline}
\dim_H\text{Higgs branch}\equiv n_h-n_v\equiv 24(c-a)\leq\\
\leq  \#\text{(positive roots of $G$)}+r\equiv \frac{(h+2)r}{2}\in\mathbb{N},\label{th3}
\end{multline}
with equality \emph{if and only if} $h\mid p$.
\end{description}

%

We present the proofs of eqns.\eqref{th1}--\eqref{th3} which illustrate well the idea that, for the $D_p(G)$ SCFT's, the value of all physical quantities have deep Lie--theoretical meaning.
Eqn.\eqref{th1} is elementary:
\begin{equation}
c(p;G)=\frac{1}{12}r(h+1)p+O(1)
\end{equation}
and $\dim G=r(h+1)$ is a well--known identity in Lie theory due to Coxeter.\smallskip

To get eqn.\eqref{th2} we consider
\begin{equation}
u(p;G)=\frac{1}{4}\sum_{j\in E(G)}\sum_{s=1}^{p-1}\Big[j-h\,s/p\Big]_+
\end{equation}
For $p\ggg1$ the sum over $s$ may be evaluated by the Euler--McLaurin summation formula; setting $x=s/p$,  the term of order $p$ in $u(p;G)$ is then
\begin{equation}
\frac{p}{4}\sum_{j\in E(G)}\int_0^1 dx\, [j-h\, x]_+=\frac{p}{8\,h}\sum_{j\in E(G)} j^2
\end{equation}
where
\begin{equation}
\sum_{j\in E(G)}j^2=\begin{cases}N(N-1)(2N-1)/6 & SU(N)\\
n(4n-5)(n-1)/3 & SO(2n)\\
276 & E_6\\
735 & E_7\\
2360& E_8\end{cases}\label{sumj2}
\end{equation}
Therefore
\begin{equation}
a(p,G)=\frac{1}{48}\left(\frac{6}{h} \sum_{j\in E(G)}j^2+r(2h+5)\right)p+O(1)
\end{equation}
and eqn.\eqref{th2} is equivalent to the {peculiar} Lie theoretical identity
\begin{equation}\label{rr567}
4\dim G=\frac{6}{h}\sum_{j\in E(G)}j^2+r(2h+5).
\end{equation}
However unlikely it looks, this identity is actually true: indeed, plugging in eqn.\eqref{sumj2}, the \textsc{rhs} turns out
\begin{equation}
\begin{array}{c||c}
4(N^2-1)\ \text{for }SU(N) & 4n(2n-1)\ \text{for } SO(2n)\\\hline
312=4\times 78 \ \text{for }E_6 & 532=4\times 133 \ \text{for }E_7\\\hline
992=4\times 248 \ \text{for }E_8 &
\end{array}
\end{equation}
The identity \eqref{rr567} is more conveniently written 
as\footnote{\ We thank the referee for informing us that this identity was proven before, see \cite{peculiar}.}
\begin{equation}\label{rr567bis}
6\sum_{j\in E(G)}j^2=2h^2r-hr.
\end{equation}

To show eqn.\eqref{th3} we consider first the case $h\mid p$. Let $p=h\ell$; then
\begin{equation}
c-a=\frac{1}{48}\left(2h^2r-hr-6\sum_{j\in E(G)}j^2\right)\ell+\frac{1}{48}\left(6\sum_{j\in E(G)}j-2h r+2r\right).
\end{equation}
The term linear in $\ell$ vanishes by the identity \eqref{rr567bis}. The equality in Eqn.\eqref{th3} for $h\mid p$ then follows from the identity $\sum_{j\in E(G)}j=rh/2$ which, in view of the discussion after eqn.\eqref{upG}, is just 2d PCT.

Returning to the general case, we infer that, as a function of $p$, $24(c-a)$ is given by a degree--zero polynomial, $\#\Delta^+(G)+r$, plus a small Number--Theoretical modulation depending on the divisibility properties of $p$ and $h$; the modulation has two sources, from $f(p;G)$ and $u(p;G)$. For $p\geq h$, $u(p;G)$ is simply
\begin{equation}
u(p;G)=\frac{1}{8p}\sum_{j\in E(G)}\left(\frac{j^2p^2}{h}-j p\right)+\frac{h}{8p}\sum_{j\in E(G)}\left\{\frac{j p}{h}\right\}\left(1-\left\{\frac{j p}{h}\right\}\right),
\end{equation}
where $\{x\}$ denotes the fractional part. Therefore (for $p\geq h$)
\begin{equation}
\begin{split}
24(c-a)=&\big[\text{polynomial in $p$ (of degree zero)}\big]-\\
&\ -\frac{3h}{p}\sum_{j\in E(G)}\left\{\frac{j p}{h}\right\}\left(1-\left\{\frac{j p}{h}\right\}\right)-\frac{1}{2}\big(r- f(p;G)\big)
\end{split}
\end{equation}
where the second line corresponds to the modulation.
Both terms in the modulation are non--positive, and vanish if and only if $h\mid p$. This gives eqn.\eqref{th3}, in facts the more precise result ($p\geq h$)
\begin{equation}
24(c-a)=\#\Delta^+(G)+r-\Bigg(\frac{3h}{p}\sum_{j\in E(G)}\left\{\frac{j p}{h}\right\}\left(1-\left\{\frac{j p}{h}\right\}\right)+\frac{1}{2}\big(r- f(p;G)\big)\Bigg).
\end{equation}

\section{$D_2(G)$ systems and Minahan-Nemeshansky theories}

The SCFT's of period $p=2$ are expected to be particular easy since they generalize to arbitrary $G=ADE$ the $D_2$ Argyres--Douglas model which is just a free doublet. In this section we study in more detail this simple class of theories.

\subsection{Quivers and BPS spectra}

\subsubsection{Quivers and superpotentials}

The quivers $Q_{2,G}$ have $\widehat{A}(2,0)$ full subquivers over the nodes of $G$ which correspond to quadratical terms in the superpotential $\cw_\text{mat}$. Therefore the arrows of the $\widehat{A}(2,0)$ vertical subquivers get integrated out from their DWZ--reduced  quiver with superpotential, $\cd(G)$, which are particularly simple. For $G=A_n$ one gets
\begin{equation}\label{pqqq)I}
\cd(A_n)\equiv\begin{gathered}
\xymatrix{\widehat{1}\ar[rr]&&\widehat{2}\ar[rr]\ar[ddll]&&\cdots\ar[ddll]\ar[rr]&&\widehat{n-1}\ar[rr]\ar[ddll]&&\widehat{n}\ar[ddll]\\
&&\\
1\ar[rr]&&2\ar[rr]\ar[uull]&&\cdots\ar[rr]\ar[uull]&&n-1\ar[rr]\ar[uull]&&n\ar[uull]}
\end{gathered}
\end{equation}
the $\cd(G)$'s for the other simply--laced Le algebras being  represented in figure \ref{D2Gquivs}.
The  superpotential for $\c(A_n)$ is
\begin{equation}
\cw_\cd=\sum_a \big(\lambda^{(1,a)}\psi_a^{(2)}-\psi^{(1)}_a\lambda^{(1,a)}\big)\big(\lambda^{(2,a)}\psi_a^{(1)}-\psi^{(2)}_a\lambda^{(2,a)}\big).
\end{equation}

\subsubsection{A finite BPS chamber}
The quivers $\cd(G)$ contain two full Dynkin $G$ subquivers with alternating orientation and non--overlapping support. \textit{E.g.}\! the two alternating $A_n$ subquivers of $\cd(A_n)$ in \eqref{pqqq)I} are the full subquivers over the nodes
\be
\big\{1,2,\widehat{3},\widehat{4},5,6,\dots\big\}\quad \text{and}\quad \big\{\widehat{1},\widehat{2},3,4,\widehat{5},\widehat{6},\dots\big\}.
\ee
With reference to this example, let us define the following mutation sequence:
\be
{\bf m}_{2,A_n} \equiv \Bigg( \prod_{a \text{ even}} \mu_{a} \circ \mu_{\widehat{a}} \Bigg)\circ \Bigg( \prod_{a \text{ odd}} \mu_{a} \circ \mu_{\widehat{a}}\Bigg)
\ee
The mutation sequence corresponding to the full quantum monodromy associated to the $A_n\oplus A_n$ chamber is symply
\be
({\bf m}_{2,A_n})^{n+1} = \underbrace{\ {\bf m}_{2,A_n} \circ \dots \circ {\bf m}_{2,A_n} \ }_{n+1 \text{ times}}.
\ee
\begin{figure}
$$
\begin{aligned}
\cd(D_{n+1}) \quad\colon\qquad & {\footnotesize\begin{gathered}
\xymatrix@R=1.5pc@C=1.5pc{&&&&&&&\widehat{n+1}\ar[dddl]\\
\widehat{1}\ar[rr]&&\widehat{2}\ar[rr]\ar[ddll]&&\cdots\ar[ddll]\ar[rr]&&\widehat{n-1}\ar[ur]\ar[rr]\ar[ddll]&&\widehat{n}\ar[ddll]\\
&&\\
1\ar[rr]&&2\ar[rr]\ar[uull]&&\cdots\ar[rr]\ar[uull]&&n-1\ar[rr]\ar[dr]\ar[uull]&&n\ar[uull]\\
&&&&&&&n+1\ar[uuul]}
\end{gathered}}\\
\cd(E_6) \quad\colon\qquad &{\footnotesize \begin{gathered}
\xymatrix@R=1.5pc@C=1.5pc{&&&&&\widehat{6}\ar[dddl]\\
\widehat{1}\ar[rr]&&\widehat{2}\ar[rr]\ar[ddll]&&\widehat{3}\ar[ur]\ar[ddll]\ar[rr]&&\widehat{4}\ar[rr]\ar[ddll]&&\widehat{5}\ar[ddll]\\
&&\\
1\ar[rr]&&2\ar[rr]\ar[uull]&&3\ar[rr]\ar[uull]\ar[dr]&&4\ar[rr]\ar[uull]&&5\ar[uull]\\
&&&&&6\ar[uuul]}
\end{gathered}}\\
\cd(E_7) \quad\colon\qquad & {\footnotesize\begin{gathered}
\xymatrix@R=1.5pc@C=1.5pc{&&&&&\widehat{7}\ar[dddl]\\
\widehat{1}\ar[rr]&&\widehat{2}\ar[rr]\ar[ddll]&&\widehat{3}\ar[ur]\ar[ddll]\ar[rr]&&\widehat{4}\ar[rr]\ar[ddll]&&\widehat{5}\ar[rr]\ar[ddll]&&\widehat{6}\ar[ddll]\\
&&\\
1\ar[rr]&&2\ar[rr]\ar[uull]&&3\ar[rr]\ar[uull]\ar[dr]&&4\ar[rr]\ar[uull]&&5\ar[rr]\ar[uull]&&6\ar[uull]\\
&&&&&7\ar[uuul]}
\end{gathered}}\\
\cd(E_8) \quad\colon\qquad &{\footnotesize \begin{gathered}
\xymatrix@R=1.5pc@C=1.5pc{&&&&&\widehat{8}\ar[dddl]\\
\widehat{1}\ar[rr]&&\widehat{2}\ar[rr]\ar[ddll]&&\widehat{3}\ar[ur]\ar[ddll]\ar[rr]&&\widehat{4}\ar[rr]\ar[ddll]&&\widehat{5}\ar[rr]\ar[ddll]&&\widehat{6}\ar[rr]\ar[ddll]&&\widehat{7}\ar[ddll]\\
&&\\
1\ar[rr]&&2\ar[rr]\ar[uull]&&3\ar[rr]\ar[uull]\ar[dr]&&4\ar[rr]\ar[uull]&&5\ar[rr]\ar[uull]&&6\ar[rr]\ar[uull]&&7\ar[uull]\\
&&&&&8\ar[uuul]}
\end{gathered}}
\end{aligned}
$$
\caption{The DWZ--reduced quivers $\cd(G)$ for the $D_2(G)$ SCFT's.}\label{D2Gquivs}
\end{figure}
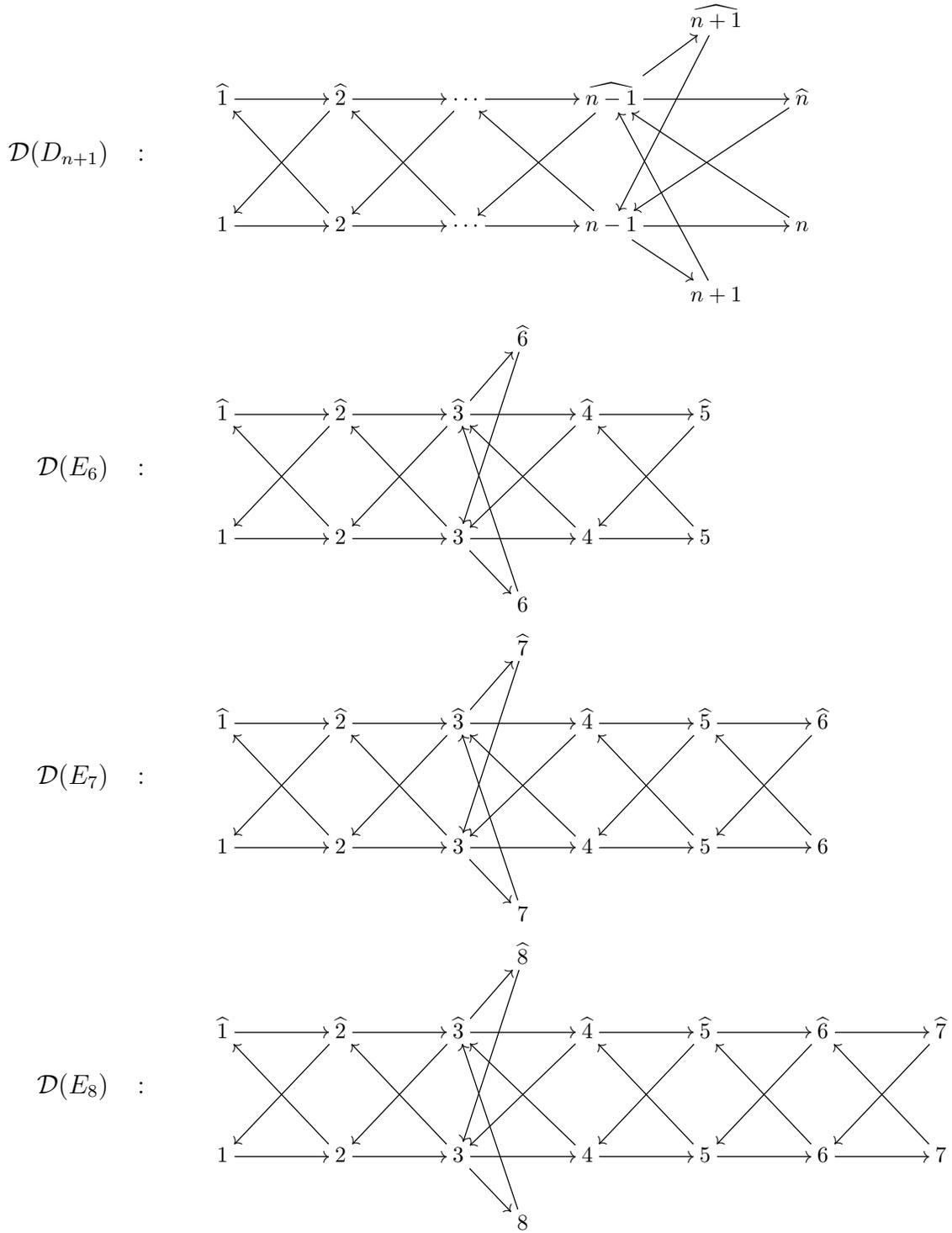

\medskip

We draw the BPS-quivers of the other $D_2(G)$ models in figure \ref{D2Gquivs}. From their structure it is clear that they  all admit Coxeter--factorized sequences of type $(G,G)$ constructed analogously. We conclude that all the $D_2(G)$ superconformal systems have a (possibly formal) finite--BPS-chamber such that the cone of particles in the charge lattice is isomorphic to the direct sum of two copies of the positive root lattice of $G$:
\be\label{D2Gchamb} 
\Gamma^+ (D_2(G)) \simeq \Delta_+(G) \oplus \Delta_+(G)
\ee
where $\Delta_+(G)$ is the set of positive roots of the Lie algebra $G$.
A few remarks are in order:
\begin{enumerate}  
\item Applying this result to $D_2(SU(2N))$ we get a refinement of the result of \cite{ACCERV2} for $SU(N)$ SQCD with $2N$ flavors. The chamber \eqref{D2Gchamb} has less BPS hypers.
\item Comparing with the previous section we see that
\be
c_{\text{eff}} = r(G)h(G) =  \# \big\{ \text{BPS-particles in the chamber \eqref{D2Gchamb}}\big\}
\ee
in agreement with a conjecture by Xie and Zhao \cite{XIE2}.
\item From the explicit BPS spectrum \eqref{D2Gchamb} one constructs (new) periodic TBA $Y$--systems whose periodicity should coincide with that of the 4d quantum monodromy, $r(2,G)$ \cite{CNV,arnold1}. We have numerically checked this prediction for the corresponding 2d solvable models along the lines of \cite{arnold1}, getting perfect agreement.
\end{enumerate}

\subsection{$D_2(G)$ theories as infrared fixed points of $\mathcal{N}=2$ SQCD}

In this section we will argue that the models with $p=2$ and $G=A,D$ 
correspond to infrared fixed points of $\mathcal{N}=2$ SQCD with gauge group $SU(N)$ and $USp(2N)$ respectively.

\subsubsection{$SU(N)$ SQCD with $N_f$ odd}

Starting from the geometry
\begin{equation}\label{CC} W_{G,s}(z,x_1,x_2,x_3)=\Lambda^{b}e^{pz}+\Lambda^be^{-z}+W_{G}(x_1,x_2,x_3),\end{equation}
one finds that the SW curve and differential associated to $D_2(SU(M))$ theories are $$\Lambda^bt^{2}+P_{M}(x)=0,\quad\lambda=\frac{x}{t}dt.$$ 
with $P_M(x)=x^{M}+u_2x^{M-2}+\dots+u_M$. For $M=2k$ these theories were identified in \cite{infinitelymany} with $SU(k)$ SQCD with $2k$ 
flavors. We will now give evidence that the models with $M$ odd correspond to an IR fixed point (more precisely the maximally 
singular point) of $SU(N)$ SQCD with $M$ flavors. The properties of this fixed point depend only on the number of flavors and not 
on the rank of the gauge group (as long as the theory is asymptotically free).

We will now briefly discuss the main features of the fixed point we are going to relate to $D_2(SU(2n+1))$ theories.
Let us write the SW curve for $SU(N)$ SQCD as ($N_f=2n+1$) $$y^2=P_{N}^2(x)-4\Lambda^{2N-2n-1}(x+m)^{2n+1},\quad \lambda=xd{\text log}\left(\frac{P-y}{P+y}\right).$$
The maximally singular point is given choosing $P_N=(x+m)^{n+1}Q(x)$. In the neighbourhood of the singular point the SW curve 
and differential can be approximated as
\begin{equation}\label{SUN1}y^2=x^{2n+1}+(u_{N-n}x^{n}+\dots+u_N)^2,\quad \lambda=\frac{y}{x^{n+1/2}}dx.\end{equation} 
This result can be found also sending $\Lambda$ to infinity and scaling accordingly the other parameters.
As long as $N_f<2N-1$, we can find a point of this kind for generic $m$. 
When $N_f=2N-1$ the formula is slightly different (and we must tune $m$ appropriately):
$$y^2=x^{2N-1}+(u_2x^{N-2}+\dots u_N)^2;\quad\lambda=\frac{y}{x^{N-1/2}}dx.$$
One can now easily determine the scaling dimensions of all the parameters appearing in the curve using the technique proposed in 
\cite{SAW}: requiring that the SW differential has dimension one we get from the above formula $[x]=1$. Imposing then that all the 
terms appearing in the curve have the same dimension we find $[u_j]=N_f/2+j-N$. Notice that all the mass parameters associated to 
the flavors have always canonical dimension due to the condition $[x]=1$, in agreement with the general constraint discovered in 
\cite{SAW} for all $\mathcal{N}=2$ SCFTs having a nonAbelian global symmetry. We will exploit this fact again later.

For $N_f=3$ we find a familiar theory: it is the $D_4$ Argyres-Douglas theory (one can check this comparing the scaling dimensions 
of the operators), consistently with the enhancement of the flavor symmetry to $SU(3)$ for the $D_4$ theory. For $N_f=5$ we find one 
of the rank two theories studied in \cite{AWV}.

In order to make contact with \cite{infinitelymany}, let us start from the $SU(n+1)\times SU(2n+1)$ theory (remember that $N_f=2n+1$) with a 
multiplet in the bifundamental as the only matter field. The SW curve for this model can be written as
$$ \Lambda^b\,t^{2}+c\,t\,P_{n+1}(x)+P_{2n+1}(x)+\frac{\Lambda^b}{t}=0;\quad\lambda_{SW}=\frac{x}{t}\,dt,$$ where $\Lambda$ is the dynamical 
scale for the $SU(2n+1)$ group and the dynamical scale for the $SU(n+1)$ group ($\Lambda_{SU(n+1)}$) is proportional to $c^{-2}$. To see this drop the term proportional 
to $t^{-1}$ (this is equivalent to turning off the $SU(2n+1)$ gauge coupling). We are then left with $SU(n+1)$ SQCD with $2n+1$ flavors. If we define 
$t'=c\,t$ the term quadratic in t becomes proportional to $t'^2/c^2$. The coefficient of this term is in turn identifyable with the dynamical scale. 
Sending c to zero thus corresponds to taking the limit $\Lambda_{SU(n+1)}\rightarrow\infty$. We then find precisely the maximally singular point 
described before.

Sending instead c to zero first we recognize the SW curve for the theory defined by equation (\ref{CC}) in the case $p=2$ (we have set $t=e^z$). Sending 
then $\Lambda$ to zero as explained in \cite{infinitelymany} (i.e. dropping the term proportional to $t^{-1}$ as before) we find the $D_2(SU(2n+1))$ theory.
Both procedures lead to the same curve so, we identify our singular point with the $D_2(SU(2n+1))$ theory. In the case $n=1$ our proposal clearly works:
the equivalence between $D_2(SU(3))$ and the $D_4$ Argyres-Douglas theory is proven explicitly in \cite{infinitelymany}. 

As a check of our claim we can compute 
the $SU(2n+1)$ flavor central charge (which gives in turn the contribution to the $SU(2n+1)$ beta function of our SCFT) and the 
central charges $(a,c)$ for the maximally singular point. These can be determined using the technique presented in \cite{ShapereTachikawa}, once the scaling dimensions of operators 
are known. Reading them from the curve as we described before we find that the contribution to the $SU(2n+1)$ beta function is 
$n+1/2$ and the central charges are $$a=\frac{7n^2+7n}{24};\quad c=\frac{n^2+n}{3}.$$ These values are precisely in agreement 
with those extracted from the BPS quiver for the $D_2(SU(2n+1))$ theory.

\subsubsection{The $USp(2N)$ theory with odd number of flavors}


In the case $G=SO(2N)$ equation (\ref{CC}) describes a 4d theory whose SW curve and differential are 
\begin{equation}\label{SAN} \Lambda^bt^{p}+\frac{P_{N}(x^2)}{x^2}+\frac{\Lambda^b}{t}=0;\quad\lambda_{SW}=\frac{x}{t}dt,\end{equation}
where $P_{N}(x^2)=x^{2N}+u_2x^{2N-2}+\dots+u_{N}^2$ (see the discussion in \cite{CV11}, section 6.3.1)\footnote{one can identify $X$ in that paper 
with our $x^2$ and $\lambda$ corresponds (modulo a coefficient) to $u_N$.}. For $p=1$ this corresponds to SYM theory with gauge group $SO(2N)$.  

In order to see the connection between these models (with $p=2$) and the singular points of $USp$ SQCD let us start from the following model: 
$USp(2k)\times SO(2N)$ gauge theory ($2k=N-1$ if $N$ is odd, $2k\geq N$ otherwise) with a half-hypermultiplet in the bifundamental and a (massless) 
hypermultiplet in the fundamental of $USp$. The SW curve and differential for this model are \cite{SOP} 
\begin{equation}t^2x^2+ctx^2Q_{k}(x^2)+P_{N}(x^2)+\frac{\Lambda^bx^2}{t}=0;\quad\lambda_{SW}=\frac{x}{t}dt,\end{equation}
where $Q_k$ is a generic monic polynomial of degree k. As in the previous section $\Lambda$ can be identified with the $SO(2N)$ dynamical scale, 
whereas the $USp(2k)$ scale is proportional to $c^{-2}$. Actually, with the relation given above 
between k and N, when $N$ is even the $USp$ SQCD is asymptotically free and $c^{-2}$ is proportional to the dynamical scale. When N is odd the 
$USp$ theory is scale invariant and c is related to the marginal coupling (the flavor symmetry is $SO(2N+2)$ in both cases). 
If we send $\Lambda$ to zero, thus decoupling the $SO(2N)$ gauge multiplet, we are left with 
$USp(2k)$ SQCD with $N+1$ hypermultiplets in the fundamental. Sending now c to zero, in 
the first case this corresponds to sending the dynamical scale to infinity (thus scaling towards the singular point); in the second it is equivalent 
to taking the weak-coupling limit for the scale invariant theory. 

Now we reverse the order of the limits as before. When we set c to zero we recover equation (\ref{SAN}) in the case $p=2$. Turning then off the $SO(2N)$ 
gauge coupling ($\Lambda\rightarrow0$) we are left with the $D_2(SO(2N))$ models of \cite{infinitelymany}. For $N$ odd the theory is Lagrangian 
as we noted above and describes $USp(N-1)$ SQCD with $N+1$ flavors, which is the expected result.  
For N even it describes the maximally singular point of $USp(2k)$ SQCD with $N_f=N+1$. Once again the properties of the IR fixed point 
depend only on $N_f$. Using the technique of \cite{ShapereTachikawa} we can easily compute the central charges:
When $N$ is odd we get the scale invariant theory $USp(N-1)$ with $N_f=N+1$, so $$a=\frac{7N^2-5N-2}{48};\quad c=\frac{2N^2-N-1}{12}.$$
For $N$ even we find instead our singular point and the central charges are $$a=\frac{7N^2-5N-10}{48};\quad c=\frac{2N^2-N-2}{12}.$$

\medskip

A similar analysis can be carried over for all $D_p(G)$ systems, see appendix \ref{DpGAD}.

\subsection{The exceptional Minahan--Nemeshansky theories.}\label{exceptttt}
\subsubsection{$E_6$ MN $\equiv$ $D_2(SO(8))$.}
Notice that $D_2(SO(8))$ theory coincides with the $E_6$ MN theory: The $D_2(SO(8))$ quiver is mutation equivalent to the quiver in figure (6.11) of \cite{ACCERV2} we reproduce below on the \textsc{lhs}: It is sufficient to mutate it on one of the white nodes to show this result is true.
\be
\begin{gathered}
\begin{xy} 0;<0.3pt,0pt>:<0pt,-0.3pt>:: 
(200,0) *+{\bullet} ="0",
(350,440) *+{\bullet} ="1",
(500,440) *+{\bullet} ="2",
(0,300) *+{\circ} ="3",
(200,440) *+{\bullet} ="4",
(500,0) *+{\bullet} ="5",
(350,0) *+{\bullet} ="6",
(0,150) *+{\circ} ="7",
"0", {\ar"1"},
"0", {\ar"2"},
"3", {\ar"0"},
"0", {\ar"4"},
"7", {\ar"0"},
"1", {\ar"3"},
"5", {\ar"1"},
"6", {\ar"1"},
"1", {\ar"7"},
"2", {\ar"3"},
"5", {\ar"2"},
"6", {\ar"2"},
"2", {\ar"7"},
"4", {\ar"3"},
"3", {\ar"5"},
"3", {\ar"6"},
"5", {\ar"4"},
"6", {\ar"4"},
"4", {\ar"7"},
"7", {\ar"5"},
"7", {\ar"6"},
\end{xy}
\end{gathered}\xrightarrow{ \quad\mu_{\circ}\quad }\begin{gathered}
\xymatrix@R=1.5pc@C=1.5pc{&&&\bullet\ar[dddl]\\
\bullet\ar[rr]&&\circ\ar[ur]\ar[ddll]\ar[rr]&&\bullet\ar[ddll]\\
&&&&\\
\bullet\ar[rr]&&\circ\ar[rr]\ar[uull]\ar[dr]&&\bullet\ar[uull]\\
&&&\bullet\ar[uuul]}
\end{gathered}
\ee 
As a consistency check, let us now show that all the invariants of the $D_2(SO(8))$ model agrees with the ones of the $E_6$ MN theory. First of all notice that
\be
\d(2,D_4) = 2 \Longrightarrow  f(2,D_4) = \d(2,D_4) \cdot \varphi(2) = 2 
\ee
Therefore the rank of the flavor group of the $D_2(SO(8))$ model is $4+2 = 6$. From this we can recover the rank of $B$:
\be
\text{rank}\, B = 2 \cdot 4 - 6 = 2 \Longrightarrow \text{dimension of the Coulomb branch } = 1.
\ee
Now,
\be
c = \frac{1}{12} \big(4\cdot 6 + 2\big) = \frac{13}{6}
\ee
and
\be
4 \cdot u(2,D_4) = \Big[5-\frac{6}{2}\Big]_+ + 2 \Big[3-\frac{6}{2}\Big]_+ \Big[1-\frac{6}{2}\Big]_+ = 2.
\ee
Therefore,
\be
a = \frac{13}{12} + \frac{1}{8} + \frac{1}{2} = \frac{41}{24}.
\ee
Moreover, the Coulomb branch coordinate has dimension
\be
b_{2,D_4} = \frac{1}{2} \cdot 6 = 3,
\ee
in perfect agreement. 

\medskip

As a byproduct of this analysis our observations in the previous section give realizations of the $E_6$ MN theory as an IR fixed point of the following Lagrangian theories
\be
\framebox{$SO(8)$} - USp(2k) - \framebox{$SO(2)$} \qquad k \geq 2
\ee
where the boxes represent ungauged flavor groups ($N_f = 4 + 1$).

\medskip

In order to see that the SW curves match, one can proceed as follows: in the Gaiotto setting the $E_6$ theory is realized by compactifying the $A_2$ six-dimensional theory on a sphere with three maximal punctures (located at let's say $z=0,\lambda,\infty$). 
The SW curve is then 
\be
x^3=-\frac{uz}{(z-\lambda)^2};\quad\lambda_{SW}=\frac{x}{z}dz.
\ee
If we now take the limit $\lambda\rightarrow\infty$, we end up with a two-punctured sphere. The puncture at $z=0$ is unchanged whereas the puncture at infinity is now irregular, with a pole of order four for the cubic differential. The curve now becomes 
\be
x^3+uz=0;\quad \lambda_{SW}=\frac{x}{z}dz.
\ee
If we now multiply everything by $x^3$ and set $t=zx^3$ the curve and differential become 
\be
x^6+ut=0;\quad\lambda_{SW}=\frac{x}{t}dt.
\ee
The scaling dimensions of $x$, $u$ and $t$ are now respectively one, three and three. Since both terms appearing in the curve have dimension six we can add a term quadratic in $t$. The complete curve is then 
$$x^6+ut+t^2=0.$$ Setting $u$ to zero we recognize the curve describing  $D_2(SO(8))$  at the conformal point.

\subsubsection{$D_2(SO(10))$ is $E_7$ MN coupled to $SU(2)$ SYM}
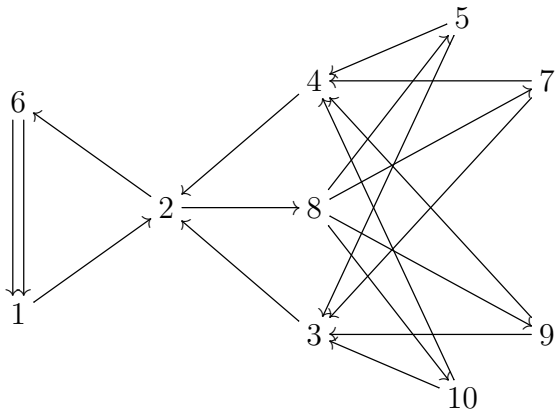
\begin{figure}
$$
\begin{gathered}
\begin{xy} 0;<0.8pt,0pt>:<0pt,-0.8pt>:: 
(0,140) *+{1} ="0",
(70,90) *+{2} ="1",
(140,150) *+{3} ="2",
(140,30) *+{4} ="3",
(210,0) *+{5} ="4",
(0,40) *+{6} ="5",
(250,30) *+{7} ="6",
(140,90) *+{8} ="7",
(250,150) *+{9} ="8",
(210,180) *+{10} ="9",
"0", {\ar"1"},
"5", {\ar@<0.4ex>"0"\ar@<-0.4ex>"0"},
"2", {\ar"1"},
"3", {\ar"1"},
"1", {\ar"5"},
"1", {\ar"7"},
"4", {\ar"2"},
"6", {\ar"2"},
"8", {\ar"2"},
"9", {\ar"2"},
"4", {\ar"3"},
"6", {\ar"3"},
"8", {\ar"3"},
"9", {\ar"3"},
"7", {\ar"4"},
"7", {\ar"6"},
"7", {\ar"8"},
"7", {\ar"9"},
\end{xy}
\end{gathered}
$$
\caption{Element in the quiver mutation class of $D_2(SO(10))$ describing the $S$-duality frame of Argyres--Seiberg \cite{argyresseiberg}. The full subquiver of type $A(1,1)$ on the nodes $\{ \, 1 , 6 \, \}$ represents the $SU(2)$ SYM subsector, and the node 2 represents the gauged $SU(2)$ flavor symmetry of $E_7$ MN.}\label{E7MNquiv}
\end{figure}

From our results of section \ref{SOSplagR} it follows that the theory $D_2(SO(10))$ is Lagrangian. In an $S$-duality frame, this is just the model $USp(4)$ coupled to 6 hypermultiplets in the fundamental representation, \emph{i.e.}
\be
\framebox{$SO(10)$} - USp(4) - \framebox{$SO(2)$}
\ee
where the boxes represents ungauged flavor groups ($N_f = 5 + 1$). One of the most famous examples of $S$--duality \cite{argyresseiberg} relates precisely this model with an $SU(2)$ SYM sector weakly gauging an $SU(2)$ subgroup of the flavor group of the $E_7$ MN model. By quiver mutations we are able to give an \emph{explicit proof} of this statement: In the mutation class of the BPS-quiver of $D_2(SO(10))$ there is an element that clearly describes an $SU(2)$ SYM sector weakly gauging the flavor symmetry of a subsystem that we identify with the $E_7$ MN one. We draw such quiver in figure \ref{E7MNquiv}. See appendix \ref{mutationMN} for the explicit sequence of mutations.
 
\medskip

From this result, it is easy to obtain the quiver for $E_7$ MN and from such quiver to prove that the $E_7$ theory has a finte BPS-spectrum, but this is out of the scope of the present project: The explicit computation can be found in \cite{CDZtoappear}.

\subsubsection{$D_2(E_6)$ is $E_8$ MN coupled to $SU(3)$ SYM}

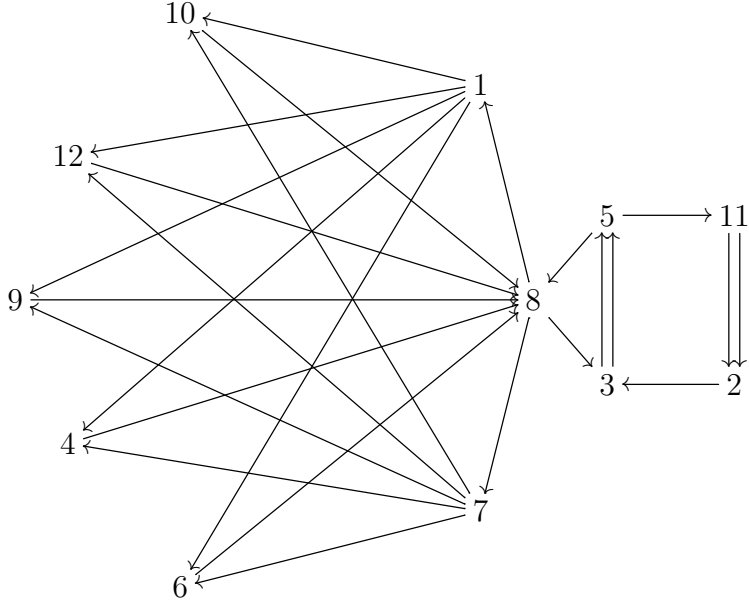
\begin{figure}
$$
\begin{gathered}
\begin{xy} 0;<0.8pt,0pt>:<0pt,-0.8pt>:: 
(220,34) *+{1} ="0",
(340,176) *+{2} ="1",
(280,176) *+{3} ="2",
(25,204) *+{4} ="3",
(280,96) *+{5} ="4",
(78,272) *+{6} ="5",
(220,236) *+{7} ="6",
(245,136) *+{8} ="7",
(0,136) *+{9} ="8",
(78,0) *+{10} ="9",
(340,96) *+{11} ="10",
(25,68) *+{12} ="11",
"0", {\ar"3"},
"0", {\ar"5"},
"0", {\ar"8"},
"0", {\ar"9"},
"0", {\ar"11"},
"7", {\ar"0"},
"1", {\ar"2"},
"10", {\ar@<0.4ex>"1"\ar@<-0.4ex>"1"},
"2", {\ar@<0.4ex>"4"\ar@<-0.4ex>"4"},
"7", {\ar"2"},
"6", {\ar"3"},
"3", {\ar"7"},
"4", {\ar"7"},
"4", {\ar"10"},
"5", {\ar"7"},
"7", {\ar"6"},
"6", {\ar"5"},
"6", {\ar"8"},
"6", {\ar"9"},
"6", {\ar"11"},
"8", {\ar"7"},
"9", {\ar"7"},
"11", {\ar"7"},
\end{xy}
\end{gathered}
$$
\caption{An element in the mutation-class of the quiver of the model $D_2(E_6)$. This quiver clearly represents an $S$-duality frame in which we have an explicit $SU(3)$ SYM sector coupled to the $E_8$ MN theory. The $SU(3)$ SYM full subquiver is on the nodes $\{ 2, 3, 5, 11 \}$. The node 8 represents the gauged $SU(3)$ flavor symmetry of $E_8$ MN.}\label{E8MNgauged}
\end{figure}

We have found how the exceptional MN theories of type $E_6$ and $E_7$ appear in between the $D_p(G)$ systems: The question, now, is if we can find also the $E_8$ MN theory. Such theory has rank 1 and it has $E_8$ flavor symmetry, therefore its charge lattice has dimension $10$. Since there are no $D_p(E_8)$ theories with such a small charge lattice, we expect that if the $E_8$ MN theory appears in between the $D_p(G)$ systems, it will manifest itself with part of its flavor symmetry weakly gauged. The first possibility we have is  the $E_7 \otimes SU(2) \subset E_8$ group, but gauging the $SU(2)$ symmetry will leave us with an $E_7$ flavor symmetry, and there is no $D_p(E_7)$ theories with $10 - 1 + 2 = 11$ nodes. The next possibility is the $E_6 \otimes SU(3) \subset E_8$ group, here we would be gauging the $SU(3)$ flavor symmetry subgroup, remaining with a $E_6$ flavor group. The theory would have $10 - 2 + 4 = 12$ nodes. And we have a theory with 12 nodes and $E_6$ flavor symmetry: It is precisely the $D_2(E_6)$ system!!

\medskip

Let us use our results about the central charges of the $D_2(E_6)$ theory to check if this prediction makes sense. Looking at table \ref{ADEcharacteristic} one obtains that $\d(2,E_6) = 0$ and therefore $f(2,E_6) = 0$. The rank of the quiver exchange matrix is simply
\be
\text{rank}\, B = 2 \cdot 6 - 6 = 6
\ee
Moreover $c_\text{eff} (D_2(E_6))= 6 \cdot 12 = 72$, therefore
\be
c(2,E_6) = \frac{1}{12} \big(72 + 6\big) = \frac{39}{6}.
\ee
Now, by additivity
\be
c(E_8 \text{ MN}) = c(2,E_6) - \frac{1}{6} \text{dim}\, SU(3) = \frac{39}{6} - \frac{8}{6}=\frac{31}{6}
\ee
which is the correct result!! Let us check that also $a$ is correct. We have
\begin{equation}\begin{split}\label{wwwpp}
u(2;E_6)&=\frac{1}{4}\Big([1-6]_++[4-6]_++[5-6]_++[7-6]_++[8-6]_++[11-6]_+\Big)=\\
&=\frac{1}{4}(1+2+5)=2\\
\end{split}\end{equation}
Then
\begin{equation}
a(2;E_6)=2+\frac{31}{12} + \frac{1}{16} \cdot 6 = \frac{45}{8}
\end{equation}
so,
\begin{equation}
a(E_8 \text{ MN})=\frac{45}{8}-\frac{5}{24}\,\dim SU(3)=\frac{95}{24}
\end{equation}
which is again the correct result !!!

\medskip

Based on these very strong evidences, we may try to find a representative of the mutation class of the BPS-quiver of $D_2(E_6)$ such that this result is manifest: We draw in figure \ref{E8MNgauged} such representative. This concludes our proof of the identification. The explicit mutation sequence is given in appendix \ref{mutationMN}.

\medskip

Again starting from our result one can easily obtain the explicit quiver for the $E_8$ MN theory. In reference \cite{CDZtoappear} starting from this result, a finite chamber for the BPS-spectrum of the $E_8$ MN theory was obtained.


\section*{Acknowledgments}
We thank Cumrun Vafa, Bernhard Keller, Clay Cordova, and Dan Xie for useful discussions. SC and MDZ thanks the center for the fundamental laws of nature of Harvard University for hospitality during the completion of this work.

\appendix

\section{Mutation sequences for the MN theories}\label{mutationMN}
The quiver for the model $D_2(SO(10))$ is simply
\be
\begin{gathered}
\begin{xy} 0;<0.8pt,0pt>:<0pt,-0.8pt>:: 
(0,160) *+{1} ="0",
(0,60) *+{2} ="1",
(70,160) *+{3} ="2",
(70,60) *+{4} ="3",
(140,160) *+{5} ="4",
(140,60) *+{6} ="5",
(170,0) *+{7} ="6",
(210,60) *+{8} ="7",
(210,160) *+{9} ="8",
(170,220) *+{10} ="9",
"0", {\ar"2"},
"3", {\ar"0"},
"2", {\ar"1"},
"1", {\ar"3"},
"2", {\ar"4"},
"5", {\ar"2"},
"4", {\ar"3"},
"3", {\ar"5"},
"6", {\ar"4"},
"7", {\ar"4"},
"4", {\ar"8"},
"4", {\ar"9"},
"5", {\ar"6"},
"5", {\ar"7"},
"8", {\ar"5"},
"9", {\ar"5"},
\end{xy}
\end{gathered}
\ee
By performing the sequence of mutations
\begin{center}
9 4 3 4 10 3 10 8 3 10 5 8 1 2 5
\end{center}
We obtain the element of the mutation class of the quiver of $D_2(SO(10))$ in figure \ref{E7MNquiv}. 

\medskip

The quiver for $D_2(E_6)$ is
\be
\begin{gathered}
\begin{xy} 0;<0.7pt,0pt>:<0pt,-0.7pt>:: 
(0,170) *+{1} ="0",
(80,170) *+{2} ="1",
(160,170) *+{3} ="2",
(240,170) *+{4} ="3",
(300,170) *+{5} ="4",
(240,210) *+{6} ="5",
(0,50) *+{7} ="6",
(80,50) *+{8} ="7",
(160,50) *+{9} ="8",
(240,50) *+{10} ="9",
(300,50) *+{11} ="10",
(240,0) *+{12} ="11",
"0", {\ar"1"},
"7", {\ar"0"},
"1", {\ar"2"},
"1", {\ar"6"},
"8", {\ar"1"},
"2", {\ar"3"},
"2", {\ar"5"},
"2", {\ar"7"},
"9", {\ar"2"},
"11", {\ar@<-0.6ex>"2"},
"3", {\ar"4"},
"3", {\ar"8"},
"10", {\ar"3"},
"4", {\ar"9"},
"5", {\ar@<0.6ex>"8"},
"6", {\ar"7"},
"7", {\ar"8"},
"8", {\ar"9"},
"8", {\ar"11"},
"9", {\ar"10"},
\end{xy}
\end{gathered}
\ee
The following sequence of mutations
\begin{quote}
9 12 4 9 10 12 6 12 3 11 1 9 10 9 12 10 9 12 6 3 12 9 6 5 6 9 10 12 1 7 3 1 7 8 4 8 1 2 1 8 1 4 8 2 8 4 3 5 7 4 7 5 7 1 7 11 7 4 1 8
\end{quote}
gives the quiver in figure \ref{E8MNgauged}.

\section{The BPS spectrum of the $\widehat{H}\boxtimes G$ models}\label{BPSpec}

The aim of the following section is to give a proof of the

\medskip
{\bf Main Claim.} All $\widehat{H}\boxtimes G$ models have a finite chamber at strong $G$ coupling consisting only of hypers. Moreover, in such a chamber, the cone of particles of the theory has the structure
\begin{equation}
\Gamma^+ = \bigoplus_{i = 1}^{\text{rk}(\widehat{H})} \Delta^+(G),
\end{equation}
and therefore it consists of $\tfrac{1}{2} (\text{rk}(\widehat{H})\times\text{rk}(G) \times h (G))$ hypermultiplets as stated in section 3.5 of \cite{infinitelymany}.

\medskip

\subsection{Sink--source sequences \cite{BGP}\!\!\cite{CNV,arnold1}: a lightning review}
Here we present a version of the sink-source sequence technique \emph{ad usum Delphini}: we are interested only in the computation of the quantum monodromy, $i.e.$ of the charges of BPS-particles\footnote{\ The charges of the BPS-antiparticles follows from PCT.} for a finite chamber consisting only of hypermultiplets. The interested reader is referred to the papers \cite{CNV,arnold1,ACCERV2} for a complete discussion and for the proof of our statements here.
\medskip

Let $\Gamma$ be the charge lattice of a $\cn=2$ theory with BPS-quiver property. Let $\{e_i\}_{i = 1, ... , 2r + f}$ be the basis of the charge lattice in the chamber we are computing the spectrum. Let $(Q,\cw)$ be the associated 2-acyclic quiver with potential. Following \cite{arnold1}, given a subset $S$ of the set of nodes $Q_0$, we introduce the notation $Q|_S$ to denote the full subquiver of $Q$ over the nodes $S$. Consider the node set $Q_0$ as the disjoint union of a family of sets $\{q_\a\}_{\a\in A}$:
\begin{equation}
Q_0 = \coprod_{\a \in A} q_\a
\end{equation}
To each subset of nodes $q_\a$ we associate the full subquiver $Q|_{q_\a}$ of $Q$. Given a node $i \in Q_0$, we will denote $q_{\a(i)}$ the unique element in the family that contains node $i$. Now, consider a finite sequence of nodes $\Lambda =  \{i(1),i(2),\dots, i(m)\}, i(\ell) \in Q_0$ such that:
\begin{itemize}
\item[$i)$] We allow repetitions in the node list $ \{i(1),i(2),\dots, i(m)\}$, but each node of $Q_0$ is present at least once.
\item[$ii)$] $i(\ell) \neq i(\ell+1)$
\item[$iii)$] Let $\mathbf{m}_{\Lambda} \equiv \mu_{i(m)} \circ \cdots \circ \mu_{i(1)}$ be the sequence of \emph{left} mutations associated to the sequence $\Lambda$. We require that
\begin{itemize}
\item[$a)$] At the quiver level, $\mathbf{m}_{\Lambda}(Q) = Q$, up to an order two permutation of the nodes $\sigma: Q_0 \to Q_0$, that may be the identity;
\item[$b)$] The action of $\mathbf{m}_{\Lambda}$ on the basis of the charge lattice is
\begin{equation}
\mathbf{m}_{\Lambda}(e_i)  = - e_{\sigma(i)}.
\end{equation}
\end{itemize}
\end{itemize}
By \cite{CNV,ACCERV2} any sequence of mutations satisfying these criteria computes the BPS spectrum in a finite chamber made only of hypermultiplets. Indeed, an elementary left mutation at node $k$ acts on the basis of charge lattice $\{e_i\}_i$ sending it to a new basis $\{\mu_k(e)_i\}_i$ such that 
\begin{equation}\label{lmutz}
\begin{aligned}
&\mu_k(e)_k \equiv - e_k\\
 &\mu_k(e)_j \equiv \begin{cases}
 e_j &\text{ if there are no arrows } k \to j \\ 
 e_j + m e_k & \text{ if there are } m \text{ arrows } k \to j
 \end{cases}
\end{aligned}
\end{equation}
And the phase-ordered charges of the BPS hypermultiplets in the chamber are:
\begin{equation}\label{ppppqty}
\begin{aligned}
& e_{i(1)} ,\\
& \mu_{i(1)}(e)_{i(2)} , \\
& \mu_{i(2)} \circ \mu_{i(1)} (e)_{i(3)} ,\\
&\,\,\, \vdots \\
& \mu_{i(m-1)}\circ\dots\circ \mu_{i(2)}\circ \mu_{i(1)}(e)_{i(m)}.
\end{aligned}
\end{equation}
A sequence $\Lambda$ that satisfies properties $i)-iii)$ is said to be \emph{source-factorized} of type $\{Q|_{q_\a}\}_{\a \in A}$ if
\begin{itemize}
\item[$iv)$] For all $\ell = 1,2,...,m$, the $\ell$-th node in the sequence $i(\ell)$ is a sink in 
\begin{equation}
\mu_{i(\ell-1)} \circ \cdots \circ \mu_{i(1)}(Q)\big|_{\{i(\ell)\} \cup Q_0 \setminus q_{\a(i(\ell))}}
\end{equation}
\item[$v)$] For all $\ell = 1,2,...,m$ the $\ell$-th node in the sequence $i(\ell)$ is a source in
\begin{equation}
\mu_{i(\ell-1)} \circ \cdots \circ \mu_{i(1)}(Q)\big|_{q_{\a(i(\ell))}}
\end{equation}
\end{itemize}
A source-factorized sequence is in particular \emph{Coxeter-factorized} of type $(Q|_{q_\a})_\a$, provided all $Q|_{q_\a}$ are Dynkin $ADE$ quivers with alternating orientation. In such a case, by \eqref{lmutz} combined with $iv),v)$, each element of the sequence corresponds to the action of the simple Weyl reflection 
\begin{equation}
s_{i(\ell)} \in \text{Weyl}(Q\big|_{q_{\a(i(\ell))}})
\end{equation}
on the charges on nodes $i \in q_{\a(i(\ell))}$, and as the identity operation on all other charges! Combining \eqref{ppppqty} with $iii)-v)$ and with proposition VI.\S.\,1.33 of \cite{BOURB:LIE}, in such a chamber, the cone of particles of the charge lattice results to be of the form
\begin{equation}
\Gamma^+ = \bigoplus_{\a \in A} \Delta^+(Q\big|_{q_\a})
\end{equation}
where by $\Delta^+(G)$ is meant the set of positive roots of  $G$. It is useful to remark that any quiver that admits in its mutation class a square product form of type $Q \square G$ admits Coxeter-factorized sequences of type $\{G^{\# Q_0}\}$. Our {\bf Main Claim} is reduced to the equivalent statement that

\medskip

{\bf Main Claim} (equivalent): All $\widehat{H}\boxtimes G$ models admit Coxeter-factorized sequences of type
\begin{equation}
\big( \underbrace{G, \dots , G}_{\text{rk}(\widehat{H}) \ \text{times}} \big).
\end{equation}

\subsection{All $\widehat{H}\boxtimes G$ models admit Coxeter-factorized sequences.}

\begin{figure}
\begin{center}
\begin{equation}\label{Apqquivs}
A(p,q) \colon \quad\begin{gathered}
\xymatrix@R=0.7pc@C=0.7pc{
&2\ar[r]&3\ar[r]&\dots\ar[r]&p\ar[dr]&\\
1\ar[ur]\ar[dr]&&&&&p+q\\
&p+1\ar[r]&\dots\ar[r]&p+q-2\ar[r]&p+q-1\ar[ur]&\\}
\end{gathered}
\end{equation}
\begin{equation}
\widehat{D}_p \colon \quad\begin{gathered}
\xymatrix{
1\ar[dr]&&&&&& p\\
&3\ar[r]&4\ar[r]&\dots\ar[r]&p-2\ar[r]&p-1\ar[ur]\ar[dr]&\\
2\ar[ur]&&&&&&p+1\\}
\end{gathered} 
\end{equation}
\begin{equation}
\widehat{E}_6\colon \quad\begin{gathered}
\xymatrix{
&&6&&\\
&&5\ar[u]&&\\
0\ar[r]&1\ar[r]&2\ar[r]\ar[u]&3\ar[r]&4}
\end{gathered}
\end{equation}
\begin{equation}
\widehat{E}_7\colon \quad\begin{gathered}
\xymatrix{
&&&7&&\\
0\ar[r]&1\ar[r]&2\ar[r]&3\ar[r]\ar[u]&4\ar[r]&5\ar[r]&6}
\end{gathered}
\end{equation}
\begin{equation}
\widehat{E}_8\colon \quad\begin{gathered}
\xymatrix{
&&&&&8&&\\
0\ar[r]&1\ar[r]&2\ar[r]&3\ar[r]&4\ar[r]&5\ar[u]\ar[r]&6\ar[r]&7}
\end{gathered}
\end{equation}
\end{center}
\caption{Our conventions on the nodes of the affine quivers.}\label{conventsaff}
\end{figure}

First we need to fix the notation. We collect in figure \ref{conventsaff} our conventions about the labelings of the nodes and orientations of the affine quivers. 

\medskip

{\bf Remark}: Let $a\searrow$ denote the sequence of nodes of $\widehat{H}$
$$ a\searrow \ \equiv \big\{1 \ ,\ 2 \ , \ \dots \ , \ \text{rk}(\hat{H}) - 2\ , \text{rk}(\hat{H})-1 \ ,\ \text{rk}(\hat{H}) \big\}.$$
The corresponding sequence of mutations 
\begin{equation}
\mathbf{\mu}_{a\searrow}(\hat{H}) \equiv \prod_{a\searrow} \mu_a = \mu_1 \circ \mu_2 \circ \cdots \circ \mu_{\text{rk}(\hat{H})-2} \circ \mu_{\text{rk}(\hat{H})-1} \circ \mu_{\text{rk}(\hat{H})}
\end{equation}
is a sequence of mutations on \emph{sinks} that satisfies properties $i),ii),iii)$. If we interpret them as right mutations, we obtain the minimal BPS chamber of the models associated to the affine quivers.

\medskip

Label the nodes of the $\widehat{H}\boxtimes G$ quiver, as 
\begin{equation}
(i,a), \qquad i=1,...,\text{rk}(\widehat{H}), \quad a=1,...,\text{rk}(G).
\end{equation}
Let
\begin{equation}
q_k \equiv \{ (k,a) \ \ | \  a=1,...,\text{rk}(G)\}, \qquad v_j \equiv \{ (i,j) \ \ | \ i = 1,...,\text{rk}(\widehat{H})  \}.
\end{equation}
We have
\begin{equation}
(\widehat{H}\boxtimes G)\big|_{q_k} =G\quad\text{and}\quad(\widehat{H}\boxtimes G)\big|_{v_j} = \widehat{H}.
\end{equation}

\begin{figure}

\begin{center}
\begin{equation}
\begin{aligned}
&\overleftrightarrow{A}_{2k+1} \colon \quad
\begin{gathered}
\xymatrix{1 \ar[r]&2&3\ar[r]\ar[l]&\dots&2k-1\ar[l]\ar[r]&2k&2k+1\ar[l]\\}
\end{gathered}\\
& \\
&\overleftrightarrow{A}_{2k} \colon \quad\begin{gathered}
\xymatrix{1 \ar[r]&2&3\ar[r]\ar[l]&\dots&2k-1\ar[l]\ar[r]&2k\\}
\end{gathered}\\
& \\
&\overleftrightarrow{D}_{2k+1}\colon \quad \begin{gathered}
\xymatrix{
&&&&2k+1&\\
1 \ar[r]&2&3\ar[r]\ar[l]&\dots&2k-1\ar[l]\ar[u]\ar[r]&2k}
\end{gathered}\\
& \\
&\overleftrightarrow{D}_{2k}\colon \quad \begin{gathered}
\xymatrix{
&&&&&2k\ar[d]&\\
1 \ar[r]&2&3\ar[r]\ar[l]&\dots&2k-3\ar[l]\ar[r]&2k-2&\ar[l]2k-1}
\end{gathered}
& \\
&\overleftrightarrow{E}_6 \colon \quad \begin{gathered}
\xymatrix{
&&6&&\\
1\ar[r]&2&3\ar[r]\ar[l]\ar[u]&4&\ar[l]5}
\end{gathered}
& \\
&\overleftrightarrow{E}_7 \colon \quad \begin{gathered}
\xymatrix{
&&&7\ar[d]&&\\
1\ar[r]&2&3\ar[r]\ar[l]&4&5\ar[l]\ar[r]&6}
\end{gathered}
& \\
&\overleftrightarrow{E}_8 \colon \quad \begin{gathered}
\xymatrix{
&&&&8&&\\
1\ar[r]&2&3\ar[r]\ar[l]&4&5\ar[l]\ar[r]\ar[u]&6&7\ar[l]}
\end{gathered}
\end{aligned}
\end{equation}

\end{center}

\caption{Our conventions about the alternating orientations of the Dynkin subquivers.}\label{conventsdyn}

\end{figure}

\medskip

{\bf Lemma 1.} \emph{Let $\overleftrightarrow{G}$ be the alternating orientation of the Dynkin quiver of type $G$ such that the first node is a source, all odd nodes are sources, and all even nodes are sinks --- see figure \ref{conventsdyn}. All quivers $\widehat{H} \boxtimes G$ are mutation equivalent to $\widehat{H} \boxtimes \overleftrightarrow{G}$.}

\medskip

{\bf Proof.} The proof is organized as follows: First we are going to consider the quivers of type $A_n$, then the quivers of type $D_n$ and finally the exceptionals. 

\medskip

Let $G=A_n$. The orientation for the $A_n$ quivers that we have used elsewhere is simply $(i) \to (i+1)$: When we will speak about the ``$A_n$ quiver'' or simply ``$A_n$'' we will always mean the $A_n$ quiver with this orientation. Let $\mathbf{m}_n$ denote the following sequence of elementary mutations of the $A_n$ quiver:
\begin{equation}
\mathbf{m}_k \equiv \mu_{k} \circ \mu_{k-1} \circ \dots \circ \mu_{2} \circ \mu_{1}.
\end{equation}
Notice that $\mathbf{m}_n = \text{id}_{A_n}$. The sequences of mutations
\begin{equation}
\begin{aligned}
\mathbf{m}_2 \circ \mathbf{m}_{4} \circ \dots \circ \mathbf{m}_{2(k-2)} \circ \mathbf{m}_{2(k-1)}\circ \mathbf{m}_{2k} &\qquad \text{for } n = 2k+1\\
\mathbf{m}_2 \circ \mathbf{m}_{4} \circ \dots \circ \mathbf{m}_{2(k-2)} \circ \mathbf{m}_{2(k-1)} &\qquad \text{for } n = 2k
\end{aligned}
\end{equation}
are sequences of mutations mapping $A_n$ to $\overleftrightarrow{A}_n$ involving only mutations on sources. By construction of the $\boxtimes$ operation on quivers, the sequence of mutations
\begin{equation}
\mathbf{\mu}_{a\searrow}(1) \equiv \prod_{(a,1)\searrow} \mu_{(a,1)} = \mu_{(1,1)} \circ \mu_{(2,1)} \circ \cdots \circ \mu_{(\text{rk}(\hat{H})-2,1)} \circ \mu_{(\text{rk}(\hat{H})-1,1)} \circ \mu_{(\text{rk}(\hat{H}),1)}
\end{equation}
is a mutation on sources for the dynkin subquivers of type $A_n$, because by definition each $(i,1)$ is a source of the Dynkin on the nodes $q_i$, and a mutation on sinks for the affine subquiver of type $\widehat{H}$ on the nodes $v_1$. Analogously, one can define
\begin{equation}
\mathbf{\mu}_{a\searrow}(i) \equiv \prod_{(a,i)\searrow} \mu_{(a,i)} = \mu_{(1,i)} \circ \mu_{(2,i)} \circ \cdots \circ \mu_{(\text{rk}(\hat{H})-2,i)} \circ \mu_{(\text{rk}(\hat{H})-1,i)} \circ \mu_{(\text{rk}(\hat{H}),i)},
\end{equation}
and
\begin{equation}
\mathbf{m}(\hat{H})_k \equiv \mu_{a\searrow}(k) \circ  \mu_{a\searrow}(k-1)\circ  \dots \circ \mu_{a\searrow}(2) \circ \mu_{a\searrow}(1).
\end{equation}
Then, by construction of the $\boxtimes$ operation on quivers, $\mathbf{m}(\hat{H})_k$ involves only mutations on sources with respect to the subquivers on the nodes $q_i$, and on sinks with respect to subquivers on the nodes $v_j$. Notice that
\begin{equation}
\mathbf{m}(\hat{H})_n = \text{id}_{\widehat{H}\boxtimes A_n}
\end{equation}
Our lemma for the case $G = A_n$ is equivalent to the following

\medskip

{\bf Claim. } \emph{The sequences of mutations 
\begin{equation}
\begin{aligned}
\mathbf{m}(\hat{H})_2 \circ \mathbf{m}(\hat{H})_{4} \circ \dots \circ \mathbf{m}(\hat{H})_{2(k-2)} \circ \mathbf{m}(\hat{H})_{2(k-1)}\circ \mathbf{m}(\hat{H})_{2k} &\qquad \text{for } n = 2k+1\\
\mathbf{m}(\hat{H})_2 \circ \mathbf{m}(\hat{H})_{4} \circ \dots \circ \mathbf{m}(\hat{H})_{2(k-2)} \circ \mathbf{m}(\hat{H})_{2(k-1)} &\qquad \text{for } n = 2k
\end{aligned}
\end{equation}
maps the quiver $\widehat{H}\boxtimes A_n$ into the quiver $\widehat{H}\boxtimes \overleftrightarrow{A}_n$.}

\medskip

We will proceed by induction on $n$. For $n=1,2$ there is nothing to prove:  Let us now show that $(n-1) \Rightarrow (n)$. The case $n=2k$ is trivial. One considers the subquiver $\widehat{H}\boxtimes A_{2k-1}$, on the nodes
$$\big\{(i,a) \ | \ i = 1,\dots, \text{rk}(\widehat{H}) \ , \ a = 1, \dots , 2k-1\big\}.$$
By inductive hypothesis the sequence of mutations 
\begin{equation}
\mathbf{m}(\hat{H})_2 \circ \mathbf{m}(\hat{H})_{4} \circ \dots \circ \mathbf{m}(\hat{H})_{2(k-2)}\circ \mathbf{m}(\hat{H})_{2(k-1)}
\end{equation}
maps this subquiver into $\widehat{H}\boxtimes \overleftrightarrow{A}_{2k-1}$. But then each subquiver on the nodes $q_i$ has the form
\begin{equation}
\xymatrix{1 \ar[r]&2&3\ar[r]\ar[l]&\dots&2k-1\ar[l]\ar[r]&2k}.
\end{equation}
And we are done. If $n = 2k+1$, analogously, consider the full $\widehat{H}\boxtimes A_{2k}$ subquiver of $\widehat{H}\boxtimes A_{2k+1}$ on the nodes
$$\big\{(i,a) \ | \ i = 1,\dots, \text{rk}(\widehat{H}) \ , \ a = 1, \dots , 2k\big\}.$$
By inductive hypothesis the sequence of mutations 
\begin{equation}
\mathbf{m}(\hat{H})_2 \circ \mathbf{m}(\hat{H})_{4} \circ \dots \circ \mathbf{m}(\hat{H})_{2(k-2)} \circ \mathbf{m}(\hat{H})_{2(k-1)}
\end{equation}
maps this subquiver into $\widehat{H}\boxtimes \overleftrightarrow{A}_{2k}$. Therefore each subquiver on the nodes $q_i$, now, looks like:
\begin{equation}
\xymatrix{1 \ar[r]&2&3\ar[r]\ar[l]&\dots&2k-1\ar[l]\ar[r]&2k\ar[r]&2k+1}
\end{equation}
Clearly, if we apply to the mutated quiver the sequence of mutations
\begin{equation}
\begin{aligned}
&\mathbf{\mu}_{a\nearrow}(2k+1)\equiv \prod_{(a,2k+1)\nearrow} \mu_{(a,2k+1)}\\
&= \mu_{(\text{rk}(\hat{H}),2k+1)} \circ \mu_{(\text{rk}(\hat{H})-1,2k+1)} \circ \mu_{(\text{rk}(\hat{H})-2,2k+1)} \circ \cdots \circ \mu_{(2,2k+1)} \circ \mu_{(1,2k+1)},
\end{aligned}
\end{equation}
we obtain the quiver of $\widehat{H}\boxtimes \overleftrightarrow{A}_{2k+1}$. Notice that, by construction, the sequence $\mathbf{\mu}_{a\nearrow}(2k+1)$ is on sources with respect to the $v_j$ subquivers and on sinks with respect to the $q_i$ subquivers. Our claim follows if we are able to show that
\begin{equation}
\begin{aligned}
& \mathbf{\mu}_{a\nearrow}(2k+1) \circ  \mathbf{m}(\hat{H})_2 \circ \mathbf{m}(\hat{H})_{4} \circ \dots \circ \mathbf{m}(\hat{H})_{2(k-2)} \circ \mathbf{m}(\hat{H})_{2(k-1)}\\
&= \mathbf{m}(\hat{H})_2 \circ \mathbf{m}(\hat{H})_{4} \circ \dots \circ \mathbf{m}(\hat{H})_{2(k-2)} \circ \mathbf{m}(\hat{H})_{2(k-1)}\circ \mathbf{m}(\hat{H})_{2k}.
\end{aligned}
\end{equation}
This equality follows easily from the fact that
\begin{equation}
\begin{aligned}
 &\mathbf{\mu}_{a\searrow}(2k+1) \circ  \mathbf{m}(\hat{H})_2 \circ \mathbf{m}(\hat{H})_{4} \circ \dots \circ \mathbf{m}(\hat{H})_{2(k-2)} \circ \mathbf{m}(\hat{H})_{2(k-1)}\circ \mathbf{m}(\hat{H})_{2k}\\
&= \mathbf{m}(\hat{H})_2 \circ \mathbf{m}(\hat{H})_{4} \circ \cdots \circ\mathbf{m}(\hat{H})_{2(k-2)} \circ\mathbf{m}(\hat{H})_{2(k-1)}\circ \mathbf{\mu}_{a\searrow}(2k+1) \circ \mathbf{m}(\hat{H})_{2k}\\
 &= \mathbf{m}(\hat{H})_2 \circ \mathbf{m}(\hat{H})_{4} \circ \dots \circ \mathbf{m}(\hat{H})_{2(k-2)} \circ \mathbf{m}(\hat{H})_{2(k-1)}\circ \mathbf{m}(\hat{H})_{2k+1}\\
 &=  \mathbf{m}(\hat{H})_2 \circ \mathbf{m}(\hat{H})_{4} \circ \dots \circ \mathbf{m}(\hat{H})_{2(k-2)} \circ \mathbf{m}(\hat{H})_{2(k-1)}.
 \end{aligned}
\end{equation}
Combined with 
\begin{equation}
\mathbf{\mu}_{a\nearrow}(2k+1) \circ \mathbf{\mu}_{a\searrow}(2k+1) = \text{id}_{\widehat{H}\boxtimes A_{2k+1}}
\end{equation}
that, in turn, follows by our definitions using the fact that $\mu_i \circ \mu_i = \text{id}_Q$ for elementary mutations.

\medskip

For $G=D_n$, the proof is similar. Consider the $\widehat{H}\boxtimes A_{n-3}$ subquiver on the nodes
\begin{equation}
\big\{(i,a) \ | \ a=1,...,n-3, \ i=1,..,\text{rk}(\widehat{H})\big\}.
\end{equation}
By our result about $G=A_n$, we know these are mutation equivalent to $\widehat{H}\boxtimes \overleftrightarrow{A}_{n-3}$, and by locality of mutations, the mutations sequences are the same as the one we have obtained previously. If $n=2 \ell +1$, then one has just to use
\begin{equation}
\mathbf{m}(\hat{H})_2 \circ \dots \circ \mathbf{m}(\hat{H})_{2(\ell-1)}
\end{equation}
If, instead, $n=2\ell$, then one needs to apply the mutation sequence
\begin{equation}
\mathbf{\mu}_{a\nearrow}(n)\circ \mathbf{\mu}_{a\nearrow}(n-1) \circ \mathbf{m}(\hat{H})_2 \circ \dots \circ \mathbf{m}(\hat{H})_{2(\ell-2)}.
\end{equation}
For $E_6$, $E_7$, and $E_8$ we have, instead, that the sequences are
\begin{equation}
\begin{aligned}
&E_6 \colon \quad \mathbf{\mu}_{a\searrow}(6)\circ\mathbf{m}(\hat{H})_2 \circ \mathbf{m}(\hat{H})_4\\
&E_7 \colon \quad \mathbf{\mu}_{a\searrow}(7)\circ\mathbf{m}(\hat{H})_2 \circ \mathbf{m}(\hat{H})_4\\
&E_8\colon \quad\mathbf{\mu}_{a\searrow}(8)\circ\mathbf{m}(\hat{H})_2 \circ \mathbf{m}(\hat{H})_4\circ \mathbf{m}(\hat{H})_6.
\end{aligned}
\end{equation}

\begin{flushright}
$\square$
\end{flushright}

{\bf Lemma 2.} All $\widehat{H}\boxtimes G$ models admit Coxeter-factorized sequences of type
\begin{equation}
\big( \underbrace{G, \dots , G}_{\text{rk}(\widehat{H}) \ \text{times}} \big).
\end{equation}

\medskip

{\bf Proof.} Take the representative $\widehat{H}\boxtimes \overleftrightarrow{G}$ in the mutation class. With the notations of the previous proof, the mutation sequences
\begin{equation}
\mathbf{Cox}_{\widehat{H},G} \equiv \prod_{j \text{ even}} \mu_{a\searrow}(j) \circ \prod_{k \text{ odd}} \mu_{a\searrow}(k) \qquad\text{for } G=A_n,E_6,E_7,E_8
\end{equation}
\begin{equation}
\mathbf{Cox}_{\widehat{H},D_{2\ell+1}} \equiv\mu_{a\searrow}(2\ell+1) \circ \prod_{j \text{ even}} \mu_{a\searrow}(j) \circ \prod_{k \text{ odd} \leq 2\ell-1} \mu_{a\searrow}(k)
\end{equation}
\begin{equation}
\mathbf{Cox}_{\widehat{H},D_{2\ell}} \equiv \prod_{j \text{ even}\leq 2\ell-2} \mu_{a\searrow}(j)\circ \mu_{a\searrow}(2\ell) \circ \prod_{k \text{ odd}} \mu_{a\searrow}(k)
\end{equation}
acts as the identity on the quiver $\widehat{H}\boxtimes \overleftrightarrow{G}$, and as the Coxeter element of $G$ on the charges of each $q_i$ subquiver for the lattice $\Gamma$. Indeed, the sequences are source-sink factorized: Each mutation in the sequences is on a node that is a sink with respect to the full subquivers on the nodes $v_j$, and a source on the subquivers on the nodes $q_i$. The full quantum monodromy of the $(\widehat{H}, G)$ model is
\begin{equation}
\mathbb{M}(q) = \text{Ad} (\widehat{\mathbf{Cox}}_{\widehat{H},G})^{h(G)}.
\end{equation}
where the hat means that we are considering the corresponding quantum mutations. 
\begin{flushright}
$\square$
\end{flushright}

\section{$D_p(G)$ theories and linear quiver gauge theories}\label{DpGAD}
\subsection{$D_p(SU(N))$ theories}

In order to generalize the above argument to the case $p>2$ it is convenient to slightly change perspective as follows:
consider the SW curve for $D_2(SU(N))$ theories $t^2+P_N(x)=0$. The coefficients of the polynomial $P_N$ are the mass parameters 
associated to the $SU(N)$ flavor symmetry of the theory and have canonical dimension, as remarked above. This implies in particular 
that $x$ has scaling dimension one and then, since all the terms appearing in the curve describing a SCFT should have the 
same scaling dimension, we deduce that $t^2$ should have dimension $N$. We can actually deform the above curve adding terms of 
the form $u_ktx^k$. The scaling dimension of the parameters $u_k$ can then be fixed imposing the condition $[u_k]+[t]+k[x]=N$, 
leading to the relation $[u_k]=N/2-k$. If $N$ is even the ``complete'' curve becomes $$t^2+t(x^{N/2}+\dots+u_{0})+P_N(x)=0,$$ 
which is precisely the SW curve for the scale invariant $SU(N/2)$ theory with $N$ flavors. For $N$ odd we get instead 
$$t^2+t(u_{(N-1)/2}x^{(N-1)/2}+\dots+u_0)+P_N(x)=0.$$ Assuming that all the coefficients 
$u_k$ have positive scaling dimension, terms involving higher powers of x necessarily have scaling dimension greater than N and can 
thus be discarded. This curve precisely coincides with (\ref{SUN1}) (modulo a trivial reparametrization) and correctly describes 
the infrared fixed point of $SU(k)$ SQCD ($k\geq\frac{N+1}{2}$) with $N$ flavors studied in the previous section, which coincides 
in turn with $D_2(SU(N))$. 

The above argument can be easily generalized to the $p>2$ case: the SW curve coming from (\ref{CC}) is $t^p+P_N(x)=0$ 
and we can turn on all possible deformations of the form $u_{ij}t^{p-j}x^{i}$. The only restriction comes from the requirement 
$[u_{ij}]\geq0$. In this way we will relate $D_p(SU(N))$ theories to IR fixed points of linear quivers of unitary gauge groups.
The scaling dimension of $u_{ij}$ can be easily evaluated: $t^p$ has dimension N, so $[t]=\frac{N}{p}$. Imposing 
then the condition $[u_{ij}]+(p-j)[t]+i[x]=N$ we immediately find
\begin{equation}\label{dim}
 [u_{ij}]=\frac{N}{p}j-i.
\end{equation}
We can now readily compute the quantity
$8a-4c$, which gives the effective number of vectormultiplets for the SCFT and can be evaluated using the formula
\begin{equation}\label{cariche}4(2a-c)=\sum_{ij}(2[u_{ij}]-1),\end{equation} where the sum involves all operators whose scaling dimension is larger than one. Using 
(\ref{dim}) this can be rewritten as $$\sum_{j=1}^{p-1}\sum_{i}\left(2\frac{N}{p}j-2i-1\right).$$ We have discarded the contribution 
from $u_{i0}$'s because they are mass parameters associated to the $SU(N)$ symmetry, rather than Coulomb branch operators. Let 
us first evaluate the second summation at fixed j from $i=0$ to $i=\lfloor\frac{N}{p}j\rfloor-1$ (we denote with $\lfloor\;\rfloor$
the integer part): 
$$\sum_{i=0}^{\lfloor\frac{N}{p}j\rfloor-1}\left(2\frac{N}{p}j-2i-1\right)=\left\lfloor\frac{N}{p}j\right\rfloor\left(2\frac{N}{p}j-
\left\lfloor\frac{N}{p}j\right\rfloor\right).$$ Performing now the sum over j does not always lead to the right result for the 
following reason: when $\frac{N}{p}j$ is integer the above summation includes the contribution from a parameter $u_{ij}$ whose 
scaling dimension is one. This will happen whenever N and p are not coprime. 
These should always be regarded as mass parameters, implying the enhancement of the flavor symmetry from the naive $SU(N)$
and their contribution should be discarded in the summation. This can be done simply subtracting $\text{gcd}\{N,p\}-1$ from the 
above formula. The final result is then 
\begin{equation}
8a-4c=\sum_{j=1}^{p-1}\left\lfloor\frac{N}{p}j\right\rfloor\left(2\frac{N}{p}j-
\left\lfloor\frac{N}{p}j\right\rfloor\right)-\text{gcd}\{N,p\}+1. 
\end{equation}
This formula clearly reproduces the expected result for $D_p(SU(N))$ theories, since the scaling dimension of Coulomb branch operators 
are exactly the same.
We also recover the result found in \cite{infinitelymany} that the rank of the flavor symmetry for $D_p(SU(N))$ theories is $N+\text{gcd}\{N,p\}$.

We would now like to make some comments about the linear quiver theories associated to $D_p(SU(N))$ SCFTs. As already remarked, 
there is not a unique choice for the rank of the gauge groups and obviously, once a candidate linear quiver has been found, we 
are free to enlarge the rank of the gauge groups since the Coulomb branch of the first theory can be regarded as a submanifold of 
the Coulomb branch of the second. One natural question is then: What is the minimal \footnote{We mean that the sum 
of the ranks of the gauge groups attains the minimum value.} choice? In order to answer this question we must distinguish two cases: 
p greater and smaller than N.

\begin{figure}
\centering{\includegraphics[width=10cm]{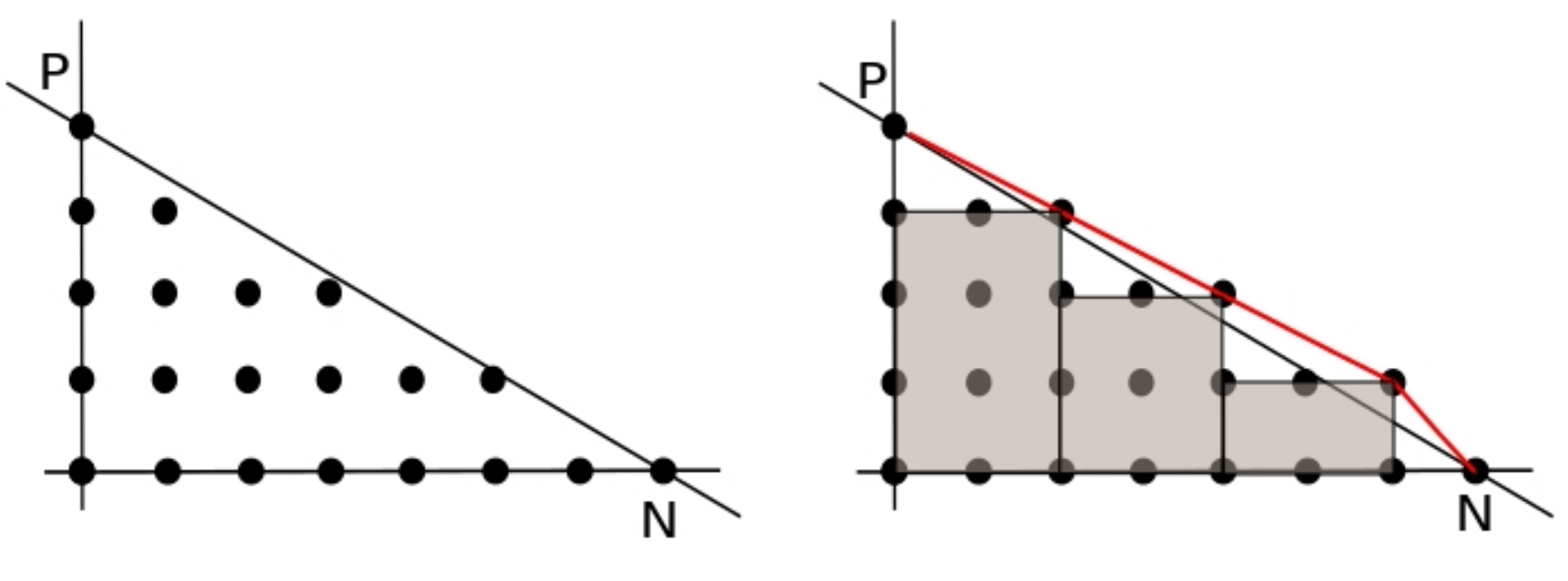}} 
\caption{\label{dotdiagram}\emph{On the left we have the dot diagram associated to $D_{p}(SU(N))$ theory (with $p=4$ and $N=7$). 
The black dots indicate all terms appearing in the SW curve. On the right we have the dot diagram associated to the minimal linear 
quiver having $D_p(SU(N))$ as an infrared fixed point. We can easily see that the linear quiver is $SU(2)-SU(4)-SU(6)-\boxed{7}$. 
Only the $SU(6)$ group is asymptotically free.}}
\end{figure}

For $p<N$ the linear quiver has $p-1$ gauge groups. The first n groups, where $n=N(\text{mod}\, p)$, are 
$$SU\left(\left\lfloor\frac{N+p}{p}\right\rfloor k\right),\quad k\leq n.$$ The remaining $p-n-1$ gauge groups are 
$$SU\left(\left\lfloor\frac{N+p}{p}\right\rfloor n+\left\lfloor\frac{N}{p}\right\rfloor j\right),\quad j\leq p-n-1.$$ 
In order to show this let us draw a diagram on the plane as in Figure \ref{dotdiagram} on the left, in which the term $x^it^j$ 
is represented by a dot located at the point with coordinates $(i,j)$. This is very similar to the Newton polygon used in 
\cite{XIE2}, which is not surprising since we have identified our theories (for $G=A_N$) with the models discussed in that paper. 
It is easy to see that all the terms associated with points lying on the straight line passing through $(N,0)$ and $(0,p)$ have 
dimension $N$ (we mean that $i[x]+j[t]=N$), and that the straight lines parallel to it identify lines of constant dimension in the above 
sense. It is then clear that the dots located at points with integer coordinates in the interior of the triangle depicted 
in Figure \ref{dotdiagram} correspond to all the terms entering in the SW curve associated to $D_p(SU(N))$ theory. 

The SW curve of any linear quiver of $SU(N)$ gauge groups has the form $$t^k+\sum_{i,j}u_{ij}t^{k-j}x^{M-i}+x^M=0,$$ and indeed can be 
represented on our diagram. The x-coordinate of the rightmost dot of each row just counts the number of colors of the corresponding 
gauge group. If we connect all rightmost dots as in Figure \ref{dotdiagram} (on the right) we obtain a polygon, and 
the requirement that all the gauge groups in the theory are asymptotically free or conformal is simply equivalent to its 
convexity (just because the number of flavors for any gauge group can be at most twice the number of colors). A given linear quiver will have
$D_p(SU(N))$ as an infrared fixed point only if its associated polygon contains the triangle identified by the straight line 
passing through $(N,0)$ and $(0,p)$ (for our purposes we can assume that $k=p$ and $M=N$) as in the figure. 

The minimal quiver can now be found simply identifying the polygon which satisfies the above requirements and has minimal area. 
In order to see this just consider the grey region in Figure \ref{dotdiagram} on the right. Its area is clearly equal to the rank of the theory 
plus $p-1$ and the area of the whole polygon can be obtained adding the contribution of the various triangles, which is equal 
to $N/2$. It is easy to see that the minimal polygon has exactly four edges (except when $p$ divides $N$) as in the figure and their slope lead to the formula 
given above. Notice that all the groups in the linear quiver but one are conformal. The only asymptotically free group is the one 
associated to the dot at which the two edges meet (see the figure).

If N is a multiple of p the theory is Lagrangian and we already know the answer. Notice anyway that our formula works in this case
as well, predicting that the quiver contains the gauge groups $SU(Nk/p)$, with $k\leq p-1$. We thus recover the expected result for 
lagrangian theories. In the limiting case $N=p$, the above rules give a quiver which formally starts with a $SU(1)$ group. The 
corresponding term in the SW curve is $t^{N-1}(x+u_{01})$ and $u_{01}$ has scaling dimension one so, as we have explained above, 
it should be interpreted as a mass parameter. This term in the curve should then be regarded as describing a hypermultiplet in the fundamental 
of the subsequent gauge group, namely $SU(2)$. We thus get the linear quiver $$\boxed{1}-SU(2)-SU(3)-\dots-SU(N-1)-\boxed{N},$$ 
as expected.

For $p>N$ we can just apply a similar argument. In this case it is important to realize that what really matters is that the polygon 
associated to the linear quiver contains all the dots associated to $D_p(SU(N))$. It is not really necessary that it contains 
the whole triangle. In the $p<N$ case these conditions are just equivalent. Since the edges of the polygon are either vertical or 
have slope smaller than one, the minimal polygon is built adding a vertical line and one with slope one.

The minimal linear quiver thus contains a tail which is identical to $D_N(SU(N))$ (apart from the doublet of $SU(2)$). The 
remaining part of the quiver is a sequence of $SU(N)$ gauge groups with a hypermultiplet in the bifundamental between neighbouring 
groups: $$SU(2)-SU(3)-\dots-SU(N-1)-SU(N)-\dots-SU(N)-\boxed{N}.$$ The doublet of $SU(2)$ at the beginning is present only if the 
SW curve admits the term $t^{p-k}(x+u_{0k})$, which occurs if and only if p is a multiple of N. In any case the number of gauge 
groups is $p-1-\lfloor p/N\rfloor$. A simple check is in order: for $N=2$ we find a linear quiver of $SU(2)$ groups. For $p=2n-1$ 
there are $n-1$ gauge groups and two doublets at one end. For $p=2n$ the number of gauge groups is the same but we also have an 
extra doublet at the other end of the quiver. Since $D_p(SU(2))$ coincides with the $D_{p}$ AD theory, we precisely 
recover the result found in \cite{Xie:generalAD}.

\subsection{$D_p(SO(2N))$ theories}

Similar considerations allow to identify $D_p(SO(2N))$ theories with IR fixed points of linear quivers with alternating SO and 
USp gauge groups and half-hypermultiplets in the bifundamental between neighbouring groups. Since the argument is anologous to 
the one given for $SU(N)$, we will be more sketchy.
For $G=D_N$ the curve can be written as $$t^{p}+\frac{P_N(x^2)}{x^2}=0,$$ where $P_N(x^2)=x^{2N}+\dots+u_N^2$. We can then 
add terms of the form $u_{ab}x^{2a}t^{p-b}$. The mass parameters associated to the $SO(2N)$ flavor symmetry will have canonical 
dimension only if $[x]=1$. This implies $$[t]=\frac{2N-2}{p};\quad [u_{ab}]=\frac{2N-2}{p}b-2a.$$ This matches precisely the 
dimension of Coulomb branch operators of $D_p(SO(2N))$. Equation (\ref{cariche}) then implies that the above curve reproduces 
the correct effective number of vectormultiplets $2a-c$.

The theory can be lagrangian only if all $u_{ab}$'s have even dimension. This constraint will be satisfied whenever
$(2N-2)/p$ is even, reproducing the expected result. We can determine precisely what the theory is simply collecting all terms 
with $[u_{ab}]=0$. All parameters satisfying this relation are simply combinations of the marginal couplings, rather than operators.
We then find the curve $$t^{p}+\sum_{k=1}^{p-1}a_kx^{km}t^{(p-k)}+x^{2N-2}=0,$$ where $pm=2N-2$. We thus precisely recover the 
curve we have found before.
A term $x^{2n-2}$ corresponds either to 
$USp(2n-2)$ or to $SO(2n)$, leading to the lagrangian theories (the number inside the boxes indicate the hypermultiplets in the fundamental)
\begin{equation}\begin{aligned}
{\bf p}\; \bf\text{{\bf even}}:\quad &{\boxed N}- USp(2N-2-m) - SO(2N-2m) -\dots-USp(m)-{\boxed 1},\nonumber\\
{\bf p}\; \bf\text{{\bf odd}}: \quad &{\boxed N}- USp(2N-2-m) - SO(2N-2m) -\dots-SO(m+2).\nonumber
\end{aligned}\end{equation}

In order to find the minimal quivers containing $D_p(SO(2N))$ as an IR fixed point we can construct the dot diagram as before, 
inserting for example a dot at the point $(i,j)$ if the curve includes the term $u_{ij}t^jx^{2i}$. The vertices of the 
triangle will then be located at $(0,p)$, $(N-1,0)$ (and obviously $(0,0)$).
Since one needs  respectively $2n+2$ or $2n-2$ hypermultiplets in the {\bf 2n} to make a $USp(2n)$ or $SO(2n)$ gauge group conformal, 
the condition for UV completeness (as opposed to IR freedom) at all the nodes is again that the associated polygon should be convex.
All the arguments given for $SU(N)$ apply also in this case, so we give directly the result.

For $p<N$ the dots lying on the perimeter of the polygon are associated to terms of the form $t^{p-i}x^{k}$ where k is 
$$2i\left(\left\lfloor\frac{N-1}{p}\right\rfloor+1\right),$$ for $i\leq (N-1)\text{mod}\, p$ and $$2i\left\lfloor\frac{N-1}{p}\right\rfloor+ 
2\left\lbrace\frac{N-1}{p}\right\rbrace p$$ otherwise (the braces indicate the fractional part). Collecting all these terms 
we find the SW curve describing a linear quiver with alternating $SO$-$USp$ gauge groups. One can easily reconstruct it explicitly from these data. 

For $p>N-1$ the polygon has a vertical edge and one of slope one as in the $SU(N)$ case. The quiver starts with a tail of the form 
$$\boxed{N}-USp(2N-2)-SO(2N)-USp(2N-2)-\dots$$ and ends in one of the following two ways
\begin{equation}\begin{aligned}
\dots-USp(10)-SO(10)-USp(6)-SO(6)-SU(2)-\boxed{1},\nonumber\\
\dots-USp(12)-SO(12)-USp(8)-SO(8)-USp(4)-SO(4).\nonumber
\end{aligned}\end{equation}
In this case the number of gauge groups in the quiver is $p-\lfloor \frac{p}{N-1}\rfloor$. If this number is odd the first option is the 
correct one, otherwise the quiver terminates as in the second sequence.
One can easily check that all the groups in the tails are conformal, except the group associated to the dot at which the two edges 
meet as for $SU(N)$.

\end{document}